\documentclass[a4paper,twocolumn,amsmath,amssymb,groupedaddress,superscriptaddress,nofootinbib,allowtoday,accepted=2026-03-17]{quantumarticle}
\pdfoutput=1

\usepackage[utf8]{inputenc}
\usepackage[english]{babel}
\usepackage[T1]{fontenc}

\usepackage[normalem]{ulem}

\usepackage{tikz}
\usepackage{amsmath,amsthm,amsfonts,amssymb}
\usepackage{mathtools}
\usepackage{verbatim}

\usepackage{hyperref}

\usepackage{dsfont}
\usepackage{xcolor}
\usepackage{graphicx}
\usepackage{caption}
\usepackage[usestackEOL]{stackengine}
\usepackage{hhline}
\usepackage{multirow}

\usepackage{xurl}
\usepackage[square,numbers]{natbib}
\hypersetup{colorlinks=true, citecolor=blue,linkcolor=blue,filecolor=blue,urlcolor=blue}

\usepackage{tikz}
\usetikzlibrary{quantikz}
\usetikzlibrary{external}

\usepackage{soul,xcolor}
\setstcolor{red}

\newcommand{\mbf}{\mathbf}
\newcommand{\mbb}{\mathbb}
\newcommand{\mc}{\mathcal}

\newcommand{\tr}[1]{\textrm{Tr}\left[ #1 \right]}
\newcommand{\ptr}[2]{\textrm{Tr}_{#1} \left[ #2 \right]}
\newcommand{\id}{\textrm{id}}
\newcommand{\ip}[1]{\langle#1\rangle}
\newcommand{\op}[2]{|#1\rangle\langle #2|}

\newcommand{\1}{\mathds{1}}
\newcommand{\conv}[1]{\text{Conv}\left(#1\right)}
\newcommand{\cost}[1]{\text{Cost}\big(#1\big)}
\newcommand{\gain}[1]{\text{Gain}\big(#1\big)}
\newcommand{\Dim}[1]{\text{Dim}\left(#1\right)}

\newcommand{\Net}{\text{Net}}

\newcommand{\Env}{\text{Env}}
\newcommand{\Rx}{\text{Rx}}
\newcommand{\Tx}{\text{Tx}}

\newcommand{\Pbf}{\mbf{P}}

\newcommand{\Gbf}{\mbf{G}}
\newcommand{\Vbf}{\mbf{V}}
\newcommand{\Fbf}{\mbf{F}}

\newcommand{\Amc}{\mc{A}}
\newcommand{\Bmc}{\mc{B}}
\newcommand{\Cmc}{\mc{C}}
\newcommand{\Emc}{\mc{E}}
\newcommand{\Hmc}{\mc{H}}
\newcommand{\Pbb}{\mbb{P}}
\newcommand{\Zbb}{\mbb{Z}}

\newcommand{\Cbb}{\mbb{C}}
\newcommand{\Qbb}{\mbb{Q}}
\newcommand{\Vbb}{\mbb{V}}
\newcommand{\Fbb}{\mbb{F}}
\newcommand{\Xmc}{\mc{X}}
\newcommand{\Ymc}{\mc{Y}}
\newcommand{\Zmc}{\mc{Z}}

\newcommand{\ANet}{\vec{\Amc}}
\newcommand{\BNet}{\vec{\Bmc}}
\newcommand{\CNet}{\vec{\Cmc}}

\newcommand{\XNet}{\vec{\Xmc}}
\newcommand{\YNet}{\vec{\Ymc}}
\newcommand{\ZNet}{\vec{\Zmc}}
\newcommand{\Av}{\vec{A}}
\newcommand{\Bv}{\vec{B}}
\newcommand{\Cv}{\vec{C}}

\newcommand{\Success}{\text{Success}}
\newcommand{\Error}{\text{Error}}

\newcommand{\av}{\vec{a}}
\newcommand{\bv}{\vec{b}}
\newcommand{\cv}{\vec{c}}
\newcommand{\dv}{\vec{d}}

\newcommand{\xv}{\vec{x}}
\newcommand{\yv}{\vec{y}}
\newcommand{\zv}{\vec{z}}
\newcommand{\reals}{\mbb{R}}
\newcommand{\Ibb}{\mbb{I}}

\newcommand{\EA}{\text{EA}}
\newcommand{\ERx}{\text{ERx}}
\newcommand{\ETx}{\text{ETx}}
\newcommand{\GEA}{\text{GEA}}

\newcommand{\IF}{\text{IF}}
\newcommand{\CIF}{\text{CIF}}
\newcommand{\BF}{\text{BF}}
\newcommand{\HG}{\text{HG}}

\newcommand{\CV}{\text{CV}}

\newcommand{\PM}{\text{PM}}

\newcommand{\BT}{\text{BT}}
\newcommand{\XtoY}{\Xmc\text{\stackon[1pt]{$\to$}{$\scriptstyle d$}} \Ymc}

\newcommand{\sigarb}[3]{#1 \text{\stackon[1pt]{$\to$}{$\scriptstyle #2$}} #3}
\newcommand{\CSigd}{\Cbb^{\XtoY}}

\newcommand{\MA}{\text{MA}}
\newcommand{\BC}{\text{BC}}

\theoremstyle{definition}

\newtheorem{theorem}{Theorem}

\newtheorem*{remark}{Remark}

\newtheorem{algorithm}{Algorithm}
\newtheorem{protocol}{Protocol}

\definecolor{prep_green}{HTML}{669933}
\definecolor{proc_red}{HTML}{CC3333}
\definecolor{meas_blue}{HTML}{3366CC}
\definecolor{quantum_purple}{HTML}{663366}
\definecolor{classical_gray}{HTML}{666666}

\usetikzlibrary{shapes,decorations,arrows,calc,arrows.meta,fit,positioning,external}
\tikzset{
    terminal/.style={},
    source/.style ={ellipse, draw, minimum width = 0.5 cm, color=classical_gray, fill=classical_gray!10, text=black},
    qsource/.style ={ellipse, draw, minimum width = 0.5 cm, color=prep_green, fill=prep_green!20, text=black},
    dev/.style={rectangle, rounded corners, draw, minimum width = 0.7 cm, color=classical_gray, fill=classical_gray!10, text=black},
    prep_dev/.style={dev, color=prep_green, fill=prep_green!20, text=black},
    proc_dev/.style={dev, color=proc_red, fill=proc_red!20, text=black},
    meas_dev/.style={dev, color=meas_blue, fill=meas_blue!20, text=black},
    el/.style = {align=left},
    meas_gate/.style={color=meas_blue, fill=meas_blue!20},
    prep_gate/.style={color=prep_green, fill=prep_green!20},
    proc_gate/.style={color=proc_red, fill=proc_red!20},
    arb_gate/.style={color=quantum_purple, fill=quantum_purple!20},
    classical_gate/.style={color=classical_gray, fill=classical_gray!20},
    meas_gate_group/.style={dashed,rounded corners,color=meas_blue},
    proc_gate_group/.style={dashed,rounded corners,color=proc_red},
    prep_gate_group/.style={dashed,rounded corners,color=prep_green},
}

\newcommand{\cedge}{edge[double, line width=1pt, double distance=1.5pt, arrows = {-Latex[length=0.5pt 2.5 0]}]}
\newcommand{\qedge}{edge[line width=2pt, arrows = {-Latex[length=6pt 1.5 0]}]}
 
\definecolor{cool_green}{rgb}{0.0, 0.5, 0.0}

\newcommand{\rev}[2]{{#1}} 

\begin{document}

\title{An Operational Framework for Nonclassicality in Quantum Communication Networks}

\author{Brian Doolittle}
\affiliation{Aliro Technologies, Inc., Brighton, Massachusetts, 02135, USA}
\affiliation{Department of Physics, University of Illinois Urbana-Champaign, Urbana, Illinois, 61801, USA}

\author{Felix Leditzky}
\affiliation{Department of Mathematics, University of Illinois Urbana-Champaign, Urbana, Illinois, 61801, USA}

\author{Eric Chitambar}
\affiliation{Department of Electrical and Computer Engineering, University of Illinois Urbana-Champaign, Urbana, Illinois, 61801, USA}


\begin{abstract}
    Quantum resources, such as entanglement or quantum communication, offer significant communication advantages in information processing.
    We develop an operational framework for realizing these communication advantages in resource-constrained quantum networks.
    The framework computes linear bounds on the input/output probabilities of classical networks with limited communication and globally shared randomness.  Since the violation of these classical bounds witnesses nonclassicality, a measurable communication advantage, the framework maximizes the violation of the classical bound using variational quantum optimization methods tailored to the communication network and quantum resources. This operational framework for nonclassicality can be scaled on quantum computers or deployed in the field to optimize noisy quantum networks for communication advantages.
    Applying this framework, we investigate the nonclassicality of communication networks that are assisted by quantum resources.  We find that entanglement between communication-constrained parties is sufficient for nonclassicality to be found, whereas in networks with multiple senders, quantum communication with no entanglement-assistance is sufficient for nonclassicality to be found. As a result, entanglement is necessary for nonclassicality when a single sender broadcasts to multiple receivers.
\end{abstract}

\maketitle

\section{Introduction}

Quantum networks promise to revolutionize science and technology by enhancing communications and distributed information processing with quantum resources, such as entanglement and quantum communication \cite{Kimble2008quantum_internet,vanmeter2012_quantum_networking,wehner2018quantum_internet,singh2021_quantum_internet}.
The \rev{communication}{distributed information processing} advantage \rev{of quantum resources}{ that a quantum communication resource provides} is often quantified by the communication complexity, which specifies how the amount of communication resources needed for an information processing task scales with the problem size.
When compared with classical protocols, quantum resources enable polynomial or even exponential communication complexity improvements \cite{Brassard2003_communication_complexity,buhrman2010_communication_complexity_nonloclaity}. However, these quantum advantages often require large amounts of fault-tolerant quantum resources, which are unavailable in practice \cite{wehner2018quantum_internet}.
Thus, it is crucial to develop practical approaches for realizing \rev{communication}{distributed information processing} advantages in resource-constrained quantum networks. 

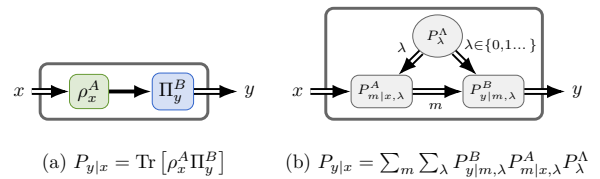
\begin{figure}[b]
    \centering
    \resizebox{\columnwidth}{!}{
    \begin{tabular}{c c}
        \begin{tikzpicture}
            \draw[rounded corners, classical_gray,style={line width=1.5pt}] (-0.125,-0.5) rectangle (2.875, 0.5);
            \node[terminal] (x1) at (-0.5,0) {$x$}; 
            \node[prep_dev] (A) at (0.75,0) {$\rho^A_{x}$};
            \node[meas_dev] (C) at (2.25,0) {$\Pi^B_y$};
            \node[terminal] (y) at (3.625, 0) {$y$};
            \path (x1) \cedge (A);
            \path (A) \qedge node[el, above=2pt, xshift=0pt] {} (C);

            \path (C) \cedge (y);
        \end{tikzpicture}&\begin{tikzpicture}
            \draw[rounded corners, classical_gray,style={line width=1.5pt}] (-0.25,-0.5) rectangle (3.625, 1.5);
            
            \node[terminal] (x1) at (-0.75,0) {$x$}; 
            \node[dev] (A) at (0.75,0) {$\scriptstyle P^A_{m|x,\lambda}$};
            \node[source] (Lambda) at (1.75,1.0) {$\scriptstyle P^{\Lambda}_{\lambda}$};
            \node[dev] (C) at (2.75,0) {$\scriptstyle P^B_{y|m,\lambda}$};
            \node[terminal] (y) at (4.25, 0) {$y$};
        
            \path (x1) \cedge (A);
            \path (A) \cedge node[el, below=2pt, xshift=-1pt] {$\scriptstyle m$} (C);
            \path (Lambda) \cedge node[el, above=2pt, xshift=-4pt] {$\scriptstyle \lambda$} (A);
            \path (Lambda) \cedge node[el, above=2pt, xshift=18pt] {$\scriptstyle \lambda\in\{0,1\dots\}$} (C);
            \path (C) \cedge (y);
        \end{tikzpicture}
        \\\\
        (a)  $P_{y|x} = \tr{\rho^A_x \Pi^B_y}$ &
        (b) $P_{y|x} = \sum_{m}\sum_{\lambda}P^B_{y|m,\lambda}P^A_{m|x,\lambda} P^\Lambda_{\lambda}$ 
    \end{tabular}
    }
    \caption{Classical communication channels with input $x\in\Xmc$, output $y\in\Ymc$, and input-output correlations $P_{y|x}$.\\
    (a) $A$ encodes $x$ into a qudit state $\rho^A_x\in D(\Hmc_d)$, sends it over a quantum channel, and $B$ applies the POVM $\{\Pi^B_y\}_{y\in\Ymc}$ to obtain $y$.
    (b) $A$ encodes $x$ into a message $m\in\{0,\dots,d-1\}$, sends it over a  classical channel, and $B$ decodes the message to obtain $y$. The globally shared randomness source $\Lambda$ correlates $A$ and $B$ with a shared random value $\lambda$.
    }
    \label{fig:sig-dim-channels}
\end{figure}

An alternative quantifier of quantum advantage in distributed processing is the classical simulation cost, which is the minimum number of bits of classical communication needed to simulate the input-output behavior of a quantum network. 
In information-theoretic terms, this problem is known as the channel simulation problem \cite{Bennett2002_rev_shannon, Winter-2002a, Bennett-2014a}, and this work focuses on the zero-error version of this problem \cite{cubit2011_nonlocal_assisted_cc}.
For instance, Holevo's celebrated result states that the classical capacity of any $d$-dimensional quantum channel cannot exceed $\log d$ \cite{holevo1973bounds}. 
More recently, Frenkel and Weiner proved a stronger statement in which the classical input-output data generated using a $d$-dimensional quantum channel (Fig.~\ref{fig:sig-dim-channels}.a) can be simulated with zero error using $\log d$ bits and shared randomness between the sender and the receiver (Fig.~\ref{fig:sig-dim-channels}.b)  \cite{Frenkel2015_classical_information_n-level_quantum_system}.

When additional senders, receivers, and processing nodes are added to the basic signaling scenario in Fig.~\ref{fig:sig-dim-channels}, the input-output correlations generated using quantum resources cannot always be reproduced by simply replacing the quantum channels with $d$-dimensional classical channels, removing all entanglement, and adding globally shared randomness. That is, classical channels with capacity greater than $d$ are necessary to simulate the quantum systems with zero error, implying that quantum resources allows certain input-output correlations to be generated more efficiently. We refer to this improved communication efficiency as a \textit{quantum communication advantage}.

This notion of quantum communication advantage is called quantum nonclassicality \cite{Bowles2015_nonclassicality_communication_networks}, which extends Bell nonlocality \cite{bell1964epr,brunner2014nonlocality} to causal models with bounded communication.
In this framework, a linear function called a nonclassicality witness is applied to the input-output
correlations produced by a quantum network. If the nonclassicality witness outputs a value that exceeds the classical bound, then the considered quantum communication network cannot be simulated with zero error using an analogous set of classical communication resources, thereby an explicit quantum advantage is witnessed.
This \rev{network}{} nonclassicality framework builds on prior \rev{investigations of }{work investigating quantum advantages in the} nonsignaling scenarios \cite{brunner2014nonlocality,Massar-2001a, bacon2003_bell_inequalities_aux_communication, Toner2003_simulating_bell_correlations, Regev-2010a, Maxwell2014_bell_inequality_aux_comm, Brask2017_bell_scenario_comm, ZambriniCruzeiro2019, Alimuddin-2023a}, point-to-point communication channels \cite{Frenkel2015_classical_information_n-level_quantum_system, Frenkel2022ea_signaling_dim_violation, DallArno_no_hypersignaling_2017, Heinosaari-2019a, heinosaari2020communication, Poderini2020_prep_meas_nonclassicality, doolittle_2021_certify_classical_cost, Tavakoli-2021a, Renner-2023a}, random-access codes \cite{Grudka-2015a, Tavakoli-2015a, Heinosaari2022, Piveteau-2022a, Sakharwade-2023a}, and more complex scenarios having multiple senders and receivers \cite{Bowles2015_nonclassicality_communication_networks,Poderini2020_prep_meas_nonclassicality,Lauand2023_causal_structure_3_variables}. 

\rev{To evaluate the communication advantage of a quantum network, we use variational quantum optimization (VQO) methods \cite{Cerezo2021_vqa} to maximize the violation of a nonclassicality witness tailored to the network. VQO methods can maximize nonclassicality in noisy communication networks and can operate in the presence of hardware noise \cite{doolittle_2023_vqo_nonlocality,doolittle_qNetVO, doolittle_phd_thesis_2023,chen_2022_inferring_network_topology,Ma2021_noisy_parameter-shift}. Since VQO methods can run on quantum hardware, they offer a practical approach for optimizing both deployed quantum networks, and simulations of quantum networks on either classical or quantum processors. }{Thus, VQO methods show promise of practical application in both simulated and deployed quantum networks. }

\rev{Our work broadly characterizes the communication advantage of quantum resources in multipoint communication networks.}{}
\rev{To this end, we develop an operational framework for maximizing the quantum communication advantages by combining the nonclassicality framework of Bowles \textit{et al.}~\cite{Bowles2015_nonclassicality_communication_networks}  with the VQO framework for quantum networks of Doolittle \textit{et al.}~\cite{doolittle_2023_vqo_nonlocality,doolittle_qNetVO, doolittle_phd_thesis_2023}.}{In this work }
\rev{Our main contributions include:
\begin{itemize}
    \item \textit{Operational Framework}: We develop a scalable and hardware-agnostic framework for optimizing quantum communication networks that are either deployed in the field, or simulated on quantum or classical processors. We introduce ``simulation games'' as a family of nonclassicality witnesses that can be efficiently derived and optimized to establish practical communication protocols.
    \item \textit{Novel Examples of Nonclassicality:} We numerically survey the nonclassicality and noise robustness of all basic communication networks and quantum resource configurations. We obtain numerous examples of nonclassicality in basic communication and distributed processing tasks. We detail interesting examples of entanglement-assisted protocols, such as dense bitwise XOR in quantum multiaccess networks (Protocol~\ref{protocol:bitwise_xor_behavior}), trit equality in classical multiaccess networks (Protocol~\ref{protocol:multiaccess_cmac_etx}), or  maximal nonclassicality in quantum broadcast networks. (Protocol~\ref{protocol:earx_broadcast_strategy}).
    \item \textit{Conditions for Nonclassicality:} We present three conditions for nonclassicality in communication-constrained networks: 1) Entanglement shared between two or more network nodes is sufficient for nonclassicality, 2) multiple senders in a quantum communication networks with no entanglement is sufficient for nonclassicality, and 3) Entanglement is necessary to achieve nonclassicality in broadcast networks (Theorem~\ref{thm:broadcast_classicality}).
\end{itemize}
}{}

Our paper is structured as follows.
In the methods (Section~\ref{section:methods}), we formalize our operational framework for nonclassicality in quantum communication networks.
In the results (Section~\ref{section:results}), we apply our framework to characterize the nonclassicality of a wide range of networks and quantum resource configurations, including bipartite communication scenarios, multiaccess networks, broadcast networks, and multipoint networks.
We conclude with a discussion of our results and open questions in Section~\ref{section:discussion}.

\section{Methods}\label{section:methods}

\rev{
We now detail our operational framework for nonclassicality in quantum communication networks.
Section~\ref{section:communication_networks} describes our approaches for modeling communication networks.
Section~\ref{section:witnessing_nonclassicality-body} formalizes nonclassicality and how quantum advantage can be witnessed. Section~\ref{section:maximizing_quantum_nonclassicality} describes our VQO methods for maximizing the nonclassicality of a quantum communication network.  Overall, our framework presents a practical approach for realizing quantum communication advantages, enabling its application in both experimental and theoretical domains of quantum information science and technology.
}{}

\subsection{Communication Networks}\label{section:communication_networks}

\rev{This section introduces our semi-device-independent framework for characterizing communication networks. We model communication networks by their stochastic behaviors (Section~\ref{section:network_behaviors}), which is constrained by their causal structure and communication resources (Section~\ref{section:causal_structure_resources}). We discuss the constraints imposed by different resource configurations, including classical communication (Section~\ref{section:classical_networks}), quantum communication (Section~\ref{section:quantum_networks}), and entanglement-assisted communication (Section~\ref{section:entanglement-assisted-networks}). }{}

\subsubsection{Communication Network Behaviors}\label{section:network_behaviors}

\begin{figure}[b!]
    \centering
    \resizebox{0.9\columnwidth}{!}{%
    \begin{tabular}{l}
        (a)  \\
        \begin{tikzpicture}
            \node[terminal] (xv) at (1.4, 3.5) {$\xv$};
            \node[terminal] (yv) at (5.4,3.5) {$\yv$};
            \node[dev, style={minimum width=2cm, minimum height=1cm}] (BB) at (3.4,3.5) {\large $P^{\Net}_{\yv|\xv}$};
            
            \node[terminal] (x1) at (-0.5,1.5) {$x_1$};
            \node[terminal] (x2) at (-0.5,-0.5) {$x_2$};    
            \node[dev] (A1) at (1.2,1.5) {$P^{A_1}_{a_1a_2|x_1}$};
            \node[dev] (A2) at (1.2,-0.5) {$P^{A_2}_{a_3a_4|x_2}$};
            \node[dev] (B) at (3.4, 0.5) {$P^B_{b_1b_2|a_2a_3}$};
            \node[dev] (C1) at (5.6,1.5) {$P^{C_1}_{y_1|a_1b_1}$};
            \node[dev] (C2) at (5.6,-0.5) {$P^{C_2}_{y_2|a_4b_2}$};
            \node[terminal] (y1) at (7.5, 1.5) {$y_1$};
            \node[terminal] (y2) at (7.5, -0.5) {$y_2$};
            \node[terminal] (Av) at (1.2, -1.5) {$\Pbf^{\Av}$};
            \node[terminal] (Bv) at (3.4,-1.5) {$\Pbf^{\Bv}$};
            \node[terminal] (Cv) at (5.6,-1.5) {$\Pbf^{\Cv}$};

            \draw[classical_gray, dashed, style={line width=1.5pt}] (2.5,3) -- (0,2.25);
            \draw[classical_gray, dashed, style={line width=1.5pt}] (4.25,3) -- (6.6,2.25);

            \draw[rounded corners, classical_gray,style={line width=1.5pt}] (0,-1.15) rectangle (6.6,2.25);
            
            \path (xv) \cedge (BB);
            \path (BB) \cedge (yv);
            \path (x1) \cedge (A1);
            \path (x2) \cedge (A2);
            \path (A1) \cedge  node[el, below=2pt, xshift=-10pt] {$a_2$} (B);
            \path (A2) \cedge node[el, below=-10pt, xshift=-10pt] {$a_3$} (B);
            \path (B) \cedge  node[el, below=2pt, xshift=8pt] {$b_1$} (C1);
            \path (B) \cedge node[el, above=2pt, xshift=8pt] {$b_2$} (C2);
            \path (C1) \cedge   (y1);
            \path (C2) \cedge (y2);
            \path (A1) \cedge node[el, above=2pt, ] {$a_1$} (C1);
            \path (A2) \cedge node[el, below=2pt] {$a_4$} (C2);
        \end{tikzpicture} \\
        (b) \\
        \begin{tikzpicture}
            \node[terminal] (xv) at (1.4, 3.5) {$\xv$};
            \node[terminal] (yv) at (5.4, 3.5) {$\yv$};
            \node[dev, style={minimum width=2cm, minimum height=1cm}] (BB) at (3.4, 3.5) {\large $P^{\Net}_{\yv|\xv}$};
            
            \node[terminal] (x1) at (-0.5,1.5) {$x_1$};
            \node[terminal] (x2) at (-0.5,-0.5) {$x_2$};    
            \node[prep_dev] (A1) at (1.2,1.5) {$\rho^{A_1}_{x_1}$};
            \node[prep_dev] (A2) at (1.2,-0.5) {$\rho^{A_2}_{x_2}$};
            \node[proc_dev] (B) at (3.4, 0.5) {$\Emc^{B}$};
            \node[meas_dev] (C1) at (5.6,1.5) {$\Pi^{C_1}_{y_1}$};
            \node[meas_dev] (C2) at (5.6,-0.5) {$\Pi^{C_2}_{y_2}$};
            \node[terminal] (y1) at (7.5, 1.5) {$y_1$};
            \node[terminal] (y2) at (7.5, -0.5) {$y_2$};

            \draw[classical_gray, dashed, style={line width=1.5pt}] (2.5,3) -- (0,2.25);
            \draw[classical_gray, dashed, style={line width=1.5pt}] (4.25,3) -- (6.6,2.25);

            \draw[rounded corners, classical_gray,style={line width=1.5pt}] (0,-1.15) rectangle (6.6,2.25);
            
            \path (xv) \cedge (BB);
            \path (BB) \cedge (yv);
            \path (x1) \cedge (A1);
            \path (x2) \cedge (A2);
            \path (A1) \qedge  node[el, below=-3pt, xshift=-16pt] {$\rho^{a_2}$} (B);
            \path (A2) \qedge node[el, above=-3pt, xshift=-16pt] {$
            \rho^{a_3}$} (B);
            \path (B) \qedge  node[el, above=-7pt, xshift=-16pt] {$\rho^{b_1}$} (C1);
            \path (B) \qedge node[el, below=-7pt, xshift=-16pt] {$\rho^{b_2}$} (C2);
            \path (C1) \cedge   (y1);
            \path (C2) \cedge (y2);
            \path (A1) \qedge node[el, above=2pt] {$\rho^{a_1}$} (C1);
            \path (A2) \qedge node[el, below=2pt] {$\rho^{a_4}$} (C2);
            \node[terminal] (Av) at (1.2, -1.5) {$\rho^{\Av}_{\xv}$};
            \node[terminal] (Bv) at (3.4,-1.5) {$\Emc^{\Bv}$};
            \node[terminal] (Cv) at (5.6,-1.5) {$\Pi^{\Cv}_{\yv}$};
        \end{tikzpicture}
        \end{tabular}%
        }
    \caption{Directed acyclic graph (DAG) depicting the causal structure of a communication network. Double-lined arrows show classical communication and single-lined arrows show quantum communication. (a) Classical network with classical devices (gray). (b) Quantum network with preparation (green), processing (red), and measurement (blue) devices.
    }
    \label{fig:communication_network_in_black_box}
\end{figure}
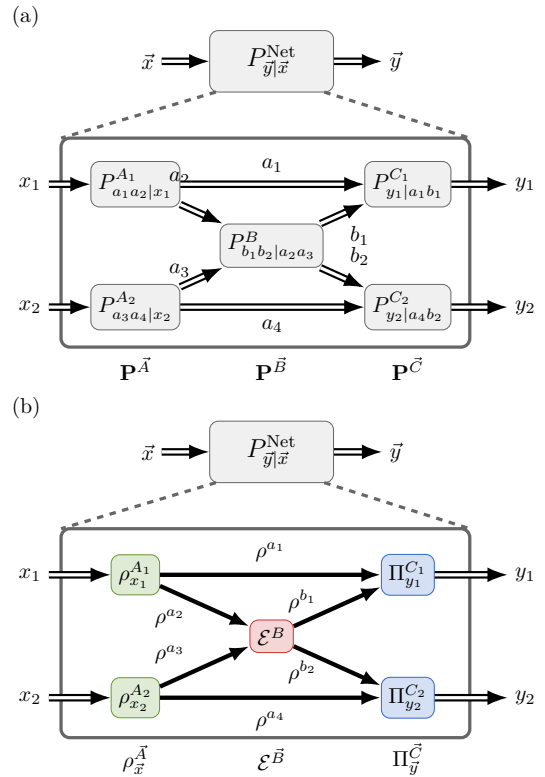

At the highest level of abstraction, a communication network is treated as a discrete and memoryless channel $\Pbf \colon \XNet \to \YNet$, which we refer to as a ``black box'' (see Fig.~\ref{fig:communication_network_in_black_box}.a).
An observer can query the black box with an input string $\vec{x}\in\vec{\mc{X}}=\Xmc_1\times \dots\times\Xmc_{M}$ to obtain the output string $\vec{y}\in\vec{\mc{Y}}=\Ymc_1\times \dots\times\Ymc_{N}$ where $\Xmc_i = \{0,1,\dots |\Xmc_i| - 1\}$ is a discrete set of finite length and (similarly for $\Ymc_i$).
Moreover, an observer can estimate the black box's transition probabilities $P_{\yv|\xv}$ by querying it many times with an input $\xv$ drawn uniformly at random.

We characterize a communication network's \textit{behavior} as the column stochastic matrix
\begin{equation}
    \Pbf \equiv \sum_{\yv\in\YNet}\sum_{\xv\in\XNet} P_{\yv|\xv}\op{\yv}{\xv}
\end{equation}
where the transition probabilities $P_{\vec{y}|\vec{x}}$ satisfy nonnegativity and normalization constraints.
For given input and output sets, $\XNet$ and $\YNet$, we refer to the set of all column stochastic matrices as the \textit{probability polytope}
\begin{align}\label{eq:full_probability_polytope}
    \Pbb_{\YNet|\XNet} \equiv\Big\{  \Pbf \in \reals^{|\YNet|\times|\XNet|}_{\geq0}\Big| \sum_{\yv\in\YNet}P_{\yv|\xv} = 1\; \forall \; \xv\in\XNet\Big\}.
\end{align}

The probability polytope in Eq.~\eqref{eq:full_probability_polytope} is equivalently defined as the convex hull of a set of extreme points or vertices, $\Pbb_{\YNet|\XNet} = \text{Conv}\big(\Vbb_{\YNet|\XNet}\big)$.
The vertices precisely correspond to deterministic behaviors 
\begin{align}\label{eq:set_of_deterministic_behaviors}
    \Vbb_{\YNet|\XNet} \equiv \big\{ \Vbf \in \Pbb_{\YNet|\XNet}\big|\; V_{\yv|\xv} = \delta_{\yv,f(\xv)} \; \forall \;  \xv\in\XNet \big\}
\end{align}
where $f : \XNet \to \YNet$ is a deterministic function and $\delta_{i,j}$ is the Kronecker delta.

\subsubsection{Causal Structure and Communication Resources}\label{section:causal_structure_resources}

A communication network's causal structure and communication resources are represented by a directed acyclic graph (DAG) that describes a set of nodes (devices) and edges (one-way communication).
For the general case, we denote the network DAG as $\Net$ where we use alternative labels when referring to specific networks.
The behavior $\Pbf^{\Net}\in \Pbb_{\XNet|\YNet}$ of a communication network is constrained by the network's causal structure and the amount of communication allowed along each edge.  

The nodes of a network DAG represent devices that are organized in a sequence of time steps or \textit{layers} $(\Av,\Bv,\Cv,\dots)$ where each layer contains a collection of nodes $\Av = (A_1,\dots,A_{M})$ that model $M$ independent devices that simultaneously process their local information.
The edges of a network DAG represent noiseless one-way $d$-dimensional communication channels $\id^{\Tx \to \Rx}_d$ from a sender device $(\Tx)$ to a receiver device ($\Rx$) where the sender's layer must precede the receiver's to maintain their causal ordering. 

A communication network is fully specified as $\Net(\sigarb{\XNet}{\vec{d}}{\YNet})$ where we fix the input and output sets, $\XNet$ and $\YNet$, and the amount of communication in each edge, $\vec{d} = (d_1,\dots,d_{K})$.   Following the language of \cite{DallArno_no_hypersignaling_2017, doolittle_2021_certify_classical_cost}, we refer to the value $d_i$ of edge $i$ in the network as its \textit{signaling dimension}. For quantum networks, $d_i$ is the Hilbert space dimension of the corresponding channel whereas for classical networks, $d_i$ refers to the number of noiseless messages that can be transmitted across the channel.  From a simulation perspective, these two values are operationally equivalent since any behavior generated using the transmission of $\log d$ qubits in a point-to-point channel can also be generated using the transmission of $\log d$ bits and shared randomness between the sender and receiver \cite{Frenkel2015_classical_information_n-level_quantum_system} (see Fig.~\ref{fig:sig-dim-channels}).

When an unbounded amount of globally shared randomness (GSR) is available to all devices, a communication network can produce any convex combination of its network behaviors
\begin{equation}\label{eq:behavior_shared_randomness}
    \Pbf^{\Net} = \sum_{\lambda}\Pbf^{\Net}_{\lambda} P^{\Lambda}_{\lambda}.
\end{equation}
In each shot of the network,  a shared random value $\lambda\in\{0,1,2,\dots\}$ is broadcast from a source $\Lambda$ to all network devices with probability $P^{\Lambda}_{\lambda}$, and then $\Pbf^{\Net}_{\lambda}$ is the network behavior performed given $\lambda$.
As a consequence of Eq.~\eqref{eq:behavior_shared_randomness}, the set of communication network behaviors is convex whenever GSR is available.

It is important that an unbounded amount of GSR is available to all devices in our framework because it allows a fair comparison between classical and quantum communication in our framework. Since classical shared randomness can be generated and distributed with ease,  the availability of GSR is the least restrictive assumption that can be placed on classical and quantum networks. GSR also plays an important role in the simulation of quantum communication networks because without it, an unbounded amount of classical communication is needed to reproduce one qubit of communication in the simple case of the point-to-point network \cite{heinosaari_unbounded_advantage2024} (see also Theorem 13 of reference \cite{doolittle_phd_thesis_2023}). It remains an active area of research to consider relaxations of globally shared randomness in networks \cite{Tavakoli-2022a}.

\subsubsection{Classical Communication Networks}\label{section:classical_networks}

In a classical communication network $\Net(\sigarb{\XNet}{\vec{d}}{\YNet})$, a classical message of length $\log d_i$ bits is communicated along the $i^{th}$ edge of the network DAG.
The network DAG gives an explicit decomposition for the network behavior $\Pbf^{\Net}$ where a classical network's transition probabilities decompose as
\begin{equation}\label{eq:classical_network_behavior_decomposition}
    P^{\Net}_{\yv|\xv}= \sum_{\vec{v}\in\vec{\mc{V}}}\dots\sum_{\cv\in\CNet}\sum_{\bv\in\BNet}\sum_{\av\in\ANet} P^{\vec{W}}_{\yv|\vec{v}} \cdots P^{\Cv}_{\cv|\bv}  P^{\Bv}_{\bv|\av}P^{\Av}_{\av|\xv}
\end{equation}
for all $\xv\in\XNet$ and $\yv\in\YNet$.
Since the devices in each layer are independent, it holds that $P^{\Av}_{\av|\xv} = \prod_{j} P^{A_j}_{a_j|x_j}$.
Each device in the DAG is modeled as a black box, which could perform any mapping on its local data, \textit{e.g.}, $\Pbf^{B_j} \in \Pbb_{\Bmc_j|\Amc_j}$ (see Eq~\eqref{eq:full_probability_polytope}).
The deterministic behaviors of the network are given by \cite{Bowles2015_nonclassicality_communication_networks},
\begin{equation} \label{eq:deterministic_behavior}
    \Vbb^{\Net} \equiv \{\Vbf \in \Vbb_{\YNet|\XNet}\; | \; V_{\yv|\xv}\;\text{satisfies Eq.~\eqref{eq:classical_network_behavior_decomposition}}  \}.
\end{equation}
A deterministic behavior $\Vbf \in \Vbb^{\Net}$ is achieved whenever all devices perform a deterministic function on its local data.  Since we permit GSR, the set of all classical network behaviors is then defined as the convex polytope
\begin{equation}\label{eq:classical_network_polytope}
    \Cbb^{\Net} = \conv{\Vbb^{\Net}},
\end{equation}
which we refer to as the \textit{classical network polytope}.

\subsubsection{Quantum Communication Networks}\label{section:quantum_networks}

The causal structure and communication resources of a quantum communication network are represented by its network DAG, $\Net(\sigarb{\XNet}{\dv}{\YNet})$.
Along the $i^{th}$ single-lined edge of the DAG, a $d_i$-dimensional quantum state is communicated over the noiseless quantum channel $\id^{\Tx_i \to \Rx_i}_{d_i}\colon D(\Hmc^{\Tx_i}_{d_i}) \to D(\Hmc^{\Rx_i}_{d_i})$ where $\Tx_i$, $\Rx_i$, and $d_i$ respectively are the sending device, receiving device, and signaling dimension. Moreover, we assume that GSR is available.

The quantum communication networks considered in this work can be decomposed into three main layers, as shown in Fig.~\ref{fig:quantum_communication_network_entanglement}.a:

\begin{enumerate}
    \item \textbf{Preparation:} $\rho_{\xv_1}^{\Av} = \bigotimes_{i=1}^{N_1} \rho_{x_{1,i}}^{A_i}$, each device locally encodes a classical input $x_{1,i}\in\Xmc_{1,i}$ into a quantum state $\rho^{A_i}_{x_{1,i}}\in D(\Hmc^{A_i})$.
    \item \textbf{Processing:} $\Emc^{\Bv}_{\xv_2} = \bigotimes_{j=1}^{N_2}\Emc^{B_j'\to B_j}_{x_{2,j}}$, each device receives a quantum state $\rho^{B_j'}\in D(\Hmc^{B_j'})$ and applies the completely-positive trace-preserving (CPTP) map, $\Emc^{B_j'\to B_j}_{x_{2,j}}\colon D(\Hmc^{B_j'}) \to D(\Hmc^{B_j})$, to obtain the output state $\rho^{B_j}_{x_{2,j}} = \Emc^{B_j'\to B_j}_{x_{2,j}}(\rho^{B_j'})$.
    \item \textbf{Measurement:} $\Pi^{\Cv}_{\yv|\xv_3} = \bigotimes_{k=1}^{N_3} \Pi^{C_k}_{y_k|x_{3,k}}$, each device receives a quantum state $\rho^{C_k}\in D(\Hmc^{C_k})$ and measures it using the positive operator-valued measure (POVM) $\{\Pi^{C_k}_{y_k|x_{3,k}}\}_{y_k\in\Ymc_k}$ to produce a classical output $y_k\in \Ymc_k$.
\end{enumerate}

Without loss of generality, the first layer contains preparation devices, the last layer contains measurement devices, while intermediate network layers can contain devices of any type, which is necessary to model the local operations and classical communication (LOCC) \rev{or local operations and quantum communication (LOQC) of entanglement-assisted networks (see Fig.~\ref{fig:quantum_communication_network_entanglement}.d-e and Section~\ref{section:entanglement-assisted-networks}).}{}

The resulting transition probabilities are calculated using the Born rule as
\begin{equation}\label{eq:quantum_network_probabilities}
    P_{\yv|\xv} = \tr{\Pi^{\Cv}_{\yv|\xv_3}\mc{S}^{\Bv\to\Cv}\circ \Emc^{\Bv}_{\xv_2}\circ \mc{S}^{\Av\to\Bv}(\rho^{\Av}_{\xv_1})}
\end{equation}
where the noiseless channels $\mc{S}^{\Av\to\Bv}$ and $\mc{S}^{\Bv\to\Cv}$ ensure that the outputs of one layer are correctly mapped to the input of the next.
Given the DAG $\Net(\sigarb{\XNet}{\dv}{\YNet})$, the set of all quantum network behaviors is defined as
\begin{equation}\label{eq:set_of_quantum_network_behaviors}
    \Qbb^{\Net} \equiv \conv{\{ \Pbf \in \Pbb_{\YNet|\XNet}\;|\; P_{\yv|\xv}\; \text{satisfies Eq.~\eqref{eq:quantum_network_probabilities}}\}}
\end{equation}
where $\Qbb^{\Net}$ is convex due to the presence of GSR.

\subsubsection{Communication Networks with Entanglement}\label{section:entanglement-assisted-networks}

\begin{figure}
    \centering
    \small
    \resizebox{\columnwidth}{!}{%
    \begin{tabular}{c}
    {\normalsize (a) Quantum Network Layers} \\
    \begin{tikzpicture}
            \node[terminal] (x1) at (-0.75,0) {$\xv_1$}; 
            \node[terminal] (x2) at (2.5, 1.2) {$\xv_2$};
            \node[terminal] (x3) at (4.5, 1.2) {$\xv_3$};
            \node[prep_dev] (A) at (0.5,0) {$\rho^{\Av}_{\xv_1}$};
            \node[terminal] (prep_dev) at (0.5,-0.75) {\small Preparation};
            \node[proc_dev] (B) at (2.5,0) {$\Emc^{\Bv}_{\xv_2}$};
            \node[terminal] (proc_dev) at (2.5,-0.75) {\small Processing};
            \node[meas_dev] (C) at (4.5,0) {$\Pi^{\Cv}_{\yv|\xv_3}$};
            \node[terminal] (meas_dev) at (4.5,-0.75) {\small Measurement};
            \node[terminal] (y) at (5.75, 0) {$\yv$};

            \path (x1) \cedge (A);
            \path (x2) \cedge (B);
            \path (x3) \cedge (C);
            \path (A) \qedge node[el, above=2pt, xshift=0pt] {$\mc{S}^{\Av \to \Bv}$} (B);
            \path (B) \qedge node[el, above=2pt, xshift=0pt] {$\mc{S}^{\Bv \to \Cv}$} (C);
            \path (C) \cedge (y);
        \end{tikzpicture}  \\
        \hfill\\
    \begin{tabular}{c c}
            {\normalsize \shortstack{(b) Entanglement-Assisted \\ Senders (ETx) }} & {\normalsize \shortstack{(c) Entanglement-Assisted \\ Receivers (ERx) } }\\
            \hfill \\
            \begin{tikzpicture}
                \node[terminal] (x1) at (-0.5,0.8) {$x_1$};
                \node[terminal] (x2) at (-0.5,-0.8) {$x_2$};
                \node[qsource] (Lambda) at (-0.5,0) {$\rho^{\Lambda}$};
                \node[proc_dev] (A1) at (1,0.8) {$\Emc^{A_1}_{x_1}$};
                \node[proc_dev] (A2) at (1,-0.8) {$\Emc^{A_2}_{x_2}$};
                \node[terminal] (y1) at (2.5, 0.8) {$\rho^{A_1}_{x_1}$};
                \node[terminal] (y2) at (2.5, -0.8) {$\rho^{A_2}_{x_2}$};

                \path (x1) \cedge (A1);
                \path (x2) \cedge (A2);
                \path (Lambda) \qedge (A1);
                \path (Lambda) \qedge (A2);
                \path (A1) \qedge   (y1);
                \path (A2) \qedge (y2);
            \end{tikzpicture} & \begin{tikzpicture}
                \node[terminal] (x1) at (-0.5,0.8) {$\rho^{M_1}$};
                \node[terminal] (x2) at (-0.5,-0.8) {$\rho^{M_2}$};
                \node[qsource] (Lambda) at (-0.5,0) {$\rho^{\Lambda}$};
                \node[meas_dev] (A1) at (1,0.8) {$\Pi^{M_1}_{y_1}$};
                \node[meas_dev] (A2) at (1,-0.8) {$\Pi^{M_2}_{y_2}$};
                \node[terminal] (y1) at (2.5, 0.8) {$y_1$};
                \node[terminal] (y2) at (2.5, -0.8) {$y_2$};
    
                \path (x1) \qedge (A1);
                \path (x2) \qedge (A2);
                \path (Lambda) \qedge (A1);
                \path (Lambda) \qedge (A2);
                \path (A1) \cedge   (y1);
                \path (A2) \cedge (y2);
            \end{tikzpicture} \\
            \hfill \\
            \hfill \\
            {\normalsize \shortstack{(d) Local Operations \& \\ Quantum Communication }} & {\normalsize \shortstack{(e) Local Operations \& \\ Classical Communication }}  \\
            \hfill \\
            \begin{tikzpicture}
                \node[terminal] (x) at (-0.5,0) {$x$};  
                \node[qsource] (src) at (-0.2,-1) {$\rho^\Lambda$};
                \node[proc_dev] (A) at (0.8,0) {$\Emc^{A}_x$};
                \node[meas_dev] (B) at (1.9,-1) {$\Pi^{B}_y$};
                \node[terminal] (y) at (3, -1) {$y$};
    
                \path (x) \cedge (A);
                \path (src) \qedge (A);
                \path (src) \qedge (B);
                \path (A) \qedge node[el, above=4pt, xshift=4pt] {$\rho^A_x$} (B);
                \path (B) \cedge (y);
            \end{tikzpicture} & \begin{tikzpicture}
                \node[terminal] (x) at (-0.5,0) {$x$};  
                \node[qsource] (src) at (-0.2,-1) {$\rho^\Lambda$};
                \node[meas_dev] (A) at (0.8,0) {$\Pi^{A}_{a|x}$};
                \node[meas_dev] (B) at (1.9,-1) {$\Pi^{B}_y|a$};
                \node[terminal] (y) at (3.1, -1) {$y$};
    
                \path (x) \cedge (A);
                \path (src) \qedge (A);
                \path (src) \qedge (B);
                \path (A) \cedge node[el, above=4pt, xshift=5pt] {$a$} (B);
                \path (B) \cedge (y);
            \end{tikzpicture}  \\
        \end{tabular}
    \end{tabular}%
    }
    \caption{DAGs for basic quantum resource configurations.}    \label{fig:quantum_communication_network_entanglement}
\end{figure}
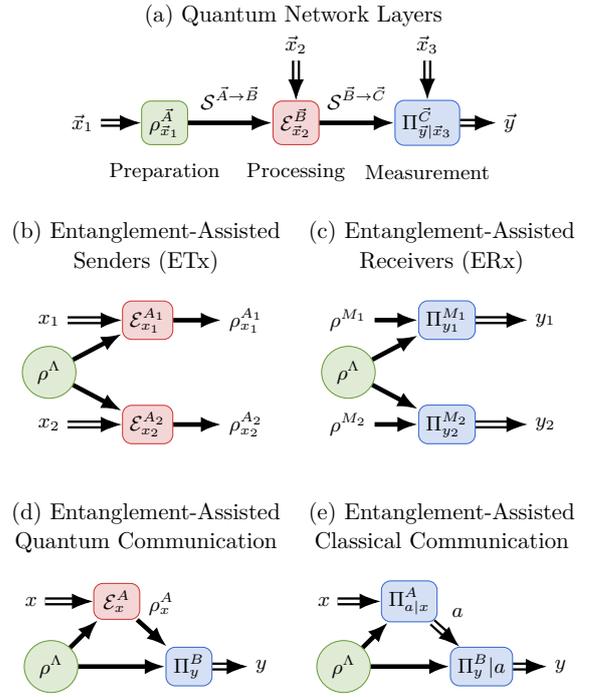

Entanglement can assist both classical and quantum communication networks. 
Entanglement is a static resource that can be distributed in the network prior to any information processing, similarly to GSR. 
Unlike shared randomness, however, entanglement is not easily produced or distributed. Therefore, given a network DAG, we consider each unique entanglement source configuration as a distinct \textit{quantum resource configuration}.
Entanglement sources can simply be treated as a preparation device with no classical input, therefore, the DAGs of entanglement-assisted communication networks decompose following the same rules as quantum communication networks.

This work focuses on entanglement-assisted senders, receivers, and communication channels (see Fig.~\ref{fig:quantum_communication_network_entanglement}.b-e), and we restrict our focus to entangled qubit states.
To distinguish between quantum resource configurations with and without entanglement we define the following sets of behaviors:
\begin{enumerate}
    \item $\Cbb^{\Net}$: The set of network behaviors using classical communication and no entanglement.
    \item $\Qbb^{\Net}$: The set of network behaviors using quantum communication and no entanglement.  
    \item $\Cbb^{\Net}_{\EA}$: The set of network behaviors using classical communication and entanglement.
    \item $\Qbb_{\EA}^{\Net}$: The set of network behaviors using quantum communication and entanglement.
\end{enumerate}
The $\EA$ subscript specifies the distinct entanglement configuration. \rev{Since quantum communication can trivially simulate classical communication, it holds true for any network that
\begin{equation}\label{eq:trivial_classical_quantum_hierarchy}
    \Cbb^{\Net} \subseteq \Qbb^{\Net} \quad \text{and}\quad \Cbb^{\Net}\subseteq \Cbb^{\Net}_{\EA} \subseteq \Qbb_{\EA}^{\Net}.
\end{equation}
When comparing networks with different entanglement structures, $\Cbb^{\Net}_{\text{E1}} \leftrightarrow \Cbb^{\Net}_{\text{E2}}$, $\Cbb^{\Net}_{\text{E1}} \leftrightarrow \Qbb^{\Net}_{\text{E2}}$, and $\Qbb^{\Net}_{\text{E1}} \leftrightarrow \Qbb^{\Net}_{\text{E2}}$, nontrivialities arise that make it difficult to determine whether or not a given entanglement configuration can simulate another. }{}

\subsection{Quantum Nonclassicality}\label{section:witnessing_nonclassicality-body}

\rev{
This section introduces the concept of nonclassicality in communication networks. Section~\ref{section:nonclassicality_advantage} formalizes the correspondence between nonclassicality and communication advantage, while Section~\ref{section:witnessing_nonclassicality} discusses how nonclassicality can be witnessed using linear functions. We then present two types of nonclassicality witnesses: simulation games (Section~\ref{section:simulation_games}) and facet inequalities (Section~\ref{section:facet_inequalities}). Simulation games are easy to derive and have a clear operational interpretation, whereas facet inequalities are more sensitive, but also more challenging to derive. Finally, Section~\ref{section:quantifying-certifying-nonclassicality} discusses how the amount of nonclassicality and its noise robustness can be quantified.}{}

\subsubsection{Nonclassicality as a Communication Advantage}\label{section:nonclassicality_advantage}

Given a network DAG, $\Net(\sigarb{\XNet}{\dv}{\YNet})$, our goal is to compare the set of classical network behaviors $\Cbb^{\Net}$ with the sets of quantum and entanglement assisted network behaviors, $\Qbb^{\Net}$, $\Cbb^{\Net}_{\EA}$, and $\Qbb^{\Net}_{\EA}$ where these sets of behaviors are convex because we assume that an unbounded amount of GSR is available. A quantum or entanglement-assisted network behavior $\Pbf$ is nonclassical if $\Pbf\not\in \Cbb^{\Net}$. 
Equivalently, a nonclassical behavior $\Pbf\in\Pbb_{\YNet|\XNet}$ cannot be  simulated with zero error by a classical network.

To formalize the concept of simulation within our framework, consider 
two behaviors $\Pbf,\Pbf'\in\Pbb_{\YNet|\XNet}$. We say that $\Pbf$ and $\Pbf'$ simulate each other with zero error if $\Pbf = \Pbf'$. Since behaviors are estimated in practice, it is important to introduce a metric for simulation error. Following the approach taken by Britto \textit{et al.} \cite{brito2018_nonlocality_trace_distance}, we use the variational distance for probability distributions to define the distance between two behaviors, $\Pbf,\Pbf'\in\Pbb_{\YNet|\XNet}$, as 
\begin{equation}\label{eq:behavior_trace_distance}
    \Delta(\Pbf,\Pbf')  = \frac{1}{2|\XNet|}\sum_{\xv\in\XNet}\sum_{\yv\in\YNet}|P_{\yv|\xv} - P'_{\yv|\xv}|,
\end{equation}
where the scalar factor of $1/|\XNet|$ results from the assumption that the inputs $\xv\in \XNet$ are uniformly random.
Two behaviors, $\Pbf$ and $\Pbf'$, achieve an $\epsilon$-\textit{approximate simulation} of each other if $\Delta(\Pbf,\Pbf') \leq \epsilon$ where $0 \leq \epsilon \ll 1$ is the simulation error tolerance.

Suppose for a given DAG, $\Net(\sigarb{\XNet}{\dv}{\YNet})$, that a classical network cannot simulate a behavior $\Pbf \in \Pbb_{\YNet|\XNet}$, implying that the behavior is nonclassical, $\Pbf \not\in \Cbb^{\Net}$.
If the signaling dimension is sufficiently increased along each edge as $d_i \to d_i'$ where $d_i' \geq d_i$, then the classical network can simulate the nonclassical behavior $\Pbf$.
Therefore, we define the \textit{classical simulation cost} of a nonclassical behavior $\Pbf\not\in \Cbb^{\Net}$ as
\begin{align}\label{eq:classical_simulation_cost}
    \kappa^{\Net}(\Pbf) &= \min_{\vec{d'} \in \Zbb^K_{\geq 1}} \sum_{i=0}^{K-1}\log d_i' \\ 
    \text{s.t.} \quad & \Pbf \in \Cbb^{\Net(\sigarb{\XNet}{\vec{d'}}{\YNet})} \label{eq:sim_cost_simulability_constraint}\\
    & d_i' \geq d_i \; \forall \; i \in \{0, \dots, K-1\}. \label{eq:reference_sig_dim}
\end{align}
The quantity $\kappa^{\Net}(\Pbf)$ is the minimum total number of bits that the classical network must communicate to simulate the nonclassical behavior $\Pbf$ with zero error.
Note that the constraint in Eq.~\eqref{eq:reference_sig_dim} requires that the signaling dimension along each edge is nondecreasing with respect to the reference communication vector $\dv$. Therefore a  nonclassical behavior $\Pbf\not \in \Cbb^{\Net(\sigarb{\XNet}{\vec{d}}{\YNet})}$ implies an increase to the classical simulation cost $\kappa^{\Net}(\Pbf)$.
More generally, quantum nonclassicality demonstrates an explicit communication advantage in which classical networks require more communication to generate the same behaviors.
More details on how to evaluate $\kappa^{\Net}(\Pbf)$ are found in Section~\ref{section:quantifying-certifying-nonclassicality}.

\begin{figure}
    \centering
    \includegraphics[width=0.9\columnwidth]{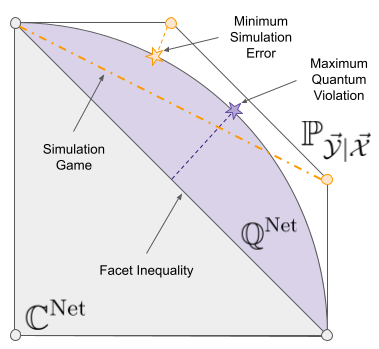}
    \caption{A qualitative view of a classical network polytope $\Cbb^{\Net}$ (gray region), the set of nonclassical quantum behaviors (purple region), and the probability polytope $\Pbb_{\YNet|\XNet}$ (outer pentagon) where the vertices correspond to deterministic behaviors. An orange dash-dotted line depicts a simulation game where the orange star corresponds to the minimum simulation error.  The purple star shows the maximal quantum violation of the facet inequality that tightly bounds the classical network polytope.  }
    \label{fig:quantum-nonclassicality}
\end{figure}

\subsubsection{Witnessing Nonclassicality}\label{section:witnessing_nonclassicality}

The nonclassicality of a behavior $\Pbf \in \Pbb_{\YNet|\XNet}$ is tested using a nonclassicality witness $\gamma \geq g(\Pbf)$ where $\gamma\in \reals$ and $g\colon \Pbb_{\YNet|\XNet}\to \reals$. Each nonclassicality witness is tailored to a particular network where the inequality $\gamma \geq g(\Pbf)$ is satisfied for all $\Pbf \in \Cbb^{\Net}$.
Nonclassicality is witnessed whenever the inequality is violated as $\gamma <g(\Pbf)$. Thus, nonclassicality witnesses serve as operational tests of communication advantage. 

Although nonclassicality witnesses can be nonlinear in general \cite{Tavakoli-2022a}, we focus on linear nonclassicality witnesses.
We define a \textit{linear nonclassicality witness} as any tuple $(\gamma, \Gbf)$ such that $\gamma \geq \ip{\Gbf, \Pbf}$
for all $\Pbf\in\Cbb^{\Net}$ where
\begin{equation}\label{eq:game_inner_product}
    \ip{\Gbf,\Pbf} \equiv \tr{\Gbf^T \Pbf} = \sum_{\yv\in\YNet}\sum_{\xv\in\XNet} G_{\yv,\xv}P_{\yv|\xv}.
\end{equation}
The matrix $\Gbf$ can be understood as a reward matrix where each element $G_{\yv|\xv}$ describes a score awarded for outputting $\yv$ given the input $\xv$. Assuming a uniformly random input, the behavior $\Pbf$ achieves the average score $\frac{1}{|\XNet|}\ip{\Gbf, \Pbf}$.

In the following sections, we introduce two types of linear nonclassicality witnesses: simulation games and facet inequalities. Simulation games are more readily derived and have clear operational interpretations, however, they are less sensitive to violations and admit false negative results. Alternatively, facet inequalities do not admit false negative results because they tightly bound the classical network polytope, however, they can be difficult to obtain as networks scale.

\subsubsection{Simulation Games}\label{section:simulation_games}

We define a \textit{simulation game} as any linear nonclassicality witness $(\gamma, \Vbf)$  where $\Vbf\in \Vbb_{\YNet|\XNet}$ is a nonclassical deterministic behavior (see Eq.~\ref{eq:set_of_deterministic_behaviors}).
A simulation game's deterministic reward matrix encodes a deterministic communication task $f\colon \XNet \to \YNet$ whose performance is measured by the average score $\frac{1}{|\XNet|}\ip{\Vbf,\Pbf}$. Since $V_{f(\xv),\xv} = \delta_{\yv,f(\xv)}$ for each input $\xv$, there is a single ``correct'' output  $f(\xv)\in\YNet$.
The average score then corresponds to a success probability
\begin{equation}\label{eq:success_probability}
    P_{\Success} = \frac{1}{|\XNet|}\sum_{\xv\in\XNet}P_{f(\xv)|\xv}= \frac{1}{|\XNet|}\ip{\Vbf, \Pbf}
\end{equation}
where $P_{\Success}\in[0,1]$ and the associated error probability is $P_{\Error} = 1 - P_{\Success}$.

The average score $\ip{\Vbf, \Pbf}$ of a simulation game relates to the variational distance in Eq.~\eqref{eq:behavior_trace_distance} as
\begin{equation}\label{eq:deterministic_behavior_simulation_error}
   \Delta(\Vbf,\Pbf) =  P_{\Error} = 1  - \frac{1}{|\XNet|}\ip{\Vbf, \Pbf}.
\end{equation}
Thus, the score of a simulation game precisely measures the simulation error $\Delta(\Vbf,\Pbf)$, and the objective of a simulation game is to simulate the deterministic behavior $\Vbf$ with zero error.
It is important to note that the general variational distance $\Delta(\Pbf',\Pbf)$ requires $|\XNet|(|\YNet| -1)$ values of $P_{\yv|\xv}$ to be estimated, whereas the simulation game error $\Delta(\Vbf, \Pbf)$ only requires $|\XNet|$ values of $P_{\yv|\xv}$ to be estimated.

\subsubsection{Facet Inequalities of Classical Network Polytopes}\label{section:facet_inequalities}

Since the classical network polytope in Eq.~\eqref{eq:classical_network_polytope} is convex, it is equivalently expressed as the intersection of linear half-spaces called \textit{facet inequalities} \cite{Ziegler-2012a_lectures_on_polytopes}. The classical network polytope is then given by
\begin{equation}
    \Cbb^{\Net} = \bigcap_{i=1}^{|\Fbb^{\Net}|} \left\{\Pbf \in \Pbb_{\YNet|\XNet}\;|\; \gamma_i \geq \ip{\Fbf_i, \Pbf}\right\},
\end{equation}
where the set of facet inequalities is
\begin{equation}\label{eq:classical_network_polytope_facets}
    \Fbb^{\Net} \equiv \left\{\left(\gamma_k \in \Zbb_{>0},\; \Fbf_k \in\Zbb_{\geq0}^{|\YNet|\times|\XNet|}\right)\right\}_{k=1}^{|\Fbb^{\Net}|},
\end{equation}
in which each tuple $(\gamma, \Fbf)$ is a linear half-space inequality $\gamma \geq \ip{\Fbf,\Pbf}$ that tightly bounds $\Cbb^{\Net}_{}$.
Furthermore, due to the symmetry of $\Cbb^{\Net}$, it is sufficient to consider $\Fbb^{\Net}$ as being a canonical subset of facet inequalities, and the complete set of facet inequalities can then be recovered through relabeling the inputs, outputs, and parties \cite{Rosset2014_classifying_bell_inequalities}. 

The facet inequalities of classical network polytopes are generally computed by first enumerating the set of vertices $\Vbb^{\Net}$ (see Eq.~\eqref{eq:deterministic_behavior}).
For the simplest classical communication networks, we can compute the complete set of facet inequalities from the vertices using Fourier-Motzkin elimination \cite{Ziegler-2012a_lectures_on_polytopes} using the standard Polytope Representation Transformation Algorithm (PoRTA) software \cite{PORTA}.
As the number of vertices scale, PoRTA fails to efficiently perform and we resort to the linear program described below in Eq.~\eqref{eq:facet_linear_program}.
We perform these facet inequality computations using the BellScenario.jl Julia package. For more details please refer to the Supplemental Code Section~\ref{section:Supplemental_Code} and our work on GitHub \cite{supplemental_software}.

Given the set of network vertices $\Vbb^{\Net}$ and a known nonclassical behavior $\Pbf\in\Pbb_{\YNet|\XNet}$, a linear nonclassicality witness can be obtained using the following linear program \cite{brunner2014nonlocality},
\begin{align}\label{eq:facet_linear_program}
     (\gamma^\star, \Gbf^\star) = \arg&\max_{\substack{\gamma\in\reals \\ \Gbf\in \reals^{|\YNet|\times|\XNet|}}}  \ip{\Gbf,\Pbf} - \gamma \\
    \text{s.t.}&\;\ip{\Gbf, \Pbf } - \gamma \leq 1\\
    & \; \ip{\Gbf,\Vbf} - \gamma \leq 0 \; \forall \; \Vbf\in\Vbb^{\Net}.\label{eq:lp-vertex-check}
\end{align}
If the input behavior $\Pbf$ is nonclassical such that $\Pbf \not\in \Cbb^{\Net}$, then the $(\gamma^\star, \Gbf^\star)$ constitutes a linear nonclassicality witness for which the input behavior $\Pbf$ achieves the violation $\ip{\Gbf^\star,\Pbf} = 1 + \gamma^\star$. Otherwise, if the input behavior $\Pbf$ is classical such that $\Pbf\in\Cbb^{\Net}$, then the  linear program outputs the trivial solution where $\gamma^\star = 0$ and $G_{\yv,\xv}^\star = 0$ for all $\xv\in\XNet$ and $\yv\in\YNet$.

The linear program in Eq.~\eqref{eq:facet_linear_program} is guaranteed to either return a linear nonclassicality witness $(\gamma^\star, \Gbf^\star)$ that is violated by the nonclassical test behavior $\Pbf\not\in\Cbb^{\Net}$, or return a trivial linear inequality that indicates $\Pbf\in\Cbb^{\Net}$. When a linear nonclassicality witness $(\gamma^\star, \Gbf^\star)$ is output, it can be verified as a facet inequality of $\Cbb^{\Net}$ if there exists at least $\Dim{\Cbb^{\Net}} = |\XNet|(|\YNet| - 1)$ affinely independent vertices $\Vbb\in \Vbb^{\Net}$ that satisfy the equality $\gamma^\star = \ip{\Gbf^\star, \Vbb}$. If the nonclassicality witness is a facet inequality, then $\gamma$ and $G^\star_{\yv,\xv}$ are rational numbers, which can be transformed to be positive integers without loss of generality. We apply the linear program throughout the Results Section~\ref{section:results}, for which we input deterministic nonclassical behaviors $\Vbf \in \Vbb_{\YNet|\XNet}$ and find that the resulting nonclassicality witnesses are facet inequalities in all cases.
However, the practicality of the linear program in Eq.~\eqref{eq:facet_linear_program} is limited  because the number of constraints in Eq.~\eqref{eq:lp-vertex-check} scales with $\Vbb^{\Net}$, which grows exponentially with the size of the input alphabet.

\subsubsection{Quantifying and Certifying Nonclassicality}\label{section:quantifying-certifying-nonclassicality}

The amount of nonclassicality that a given behavior $\Pbf\not\in \Cbb^{\Net}$ exhibits can be quantified by the amount of violation of a nonclassicality witness $(\gamma, \Gbf)$, \textit{i.e.}, $\beta = \ip{\Gbf, \Pbf} - \gamma$.
The maximal violation possible for a given linear nonclassicality witness is $\widehat{\beta} = \widehat{\gamma} - \gamma$ where the maximal possible score
\begin{equation}\label{eq:maximal_black_box_game_score}
    \widehat{\gamma} \equiv  \max_{\Pbf\in\Pbb_{\YNet|\XNet}} \ip{\Gbf, \Pbf} = \sum_{\xv\in\XNet} \max_{\yv\in\YNet} G_{\yv,\xv}
\end{equation}
is achieved by a deterministic behavior $\Pbf\in \Vbb_{\YNet|\XNet}$   that satisfies $P_{\yv|\xv} = \delta_{\yv, f(\xv)}$ where $f(\xv) = \arg\max_{\yv\in\YNet} G_{\yv,\xv}$.
Using the maximal possible violation and the classical bound, the violation of any nonclassicality witness can be rescaled to the range $\bar{\beta}\in[0,1]$ as
\begin{equation}\label{eq:rescaled_violation}
    1 \geq \bar{\beta} = \frac{\beta}{\widehat{\beta}} = \frac{\ip{\Gbf,\Pbf} - \gamma}{\widehat{\gamma} - \gamma} \geq 0
\end{equation}
where the upper bound is achieved when $\ip{\Gbf,\Pbf} = \widehat{\gamma}$ and the lower bound is achieved when $\ip{\Gbf,\Pbf} = \gamma$.

The noise robustness of a violation is a practical metric for real-world scenarios \cite{Bowles2015_nonclassicality_communication_networks, doolittle_2023_vqo_nonlocality, doolittle_phd_thesis_2023, Pal_2015_nonlocality_breaking, Zhang_2020_nonlocality_breaking}.  We define noise robustness as the minimum amount of white noise $\omega_0$ that can be added to a nonclassical behavior $\Pbf\not\in\Cbb^{\Net}$ such that $\gamma = \ip{\Gbf, \Pbf_{\omega_0}}$, where
\begin{equation}\label{eq:noise_behavior}
    \Pbf_{\omega} = (1 - \omega) \Pbf + \omega \widetilde{\Pbf}
\end{equation}
and $\widetilde{\Pbf}$ satisfying $\widetilde{P}_{\yv|\xv} = 1 / |\YNet|$ for all $\xv\in\XNet$ and $\yv\in\YNet$. Note that the  behavior $\widetilde{\Pbf}$ corresponds to white noise and can be generated using no  communication. For any linear nonclassicality witness $(\gamma, \Gbf)$, white noise achieves the 
score
\begin{equation} \label{eq:white_noise_behavior}
    \widetilde{\gamma} \equiv \ip{\Gbf,\widetilde{\Pbf}} = \frac{1}{|\YNet|}\sum_{\yv\in\YNet}\sum_{\xv\in\XNet} G_{\yv,\xv}.
\end{equation}
Given a linear nonclassicality witness $(\gamma, \Gbf)$ and a nonclassical behavior $\Pbf\not\in\Cbb^{\Net}$, the noise robustness is equal to
\begin{equation}\label{eq:noise_robustness}
    \omega_0 = \frac{\ip{\Gbf, \Pbf} - \gamma}{\ip{\Gbf, \Pbf} - \ip{\Gbf, \widetilde{\Pbf}}},
\end{equation}
which implies that the noise robustness of the maximum possible violation $\widehat{\beta}$ is equal to
\begin{equation}
    \widehat{\omega}_0 \equiv \frac{\widehat{\beta}}{\widehat{\gamma} - \ip{\Gbf, \widetilde{\Pbf}}}.
\end{equation}

The max violation and noise robustness do not precisely quantify the classical simulation cost, but they can be used in its certification. In general, it is nontrivial to obtain the classical simulation cost, $\kappa^{\Net}(\Pbf)$, as defined in Eq.~\eqref{eq:classical_simulation_cost}, for a given nonclassical behavior and network $\Net(\sigarb{\XNet}{\dv}{\YNet})$. However, upper and lower bounds can be estimated. 
In particular, a violation of a nonclassicality witness $(\gamma, \Gbf)$ that bounds a network $\Net(\sigarb{\XNet}{\dv}{\YNet})$ indicates that the communication vector $\dv$ is insufficient for simulating the nonclassical behavior. A violation therefore establishes the lower bound $\kappa^{\Net}(\Pbf) \geq \sum_{i} \log d_i$.

Interestingly, the linear program in Eq.~\eqref{eq:facet_linear_program} provides a means of estimating the classical simulation cost. If the linear program in Eq.~\eqref{eq:facet_linear_program} is solved for the given  behavior $\Pbf\in\Pbb_{\YNet|\XNet}$ and the set of vertices $\Vbb^{\Net}$, then the result $(\gamma^\star, \Gbf^\star)$ places either an upper or lower bound on $\kappa^{\Net}(\Pbf)$. Namely, if the trivial solution is returned where $\ip{\Gbf^\star,\Pbf} - \gamma^\star$ = 0, then $\Pbf\in\Cbb^{\Net}$ and $\kappa^{\Net}(\Pbf) \leq \sum_{i} \log d_i$. Otherwise, $\Pbf\not\in\Cbb^{\Net}_{}$, which implies that $\kappa^{\Net}(\Pbf) \geq \sum_{i} \log d_i$.
A binary search could be used to identify upper and lower bounds on the classical simulation cost where coarse-grained searches could also be applied where $d_i = d$ is considered to be constant for all $i$. However, such methods are limited by the fact that the number of vertices grows exponentially with the size of the network. \rev{In Section~\ref{section:operational_framework}, we discuss an alternative approach that uses our VQO methods to place a lower bound on the classical simulation cost, or equivalently, an upper bound on the classical bound for a given network and nonclassicality witness (see Eq.~\eqref{eq:classical_bound_estimation_vqo}).}{}

\subsection{Maximizing Nonclassicality in Quantum Communication Networks}\label{section:maximizing_quantum_nonclassicality}

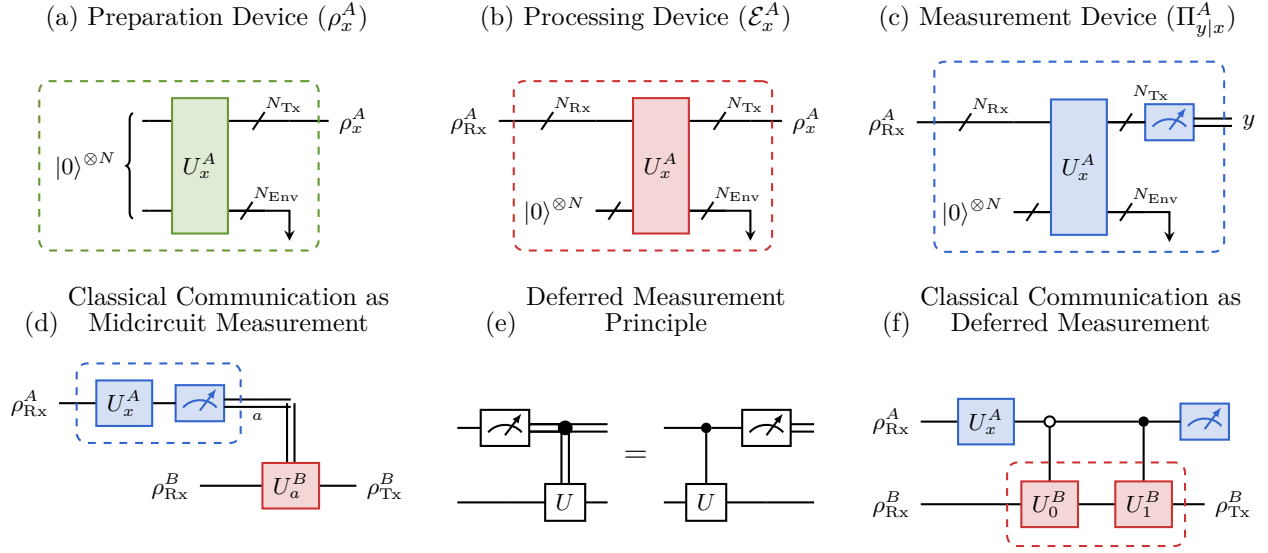
\begin{figure*}[ht]
    \centering
    \small
    \resizebox{\textwidth}{!}{%
    \begin{tabular}{c c c}
        {\normalsize (a) Preparation Device ($\rho^{A}_{x}$) }& {\normalsize (b) Processing Device ($\Emc^{A}_{x}$)} & {\normalsize (c) Measurement Device ($\Pi^{A}_{y|x}$) } \\
        \begin{tikzcd}
            \gategroup[2,style={dashed,rounded corners,color=prep_green, inner xsep=50pt, inner ysep=12pt, xshift=14pt}, label style={yshift=2pt}]{} \lstick[2]{$\ket{0}^{\otimes N}$} & [-4mm]\qw  & [-2mm]  \gate[2, style={prep_gate}]{U^{A}_{x}} & \qwbundle{N_{\Tx}} & [-3mm] \qw\rstick{$\rho^{A}_{x}$}\\
            & \qw & & \trash{} \qwbundle{N_{\Env}} & \\
        \end{tikzcd} & \begin{tikzcd}
            \gategroup[2,style={dashed,rounded corners,color=proc_red, inner xsep=46pt, inner ysep=12pt, xshift=55pt}, label style={yshift=2pt}]{}\lstick{$\rho^{A}_{\Rx}$} & [8mm]\qwbundle{N_{\Rx}} & \gate[2, style={proc_gate}]{U^{A}_{x}} & \qwbundle{N_{\Tx}} & [-4mm]\qw\rstick{$\rho^{A}_{x}$}\\
            & \lstick{$\ket{0}^{\otimes N}$} &\qwbundle{} & \trash{} \qwbundle{N_{\Env}} & \\
        \end{tikzcd} & \begin{tikzcd}
            \gategroup[2,style={dashed,rounded corners,color=meas_blue, inner xsep=52pt, inner ysep=16pt, xshift=62pt, yshift=4pt}, label style={yshift=2pt}]{} \lstick{$\rho^{A}_{\Rx}$} & [8mm]\qwbundle{N_{\Rx}} & \gate[2, style={color=meas_blue, fill=meas_blue!20}]{U^{A}_{x}} &  \meter[style={color=meas_blue, fill=meas_blue!20}]{} \qwbundle{\raisebox{0.7em}{$\scriptstyle N_{\Tx}$ }}  & \cw\rstick{$y$}  \\
            & \lstick{$\ket{0}^{\otimes N}$} &\qwbundle{} & \trash{} \qwbundle{N_{\Env}} & \\
        \end{tikzcd}  \\

        {\normalsize (d) \shortstack{Classical Communication as\\ Midcircuit Measurement}} & {\normalsize (e) \shortstack{ Deferred Measurement \\ Principle}} & {\normalsize (f) \shortstack{Classical Communication as \\ Deferred Measurement}}  \\
        \begin{tikzcd}
            \gategroup[1,style={meas_gate_group, inner xsep=28pt, inner ysep=12pt, xshift=38pt}, label style={yshift=2pt}]{}\lstick{$\rho^{A}_{\Rx}$} & \gate[style=meas_gate]{U^{A}_{x}} & [-2mm] \meter[style=meas_gate]{} & \cw{a}\vcw{1} &\\
             & & \lstick{$\rho^{B}_{\Rx}$} & [-4mm] \gate[style={proc_gate}]{U^{B}_{a}} & \qw\rstick{$\rho^{B}_{\Tx}$}\\
        \end{tikzcd} & \begin{tikzcd}
            \qw & [-2mm] \meter{} & [-3mm] \cwbend{1}\cw & [-2mm]\cw\\
            \qw & \qw & \gate{U} & \qw
        \end{tikzcd} {\large =} \begin{tikzcd}
            \qw &[-2mm] \ctrl{1} & [-3mm]\meter{} & [-2mm] \cw \\
            \qw & \gate{U} & \qw & \qw
        \end{tikzcd} & \begin{tikzcd}
            \lstick{$\rho^A_{\Rx}$} & \gate[style={meas_gate}]{U^{A}_x} &[-4mm] \octrl{1} & \ctrl{1} & [-4mm] \meter[style={meas_gate}]{} \\
            \lstick{$\rho^B_{\Rx}$} & \gategroup[1,style={proc_gate_group, inner xsep=20, inner ysep=4pt, xshift=42pt}, label style={label position=below, yshift=-16pt}]{}\qw & \gate[style={proc_gate}]{U^{B}_{0}} & [-4mm]\gate[style={proc_gate}]{U^{B}_{1}} & \qw\rstick{$\rho^B_{\Tx}$}
        \end{tikzcd}
    \end{tabular}%
    }
    \caption{\linespread{1} Quantum circuit models for quantum network devices and classical communication. }
    \label{fig:device_circuit_parameterization}
\end{figure*}

\rev{This section discusses our hardware-compatible approach for optimizing quantum network behaviors and  maximizing their nonclassicality. }{}
To obtain the maximal violation of a nonclassicality witness $(\gamma, \Gbf)$, we must solve the optimization problem
\begin{equation}\label{eq:quantum_network_optimization_problem}
    \beta^\star \equiv \max_{\Pbf \in \mbb{S}^{\Net}} \ip{\Gbf,\Pbf}
\end{equation}
for the given network where $\mbb{S}^{\Net}\in \{\Cbb^{\Net}_{\EA},\Qbb^{\Net},\Qbb^{\Net}_{\EA}\}$ denotes the quantum resource configuration.
The optimization problem in Eq.~\eqref{eq:quantum_network_optimization_problem} can be solved using the variational quantum optimization (VQO) framework for quantum networks introduced by Doolittle \textit{et al.} \cite{doolittle_2023_vqo_nonlocality}.
In this framework a quantum network DAG corresponds to a parameterized quantum circuit that explicitly encodes the communication resources and the local operations at each device (see Fig.~\ref{fig:network_circuit_parameterization}).
\rev{We build upon previous work by introducing a formalism for simulating and optimizing LOCC resources via  deferred measurements (see Fig~\ref{fig:device_circuit_parameterization}).}{}

\rev{In Section~\ref{section:simulating_quantum_networks}, we introduce our formalism for constructing parameterized quantum circuits that simulate quantum communication networks. In Section~\ref{section:variational_optimization_of_quantum_nonclassicality}, we introduce our VQO methods for our operational framework (Algorithm~\ref{alg:variational_optimization}), and how they can be applied to establish nonclassical behaviors (Algorithm~\ref{alg:vqo_nonclassicality}) and deterministic protocols (Algorithm~\ref{alg:vqo_simulation}). In Section~\ref{section:operational_framework}, we discuss the merits and practicality of the developed framework.}{}
Our VQO methods are implemented using the Quantum Network Variational Optimizer (QNetVO) Python package \cite{doolittle_qNetVO}, an extension of the PennyLane framework for quantum machine learning \cite{bergholm2018pennylane}. Details are found in the Supplemental Code Section ~\ref{section:Supplemental_Code} and on GitHub \cite{supplemental_software}.

\subsubsection{Quantum Network Variational Ans\"{a}tze}\label{section:simulating_quantum_networks}

A quantum network's causal structure and communication resources can be represented as a quantum circuit that simulates the network's behavior. 
The quantum circuit can be run using either quantum hardware or a classical simulator\footnote{This work evaluates quantum circuits on a laptop computer using Pennylane's \texttt{"default.qubit"} simulator \cite{bergholm2018pennylane}. }. 
Although quantum circuits are restricted to unitary operations, auxiliary qubits can be used to implement density matrix state preparations, CPTP maps, and POVM measurements.
Additionally, local operations and classical communication (LOCC) can be simulated using midcircuit measurements to condition future gate operations (see Fig.~\ref{fig:device_circuit_parameterization}.d).
If a hardware platform does not support midcircuit measurements, then the deferred measurement principle can be used to represent LOCC as controlled unitary operations \cite{nielsen_chuang2009} (see Fig.~\ref{fig:device_circuit_parameterization}.e-f).

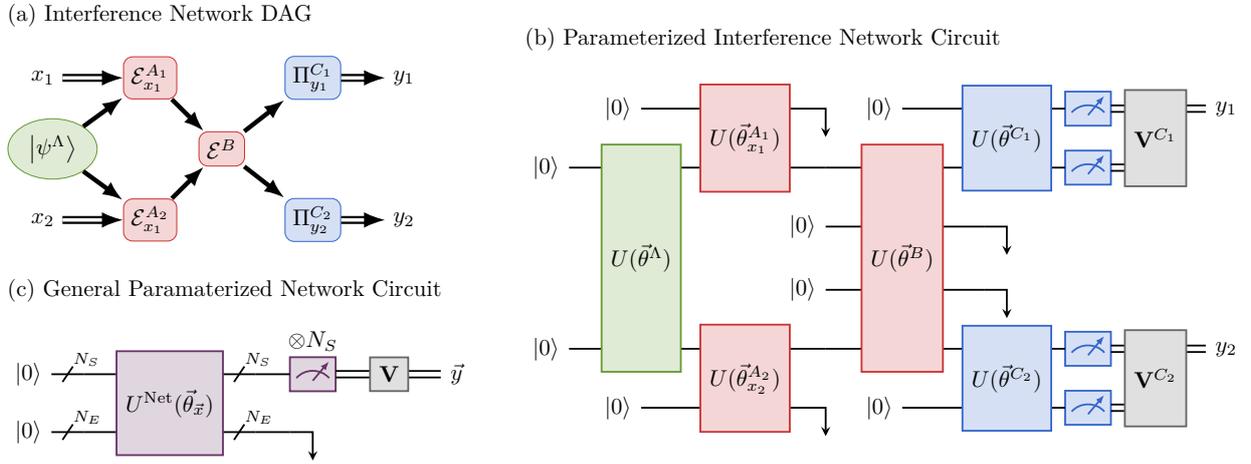
\begin{figure*}
    \centering
    \resizebox{\textwidth}{!}{%
    \begin{tabular}{c c}
        \begin{tabular}{l}
        (a) Interference Network DAG\\
        \hfill\\
        \begin{tikzpicture}
            \node[terminal] (x1) at (-0.15,1.1) {$x_1$};
            \node[terminal] (x2) at (-0.15,-1.1) {$x_2$};    
            \node[qsource] (Lambda) at (0,0) {$\ket{\psi^\Lambda}$};
            \node[proc_dev] (A1) at (1.5,1.1) {$\Emc^{A_1}_{x_1}$};
            \node[proc_dev] (A2) at (1.5,-1.1) {$\Emc^{A_2}_{x_1}$};
            \node[proc_dev] (B) at (2.6, 0) {$\Emc^B$};
            \node[meas_dev] (C1) at (4,1.1) {$\Pi^{C_1}_{y_1}$};
            \node[meas_dev] (C2) at (4,-1.1) {$\Pi^{C_2}_{y_2}$};
            \node[terminal] (y1) at (5.4, 1.1) {$y_1$};
            \node[terminal] (y2) at (5.4, -1.1) {$y_2$};


            \path (x1) \cedge (A1);
            \path (x2) \cedge (A2);
            \path (Lambda) \qedge (A1);
            \path (Lambda) \qedge (A2);
            \path (A1) \qedge (B);
            \path (A2) \qedge (B);
            \path (B) \qedge (C1);
            \path (B) \qedge (C2);
            \path (C1) \cedge (y1);
            \path (C2) \cedge (y2);
        \end{tikzpicture}    \\
        \hfill \\
        {\normalsize (c) General Paramaterized Network Circuit}  \\
        \hfill\\
        \begin{tikzcd}[row sep=0.1cm]
            \lstick{$\ket{0}$} & \qwbundle{N_S} & \gate[2, style={arb_gate}]{U^{\Net}(\vec{\theta}_{\xv})} & \qwbundle{N_S} & \meter[style={arb_gate}]{$\otimes N_S$} & \gate[nwires=1, style=classical_gate]{\Vbf}\cw & \cw\rstick{$\vec{y}$}\\
            \lstick{$\ket{0}$} & \qwbundle{N_E} & & \qwbundle{N_E} & \trash{} & & \\
        \end{tikzcd} \\
    \end{tabular} & \begin{tabular}{l}
            (b) Parameterized Interference Network Circuit \\
            \hfill\\
            \begin{tikzcd}[row sep=0.1cm]
                & \lstick{$\ket{0}$}  & [-2mm]\gate[2, style={proc_gate}]{U(\vec{\theta}^{A_1}_{x_1})} &  [-3mm]\trash{}  & [-3mm]\lstick{\ket{0}} & [-2mm]\gate[2, style={meas_gate}]{U(\vec{\theta}^{C_1})} & [-3mm]\meter[style={meas_gate}]{}  & [-3mm]\gate[2,nwires={1,2}, style=classical_gate]{\Vbf^{C_1}}\cw &  [-2mm]\cw\rstick{$y_1$} \\
                \lstick{$\ket{0}$} & \gate[4, nwires={2,3}, style={prep_gate}]{U(\vec{\theta}^\Lambda)} & & \qw & \gate[4,  style={proc_gate}]{U(\vec{\theta}^B)} &  & \meter[style={meas_gate}]{} &\cw & \\
                & & & \lstick{$\ket{0}$} & & \trash{} & & & \\
                & & & \lstick{$\ket{0}$} & & \trash{} & & & \\
                \lstick{$\ket{0}$} & & \gate[2, style={proc_gate}]{U(\vec{\theta}^{A_2}_{x_2})} & \qw & &  \gate[2, style={meas_gate}]{U(\vec{\theta}^{C_2})}& \meter[style={meas_gate}]{} & \gate[2,nwires={1,2}, style=classical_gate]{\Vbf^{C_2}}\cw & \cw\rstick{$y_2$} \\
                 & \lstick{$\ket{0}$} & & \trash{} &  \lstick{$\ket{0}$} &  &\meter[style={meas_gate}]{} & \cw & \\
            \end{tikzcd}
        \end{tabular}
    \end{tabular}%
    }
    \caption{Parameterized quantum circuits for network DAGs.}
    \label{fig:network_circuit_parameterization}
\end{figure*}

We parameterize the quantum circuit that simulates a quantum network as $U^{\Net}(\vec{\theta}_{\xv})$ where the parameters $\vec{\theta}_{\xv}\subseteq \vec{\theta}\in\reals^{M}$ vary the operations applied in the network and $\vec{\theta}_{\xv}$ denotes the network's parameters given the input $\vec{x}\in\vec{\mc{X}}$.
Since a subset of the qubits in the quantum circuit model will be measured, we can express the state prior to measurement as
\begin{equation}\label{eq:system-environment-rep}
    \rho^{\Net}(\vec{\theta}_{\xv}) = \ptr{E}{U^{\Net}(\vec{\theta}_{\xv})\op{0}{0}^{\otimes N_S\times N_E} U^{\Net}(\vec{\theta}_{\xv})^{\dagger}}
\end{equation}
where $U^{\Net}(\vec{\theta}_{\xv})$ operates on the joint Hilbert space $\Hmc^{\Net} = \Hmc^{S}\otimes \Hmc^{E}$ with $\Hmc^S$  and $\Hmc^{E}$ respectively describe the $N_S$ qubits that are measured and the $N_E$ qubit that are discarded. 
The transition probabilities of the simulated network  are then parameterized as 
\begin{equation}\label{eq:parameterized_network_behavior}
    P^{\Net}_{\yv|\xv}(\vec{\theta}_{\xv})  = \sum_{\zv\in\{0,1\}^{N_S}} V^{\text{Post}}_{\yv|\zv}\tr{|\zv\rangle\langle\zv|\rho^{\Net}_{\vec{\theta}_{\xv}}}
\end{equation}
where $\Vbf^{\text{Post}}$ designates a deterministic post-processing map that takes the $|\ZNet| = 2^{N_S}$ outputs from the computational basis measurement into the appropriate output alphabet $|\YNet|\leq |\ZNet|$ (see Fig.~\ref{fig:network_circuit_parameterization}.c). 
As a result, a quantum network's behavior is parameterized as
\begin{equation}
    \Pbf^{\Net}(\vec{\theta}) = \textstyle \sum_{\yv\in\YNet}\sum_{\xv\in\XNet}  P^{\Net}_{\yv|\xv}(\vec{\theta}_{\xv})  \op{\yv}{\xv}.
\end{equation}

The parameters and unitary operators applied at different devices are independent due to locality constraints in the communication network.
Hence, the device $A_j$ has $|\vec{\theta}^{A_j}| = |\Xmc_j|\times|\vec{\theta}^{A_j}_{x_j}|$ parameters total and a preparation device that prepares an $N$-qubit state  (Fig.~\ref{fig:device_circuit_parameterization}.a) has  $|\vec{\theta}^{A_j}_{x_j}| = 2^{N+1} - 2$ settings where we apply Pennylane's \texttt{ArbitraryStatePreparation} circuit ansatz. Both processing and measurement devices, (see Fig.~\ref{fig:device_circuit_parameterization}.b,c), have $|\vec{\theta}^{A_j}_{x_j}| = 4^N - 1$ settings where we consider Pennylane's \texttt{ArbitraryUnitary} circuit ansatz \cite{bergholm2018pennylane}. Naturally the total number of settings in the network $|\vec{\theta}|$ scales exponentially with the largest number of qubits used by any device in the network.

\rev{As shown in Fig.~\ref{fig:device_circuit_parameterization}.a-c ancillary qubits are used to model a broader set of free operations including CPTP maps and POVMs. While the number of necessary ancillary qubits depends on the application, this work follows the convention shown in Fig.~\ref{fig:network_circuit_parameterization}.b. Namely, we assume that preparation devices (green nodes) prepare pure states, such that no ancillary qubits are needed for state preparation. Processing devices (red nodes) perform CPTP maps by making use of one or more ancillary qubit for each qubit received from upstream devices.  Measurement devices (blue nodes) each perform a projective measurement on the received qubits and ancillary qubits, allowing POVM measurements via classical post-processing.}{}

\subsubsection{Variational Optimization of Quantum Networks}\label{section:variational_optimization_of_quantum_nonclassicality}

\begin{figure}
    \centering
    \includegraphics[width=\columnwidth]{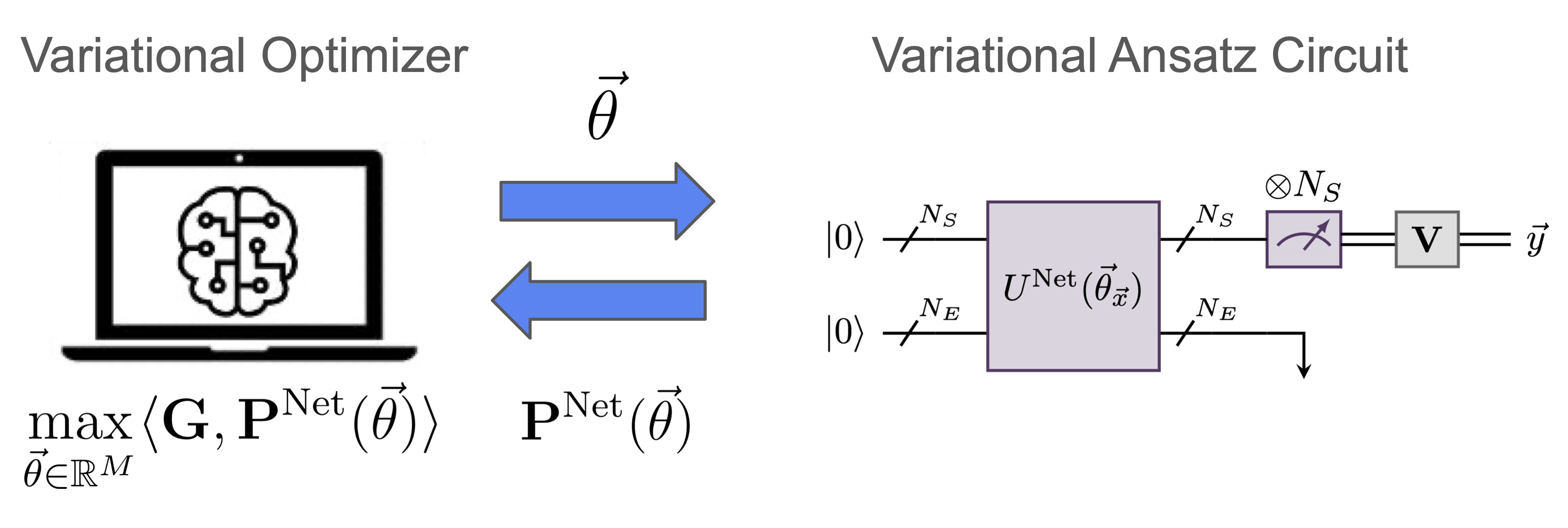}
    \caption{A classical computer performs variational optimization of a parameterized quantum circuit.}
    \label{fig:vqo_diagram}
\end{figure}

A parameterized quantum network simulation circuit can be optimized using variational optimization (see Fig.~\ref{fig:vqo_diagram}).
The goal is to solve the optimization problem $\max_{\vec{\theta} \in\reals^{M}}\gain{\vec{\theta}}$ for some function $\text{Gain}\colon \reals^M \to \reals$. Gradient ascent is used to maximize the gain where the gradient evaluated at the settings $\vec{\theta}$ is $\nabla_{\vec{\theta}}\gain{\vec{\theta}}\in \reals^{M}$ and points in the direction of steepest ascent.
The gradient can then be iteratively followed to a (local) maximum.
In practice, gradients can be calculated on classical hardware using backpropagation \cite{rumelhart1986backprop}, or evaluated on quantum hardware using parameter shift rules \cite{schuld2019_parameter-shift,Mari2021_higher-order_quantum_gradients,Wierichs2022_generall_parameter-shift_rule,Kyriienko2021_generalized_circuit_diff}.
Equivalently, the gain function can be restated as a cost function  $\gain{\vec{\theta}} = -\cost{\vec{\theta}}$ where $\min_{\vec{\theta}}\cost{\vec{\theta}} = \max_{\vec{\theta}}\gain{\vec{\theta}}$.

We use variational optimization to maximize the score of a linear black box game $(\gamma,\Gbf)$. In this case, the gain is expressed as
\begin{equation}\label{eq:nonclassicality_gain_fn}
    \gain{\vec{\theta}} = \ip{\Gbf, \Pbf^{\Net}(\vec{\theta})}
\end{equation}
where the transition probabilities of $\Pbf^{\Net}(\vec{\theta})$ are the measurement probabilities obtained when simulating the network with settings $\vec{\theta}$.  We refer to the parameterized quantum network simulation circuit as the variational ansatz. Applying gradient ascent, we maximize the gain to achieve the violation of the classical bound. We now describe our general variational optimization algorithm, which is a basic application of gradient ascent \cite{ruder2016overview}.

\begin{algorithm}\label{alg:variational_optimization}
    \textbf{Maximize a Quantum Network's Score in a Black Box Game:}
    \vspace{5pt}

    \noindent \textbf{Goal:} For a black box game having reward matrix $\Gbf$, solve $\max_{\vec{\theta}\in\reals^{M}} \ip{\Gbf, \Pbf^{\Net}(\vec{\theta})}$ for a given variational network ansatz $\Pbf^{\Net}(\vec{\theta})$. The algorithm is iterative and requires a \texttt{num\_steps} parameter specifying the number of iterations to take.

    \begin{enumerate}
        \item Select the input settings $\vec{\theta}_0\in \reals^M$ at random and initialize a log of settings-cost tuples $\texttt{LOG} = [(\vec{\theta}_0, \gain{\vec{\theta}_0})]$.
        \item For $i$ in $\{0, \dots, \texttt{num\_steps}-1\}:$\begin{enumerate}
            \item For settings $\vec{\theta}_i$, evaluate the gradient $\nabla \gain{\vec{\theta}_i}$.
            \item Update the settings by taking a step of size $\eta$ along the path of steepest ascent as $\vec{\theta}_{i+1} = \vec{\theta}_i + \eta \nabla \gain{\vec{\theta}_i}$.
            \item Append the tuple $(\vec{\theta}_{i+1}, \gain{\vec{\theta}_{i+1}})$ to $\texttt{LOG}$.
        \end{enumerate}
        \item Return the tuple $(\vec{\theta}^\star,\gain{\vec{\theta}^\star})$ that has the minimum cost in \texttt{LOG}.
    \end{enumerate}

    \begin{remark}
        In practice, we use the Adam \cite{kingma2014adam} optimizer to dynamically adjust the step-size $\eta$ in step 2.b.
    \end{remark}

    \begin{remark}
        The global optimum is not guaranteed to be found, however, the maximal gain will necessarily be a lower bound on the global maximum. It is best practice to repeatedly perform the gradient ascent procedure with randomized settings $\vec{\theta}_{\text{init}}$ each time, or apply a similar method to mitigate the effect of local optima.
    \end{remark}
\end{algorithm}

Algorithm~\ref{alg:variational_optimization} serves as a useful tool in our operation framework because when a nonclassicality witness $(\gamma, \Gbf)$ is considered as the gain function in Eq.~\eqref{eq:nonclassicality_gain_fn}, its maximized violation represents an explicit quantum advantage. In other words, Algorithm~\ref{alg:variational_optimization} can either maximize the violation of a facet inequality or minimize the error in a simulation game, and in each case, the optimal settings for the network are obtained as output.

\begin{algorithm}\label{alg:vqo_nonclassicality}
    \textbf{Establish Nonclassicality in a Quantum Network:}
    \vspace{5pt}
    
    \noindent\textbf{Goal:} Given a nonclassicality witness $(\gamma, \Gbf)$ and a variational ansatz circuit that simulates a quantum network as $\Pbf^{\Net}(\vec{\theta})$, establish a maximally nonclassical behavior.

    \begin{enumerate}
        \item Apply Algorithm~\ref{alg:variational_optimization} to obtain the optimal settings $\vec{\theta}^\star$ and the maximal score $\ip{\Gbf, \Pbf^{\Net}(\vec{\theta}^\star)}$.
        \item If $\ip{\Gbf, \Pbf^{\Net}(\vec{\theta}^\star)} > \gamma$, the variational ansatz demonstrates nonclassicality for settings $\vec{\theta}^\star$. Otherwise, $\Pbf^{\Net}(\vec{\theta}^\star)$ is classically simulable.
    \end{enumerate}
\end{algorithm}

\begin{algorithm}\label{alg:vqo_simulation}
    \textbf{Establish a Deterministic Protocol in a Quantum Network:}
    \vspace{5pt}

    \noindent \textbf{Goal:} Given a variational ansatz circuit that simulates a quantum network as $\Pbf^{\Net}(\vec{\theta})$, establish a behavior that approximately simulates the deterministic behavior $\Vbf$ within an allowed tolerance $\epsilon$ such that
    \begin{equation}\label{eq:sim_error_to_max_game_score}
        \min_{\vec{\theta} \in \reals^M} \Delta(\Vbf, \Pbf^{\Net}(\vec{\theta}))  \leq \epsilon.
    \end{equation}

    \begin{enumerate}
        \item Apply Algorithm~\ref{alg:variational_optimization} to obtain the optimal settings $\gamma^\star = \max_{\vec{\theta}}\ip{\Vbf, \Pbf^{\Net}(\vec{\theta}^\star)}$.
        \item Check the error tolerance using $\gamma^\star$ and Eq.~\eqref{eq:deterministic_behavior_simulation_error}. If the following inequality holds,
        \begin{equation}
            \Delta(\Vbf, \Pbf^{\Net}(\vec{\theta})) = 1 - \frac{1}{|\Xmc|}\gamma^\star  \leq \epsilon,
        \end{equation}
        then $\Vbf$ is approximately simulated, otherwise, the simulation fails.
        \item \textbf{Return:} The optimal settings $\vec{\theta}^\star$ and the simulation error $P^\star_{\Error} = 1 - \frac{1}{|\XNet|} \ip{\Vbf, \Pbf^{\Net}(\vec{\theta}^\star)}$.
    \end{enumerate}
\end{algorithm}

\begin{remark}
    Upon failure to surpass the specified threshold in either Algorithm~\ref{alg:vqo_nonclassicality} or Algorithm~\ref{alg:vqo_simulation}, the variational optimization in Algorithm~\ref{alg:variational_optimization} can be rerun for another set of randomized initial setting $\vec{\theta}_{\text{init}}$. After a set number of retry attempts, the algorithm exits indicating that no violation of classicality or simulation within the allowed tolerance $\epsilon$ has been found.
\end{remark}

It is important to note that Algorithm~\ref{alg:vqo_nonclassicality} places a lower bound on the quantum violation and  Algorithm~\ref{alg:vqo_simulation} places an upper bound on the simulation error.   
Therefore, it is necessary for these algorithms to operate against a user-specified tolerance of error in the established protocol or max quantum violation. 
Although the maximal possible violation of a linear nonclassicality witness is given by Eq.~\eqref{eq:maximal_black_box_game_score}, we find in practice that the maximal quantum violation does not usually approach this upper bound, meaning that our methods place upper and lower bounds on the max quantum violation.
Nevertheless, any amount of violation of a classical bound demonstrates a quantum advantage. In our results, we show that in all cases where the maximal quantum violation is known, our VQO algorithm converges to the maximal violation. Hence we provide anecdotal evidence that VQO methods obtain the maximal violations. 

The applied variational quantum optimization techniques offer many advantages. First, the network's causal structure, communication resources, and free operations are explicitly encoded into the variational ansatz circuit. As a result, a successful optimization returns the settings $\vec{\theta}^\star$ that achieve the maximal gain, thereby providing both the optimal value and a quantum circuit that achieves the desired behavior. Moreover, these methods are compatible with both quantum computing and quantum networking hardware \cite{Cerezo2021_vqa}. Although quantum hardware is generally noisy, 
the merit of our framework is that it's semi-device independent and agnostic to noise provided that a violation is found. Furthermore, the variational optimization can adaptively mitigate the effects due to noise \cite{doolittle_2023_vqo_nonlocality, chen_2022_inferring_network_topology, doolittle_qNetVO, Ma2021_noisy_parameter-shift, doolittle_phd_thesis_2023}.
\rev{VQO methods also present challenges, such as belonging to an NP-hard complexity class \cite{bittel2021_training_vqa_is_np-hard}, or exhibiting local optima and barren plateaus \cite{Cerezo2021_vqa}. As general solutions to these problems are developed, they can be incorporated into our operational framework.}{}

\rev{

\subsubsection{An Operational Framework for Realizing Communication Advantages in Quantum Networks}\label{section:operational_framework}

\begin{figure}
    \centering
    \includegraphics[width=0.8\columnwidth]{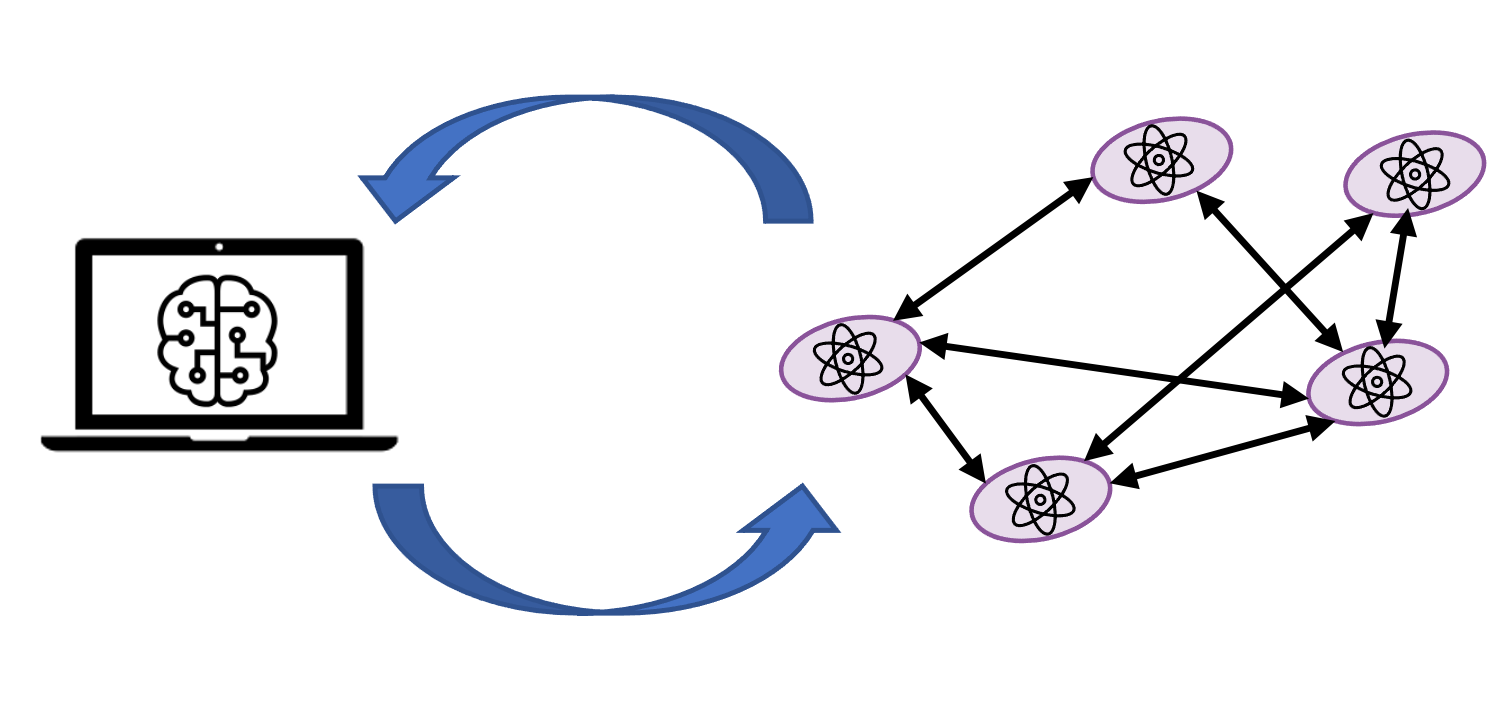}
    \caption{VQO could be used to automatically establish and maintain protocols on noisy quantum networking hardware.}
    \label{fig:variational_quantum_networking}
\end{figure}

In this section, we discuss the merits of our operational framework, including its versatility and scalability. When scaling our methods, the main challenges are the exponential growth of the Hilbert space dimension with the number of entangled qubits and the exponential growth of network behavior dimension with the number of devices and inputs. As a result, it is difficult to scale quantum network simulation and optimization, as well as nonclassicality witness derivation. Simulation games and quantum hardware offer scalable solutions to these problems.

Simulation games offer an efficient approach for obtaining nonclassicality witnesses tailored to a desired communication task. 
Recall that any deterministic behavior $\Vbf^\star\in\Vbb_{\vec{\Ymc}|\vec{\Xmc}}$ can be selected as a simulation game where Algorithm~\ref{alg:vqo_simulation} optimizes a network to exhibit the deterministic behavior $\Vbf^\star$. Thus, our framework can be used to automatically establish communication and processing protocols in quantum networks.

The classical bound for the simulation game, $\gamma$, is needed to quantify the communication advantage. Ideally, the classical bound $\gamma$ can be analytically derived, or estimated like the classical simulation cost in Section~\ref{section:quantifying-certifying-nonclassicality}. Namely, a lower bound $\gamma^-$ is estimated by considering a subset of vertices
\begin{equation}
    \gamma \geq \gamma^-= \max \{\ip{\Vbf^\star, \Vbf} | \Vbf \in \Vbb'\subset \Vbb^{\Net}\}
\end{equation}
where the subset $\Vbb'\subset \Vbb^{\Net}$ could be selected at random or by clever algorithm. Since $\Cbb^{\Net}\subseteq\Qbb^{\Net}$, an upper bound $\gamma^+$ can be estimated using Algorithm~\ref{alg:vqo_simulation} to optimize
\begin{equation}\label{eq:classical_bound_estimation_vqo}
    \gamma \leq \gamma^+ \approx \textstyle \max_{\Pbf\in\Qbb^{\Net}} \ip{\Vbf^\star, \Pbf},
\end{equation}
where we maximize over quantum communication in place of classical.
The upper bound $\gamma^+$ can  be used either to witness nonclassicality in entanglement-assisted communication networks or to lower bound the classical simulation cost (see Section~\ref{section:quantifying-certifying-nonclassicality}).

Since we simulate quantum communication networks as quantum circuits, we address the Hilbert space dimension challenge by evaluating the quantum circuit on a quantum computer. As quantum computers improve and scale, efficient simulation and optimization of quantum networks will be enabled. To accommodate network behaviors with larger dimensions, parallel simulations can be run for each distinct network input $\xv \in \XNet$, allowing computations to be accelerated by multiple quantum processors.

Our operational framework can also be deployed on quantum networks (see Fig.~\ref{fig:variational_quantum_networking}) to establish or maintain nonclassical behaviors (Algorithm~\ref{alg:vqo_nonclassicality}) or deterministic behaviors (Algorithm~\ref{alg:vqo_simulation}).
Remarkably, the applied VQO methods have demonstrated the ability to optimize quantum network protocols against uncharacterized noise models \cite{doolittle_2023_vqo_nonlocality,Ma2021_noisy_parameter-shift, chen_2022_inferring_network_topology, doolittle_phd_thesis_2023}.
These optimization approaches could be useful for automating tasks for protocol calibration and resource certification in noisy quantum networks.
}{}

\section{Results}\label{section:results}

In this section, we apply our operational framework to a wide range of communication networks and resource configurations including bipartite communication scenarios (Section~\ref{section:nonclassicality_bipartite_communication}), multiaccess networks (Section~\ref{section:nonclassicality_multiaccess_networks}), broadcast networks (Section~\ref{section:nonclassicality_broadcast_networks}), and multipoint communication networks (Section~\ref{section:nonclassicality_multipoint_networks}).
All supporting software and numerics can be found on GitHub  \cite{supplemental_software} and additional details can be found in the Supplemental Code Section~\ref{section:Supplemental_Code}.

For each communication network we obtain simulation games and facet inequalities that bound the classical network polytope $\Cbb^{\Net}$.
For each of the network's quantum resource configurations, we maximize the violation of each nonclassicality witness using the variational methods in  Algorithm~\ref{alg:variational_optimization} and  evaluate the noise robustness of these violations using Eq.~\ref{eq:noise_robustness}. Finally, we compare the relative communication advantage offered by each quantum resource configuration. 

\rev{
Our main results include a numerical survey of nonclassicality in communication networks with supporting analytical examples. For each network, we derive novel nonclassicality witnesses and their associated quantum violations. Notable examples include protocols for a dense two-bit XOR advantage in a quantum multiaccess network with entanglement-assisted senders (Protocol~\ref{protocol:bitwise_xor_behavior}), a protocol for checking trit equality in a classical multiaccess network with entanglement-assisted senders (Protocol~\ref{protocol:multiaccess_cmac_etx}), and a protocol for nonclassicality in a quantum broadcast network with entanglement-assisted receivers (Protocol~\ref{protocol:earx_broadcast_strategy}). From our numerical results, we draw three key insights: 1) entanglement is sufficient for nonclassicality in all networks, 2) quantum communication is sufficient for nonclassicality in all networks that have multiple independent senders, and 3) entanglement is necessary for nonclassicality in broadcast networks, which is proven by Theorem~\ref{thm:broadcast_classicality}. 
}{}

\subsection{Bipartite Communication Scenarios}\label{section:nonclassicality_bipartite_communication}

\begin{figure*}
    \small
    \resizebox{\textwidth}{!}{%
    \begin{tabular}{c c c c c}
        {\normalsize (a) Point-to-Point} & & {\normalsize (b) Prepare-and-Measure} & & {\normalsize (c) Bell Scenario with  Signaling}   \\
        \hfill \\
        \begin{tikzpicture}
            \node[terminal] (x) at (-0.1,1) {$x_1$};
            \node[dev] (A) at (1.4,1) {$P^{A}_{a|x_1}$};
            \node[dev] (B) at (3, -0.5) {$P^B_{y_2|a}$};
            \node[terminal] (y) at (4.5,-0.5) {$y_2$};

            \path (x) \cedge (A);
            \path (A) \cedge  node[el, above=3pt, xshift=4pt] {$a$} (B);
            \path (B) \cedge (y);
        \end{tikzpicture} & & \begin{tikzpicture}
            \node[terminal] (x1) at (-0.1,1) {$x_1$};
            \node[terminal] (x2) at (-0.1,-0.5) {$x_2$};
            \node[dev] (A) at (1.4, 1) {$P^{A}_{a|x_1}$};
            \node[dev] (B) at (3, -0.5) {$P^B_{y_2|ax_2}$};
            \node[terminal] (y) at (4.5,-0.5) {$y_2$};    
        
            \path (x1) \cedge (A);
            \path (x2) \cedge (B);
            \path (A) \cedge  node[el, above=3pt, xshift=4pt] {$a$} (B);
            \path (B) \cedge (y);
        \end{tikzpicture}  & & \begin{tikzpicture}
            \node[terminal] (x1) at (-0.1,1) {$x_1$};
            \node[terminal] (x2) at (-0.1,-0.5) {$x_2$};
            \node[dev] (A) at (1.4,1) {$P^{A}_{ay_1|x_1}$};
            \node[dev] (B) at (3.0, -0.5) {$P^B_{y_2|ax_2}$};
            \node[terminal] (y1) at (4.5, 1) {$y_1$};
            \node[terminal] (y2) at (4.5, -0.5) {$y_2$};
        
            \path (x1) \cedge (A);
            \path (x2) \cedge (B);
            \path (A) \cedge  node[el, above=3pt, xshift=4pt] {$a$} (B);
            \path (B) \cedge (y2);
            \path (A) \cedge (y1);
        \end{tikzpicture}\\
    \end{tabular}
    }
    \caption{Classical bipartite signaling scenario DAGs in which a sender device $A$ and a receiver device $B$.
    }
    \label{fig:bipartite_communication_scenarios}
\end{figure*}
\begin{figure*}
    \centering
    \scriptsize
    \resizebox{\textwidth}{!}{%
    \begin{tabular}{c c c}
        {\normalsize (a) QC } & {\normalsize (b) EACC} & {\normalsize (c) EAQC }\\
        \hfill \\
        \begin{tikzpicture}
            \node[terminal] (x) at (-0.15,1) {$x_1$};
            \node[prep_dev] (A) at (1.4,1) {$\rho^A_x$};
            \node[meas_dev] (B) at (2.8, -1) {$\Pi^B_{y|x_2}$};
            \node[terminal] (y) at (4,-1) {$y$};
            \node[terminal] (x2) at (-0.15, -1) {$x_2$};
        
            \path (x) \cedge (A);=
            \path (A) \qedge  node[el, above=3pt, xshift=4pt] {$\rho^A_x$} (B);
            \path (B) \cedge (y);
            \path (x2) \cedge (B);
        \end{tikzpicture} & \begin{tikzpicture}
            \node[terminal] (x) at (-0.15,1) {$x_1$};
            \node[qsource] (lambda) at (-0.15,0) {$\rho^{\Lambda}$};
            \node[meas_dev] (A) at (1.4,1) {$\Pi^A_{a|x_1}$};
            \node[meas_dev] (B) at (2.8, -1) {$\Pi^B_{y|a,x_2}$};
            \node[terminal] (y) at (4.2,-1) {$y$};
            \node[terminal] (x2) at (-0.15, -1) {$x_2$};
        
            \path (x) \cedge (A);
            \path (x2) \cedge (B);
            \path (lambda) \qedge node[el, below=-2pt,xshift=4pt] {$\rho^{\Lambda}_1$} (A);
            \path (lambda) \qedge node[el, below=-8pt,xshift=-24pt] {$\rho^{\Lambda}_2$} (B);
            \path (A) \cedge  node[el, above=3pt, xshift=4pt] {$a$} (B);
            \path (B) \cedge (y);
        \end{tikzpicture} &
        \begin{tikzpicture}
            \node[terminal] (x) at (-0.15,1) {$x_1$};
            \node[qsource] (lambda) at (-0.15,0) {$\rho^{\Lambda}$};
            \node[proc_dev] (A) at (1.4,1) {$\Emc^A_{x_1}$};
            \node[meas_dev] (B) at (2.8, -1) {$\Pi^B_{y|a,x_2}$};
            \node[terminal] (y) at (4.2,-1) {$y$};
            \node[terminal] (x2) at (-0.15, -1) {$x_2$};

            \path (x) \cedge (A);
            \path (x2) \cedge (B);
            \path (lambda) \qedge node[el, below=-2pt,xshift=4pt] {$\rho^{\Lambda}_1$} (A);
            \path (lambda) \qedge node[el, below=-8pt,xshift=-24pt] {$\rho^{\Lambda}_2$} (B);
            \path (A) \qedge  node[el, above=3pt, xshift=4pt] {$\rho^A_{x_1}$} (B);
            \path (B) \cedge (y);
        \end{tikzpicture}
        \hfill \\
        && \\
        \hfill \\
        \begin{tikzcd}[column sep=0.15cm] 
            \lstick{$\ket{0}$} & \gate[style={prep_gate}]{U(\vec{\theta}^A_{x_1})}   & \qw{} &  [-1mm]\gate[2, style={meas_gate}]{U(\vec{\theta}^B_{x_2})} & \meter[style={meas_gate}]{} & \cw\rstick[2]{y} \\
               \lstick{$\ket{0}$} &\qw&\qw&&\meter[style={meas_gate}]{} & \cw\\
        \end{tikzcd} &
        \begin{tikzcd}[column sep=0.15cm]
            \lstick{$\ket{0}$} & \qw & \gate[2, style={meas_gate}]{U(\vec{\theta}^A_{x_1})}  & \trash{} & &   \\
            \lstick{$\ket{0}$} & \gate[2, style={prep_gate}]{U(\vec{\theta}^\Lambda)} &  & \meter[style={meas_gate}]{a}\vcw{1} & & \\ 
            \lstick{$\ket{0}$} & & \qw &  \gate[2, style={meas_gate}]{U(\vec{\theta}^B_{a,x_2})} &  \meter[style={meas_gate}]{} & \cw   \rstick[2]{$y$}\\
            \lstick{$\ket{0}$}& \qw & \qw  &   & \meter[style={meas_gate}]{} & \cw  \\
        \end{tikzcd} & \begin{tikzcd}[column sep=0.15cm]
            \lstick{$\ket{0}$} & \gate[2, style={prep_gate}]{U(\vec{\theta}^\Lambda)}  & \gate[style={proc_gate}]{U(\vec{\theta}^A_{x_1})} & \gate[2, style={meas_gate}]{U(\vec{\theta}^B_{x_2})} & \meter[style={meas_gate}]{} & \rstick[2]{y} \cw \\
            \lstick{$\ket{0}$} & & \qw & &  \meter[style={meas_gate}]{} & \cw \\
        \end{tikzcd} \\
    \end{tabular}%
    }
    \caption{Point-to-point and prepare and measure DAGs and variational ansatz circuits. (a) Quantum communication, (b) entanglement-assisted classical communication, and (c) entanglement-assisted quantum communication. In figure (b) the classical measurement result $a$ is used to condition the applied measurement. }
    \label{fig:point-to-point_circuit_ansatzes}
\end{figure*}

\begin{figure*}
    \centering
    \scriptsize
    \resizebox{\textwidth}{!}{%
    \begin{tabular}{c c c}
        {\normalsize  (a) QC} & {\normalsize (b) EACC } & {\normalsize (c) EAQC } \\
        \hfill \\
        \begin{tikzpicture}
            \node[terminal] (x) at (-0.15,1) {$x_1$};
            \node[prep_dev] (A) at (1,1) {$\rho^A_{x_1}$};
            \node[meas_dev] (Ameas) at (2.5, 1) {$\Pi^{A'}_{y_1|x_1}$};
            \node[meas_dev] (B) at (2.5, -1) {$\Pi^B_{y_2|x_2}$};
            \node[terminal] (y1) at (4, 1) {$y_1$};
            \node[terminal] (y2) at (4,-1) {$y_2$};
            \node[terminal] (x2) at (-0.15, -1) {$x_2$};

            \draw[rounded corners, quantum_purple,dashed,style={line width=1.5pt}] (0.2, 0.6) rectangle (3.2, 1.45);
        
            \path (x) \cedge (A);
            \path (Ameas) \cedge (y1);
            \path (A) \qedge  node[el, above=2pt, xshift=4pt] {$\rho^A_{x_1}$} (B);
            \path (A) \qedge (Ameas);
            \path (B) \cedge (y2);
            \path (x2) \cedge (B);
        \end{tikzpicture} & \begin{tikzpicture}
            \node[terminal] (x) at (-0.15,1) {$x_1$};
            \node[qsource] (lambda) at (-0.15,0) {$\rho^{\Lambda}$};
            \node[meas_dev] (A) at (1.6,1) {$\Pi^A_{a,y_1|x_1}$};
            \node[meas_dev] (B) at (2.6, -1) {$\Pi^B_{y_2|a,x_2}$};
            \node[terminal] (y1) at (4.2, 1) {$y_1$};
            \node[terminal] (y2) at (4.2,-1) {$y_2$};
            \node[terminal] (x2) at (-0.15, -1) {$x_2$};
        
            \path (x) \cedge (A);
            \path (A) \cedge (y1);
            \path (x2) \cedge (B);
            \path (lambda) \qedge node[el, below=-2pt,xshift=4pt] {$\rho^{\Lambda}_1$} (A);
            \path (lambda) \qedge node[el, below=-4pt,xshift=-24pt] {$\rho^{\Lambda}_2$} (B);
            \path (A) \cedge  node[el, above=3pt, xshift=4pt] {$a$} (B);
            \path (B) \cedge (y2);
        \end{tikzpicture} &
        \begin{tikzpicture}
            \node[terminal] (x) at (-0.15,1) {$x_1$};
            \node[qsource] (lambda) at (-0.15,0) {$\rho^{\Lambda}$};
            \node[proc_dev] (A) at (1.1,1) {$\Emc^A_{x_1}$};
            \node[meas_dev] (Ameas) at (2.5, 1) {$\Pi^{A'}_{y_1|x_1}$};
            \node[meas_dev] (B) at (2.8, -1) {$\Pi^B_{y_2|x_2}$};
            \node[terminal] (y1) at (4.2, 1) {$y_1$};
            \node[terminal] (y2) at (4.2,-1) {$y_2$};
            \node[terminal] (x2) at (-0.15, -1) {$x_2$};

            \draw[rounded corners, quantum_purple,dashed,style={line width=1.5pt}] (0.3, 0.6) rectangle (3.2, 1.45);

            \path (x) \cedge (A);
            \path (A) \qedge (Ameas);
            \path (Ameas) \cedge (y1);
            \path (x2) \cedge (B);
            \path (lambda) \qedge node[el, below=-1pt,xshift=4pt] {$\rho^{\Lambda}_1$} (A);
            \path (lambda) \qedge node[el, below=-4pt,xshift=-24pt] {$\rho^{\Lambda}_2$} (B);
            \path (A) \qedge  node[el, above=3pt, xshift=4pt] {$\rho^A_x$} (B);
            \path (B) \cedge (y2);
        \end{tikzpicture}
        \hfill \\
        && \\
        \hfill \\
        \begin{tikzcd}[column sep=0.15cm]
            \lstick{$\ket{0}$} & \qw & \gate[2, style={meas_gate}]{U(\vec{\theta}^{A'}_{x_1})} & \meter[style={meas_gate}]{} & \rstick[2]{$y_1$}\cw \\
            \lstick{$\ket{0}$} & \gate[style={prep_gate}, 2]{U(\vec{\theta}^A_{x_1})}   & \qw{}  & [-1mm]\meter[style={meas_gate}]{} & \cw \\
            \lstick{$\ket{0}$} & & \qw{} & [-1mm]\gate[style={meas_gate}]{U(\vec{\theta}^B_{x_2})} & \meter[style={meas_gate}]{}  & \cw\rstick{$y_2$} \\
        \end{tikzcd} &
        \begin{tikzcd}[column sep=0.15cm]
            \lstick{$\ket{0}$} & \qw & \gate[3, style={meas_gate}]{U(\vec{\theta}^A_{x_1})}  & \meter[style={meas_gate}]{} & \cw\rstick[2]{$y_1$} & \\
            \lstick{$\ket{0}$} & \qw &   & \meter[style={meas_gate}]{} & \cw &   \\ 
            \lstick{$\ket{0}$} & \gate[2, style={prep_gate}]{U(\vec{\theta}^\Lambda_{x_1})} &  & \meter[style={meas_gate}]{a}\vcw{1} & & \\ 
            \lstick{$\ket{0}$} & & \qw &  \gate[2, style={meas_gate}]{U(\vec{\theta}^B_{a,x_2})} &  \meter[style={meas_gate}]{} & \cw   \rstick[2]{$y_2$}\\
            \lstick{$\ket{0}$}& \qw & \qw  &   & \meter[style={meas_gate}]{} & \cw  \\
        \end{tikzcd} & \begin{tikzcd}[column sep=0.15cm]
            \lstick{$\ket{0}$} & \qw{} & \qw{} & \gate[2, style={meas_gate}]{U(\vec{\theta}^{A'}_{x_1})} & \meter[style={meas_gate}]{} &  \rstick[2]{$y_1$} \cw \\
            \lstick{$\ket{0}$} & \qw{} & \gate[2, style={proc_gate}]{U(\vec{\theta}^A_{x_1})} &  & \meter[style={meas_gate}]{} & \cw{} \\
            \lstick{$\ket{0}$} & \gate[2, style={prep_gate}]{U(\vec{\theta}^\Lambda)}  &  & \gate[2, style={meas_gate}]{U(\vec{\theta}^B_{x_2})} & \meter[style={meas_gate}]{} & \rstick[2]{$y_2$} \cw \\
            \lstick{$\ket{0}$} & & \qw & &  \meter[style={meas_gate}]{} & \cw \\
        \end{tikzcd} \\
    \end{tabular}%
    }
    \caption{Bell scenario with communication DAGs and variational ansatz circuits. (a) Quantum communication, (b) entanglement-assisted classical communication, and (c) entanglement-assisted quantum communication. In figure (b) the classical measurement result $a$ is used to condition the applied measurement.}
    \label{fig:bell_scenarios_aux_communication_dags}
\end{figure*}

\rev{In this section, we apply our methods to bipartite communication scenarios. We validate our operational framework for nonclassicality against established results and obtain novel examples of nonclassicality.}{}
In a bipartite communication scenario a sender $A$ and a receiver $B$ communicate over the one-way channel $\id^{A\to B}_d$ with signaling dimension $d$. We consider three distinct signaling scenarios: the point-to-point communication scenario (Section~\ref{section:point-to-point}), the prepare-and-measure scenario (Section~\ref{section:prepare-and-measure_scenario}), and the Bell scenario with signaling (Section~\ref{section:bell_scenario_aux_signaling}).

For each scenario, we obtain nonclassicality witnesses and maximize their violation over three quantum resource configurations: unassisted quantum communication ($\Qbb^{\Net}$), entanglement-assisted classical communication ($\Cbb^{\Net}_{\EA}$), and entanglement-assisted quantum communication ($\Qbb^{\Net}_{\EA}$) (see Fig.~\ref{fig:point-to-point_circuit_ansatzes}). Note that we only consider two-qubit entanglement and/or qubit communication channels with $d=2$.

\rev{
In each bipartite communication scenario, we show that our numerical results are consistent with the results of previous works \cite{Frenkel2015_classical_information_n-level_quantum_system,Frenkel2022ea_signaling_dim_violation,bennet1992_dense_coding,chitambar_2023_cv_channel,ambainis2008quantum,Heinosaari2022,gallego2010_dim_witnessing,Tavakoli-2015a,pawlowski2012_eacc_qc_simulation,Tavakoli2022_ea_communications}. In each case, we also find examples of nonclassicality that have not been reported to our knowledge.
Each example of bipartite communication nonclassicality  demonstrates the communication advantage of quantum resources in the simplest and most fundamental settings, offering new avenues for certifying entanglement-assisted communications and developing semi-device-independent protocols.}{}

\subsubsection{Point-to-Point Networks}\label{section:point-to-point}

The point-to-point network is a bipartite communication scenario, denoted as $\sigarb{\Xmc_1}{d}{\Ymc_2}$, in which the sender's classical input $x_1\in\Xmc_1$ is encoded into a message and sent to the receiver who outputs the classical value $y_2\in\Ymc_2$. Naturally, we consider the point-to-point network polytope $\CSigd_{}$, for which $d < \min\{|\XNet_1|, |\YNet_2|\}$ must hold, otherwise $\CSigd_{} = \Pbb_{\YNet_2|\XNet_1}$\cite{doolittle_2021_certify_classical_cost}. Violations to the facet inequalities of $\CSigd_{}$ witness the communication advantage of entanglement-assisted communication channels as having a larger signaling dimension than $d$. In Ref. \cite{doolittle_2021_certify_classical_cost}, the complete set of facet inequalities bounding the signaling dimension for $d=2$ were derived (see Table~\ref{Table:d2_signaling_dimension_witness}), which we investigate using our framework.

In particular, the nonclassicality witness $(2, \Fbf^b_{4\to 4})$ of Table~\ref{Table:d2_signaling_dimension_witness} describes an important  simulation game,  $\Vbf^{\CV}_{4 \to 4}=\Fbf_{4\to 4}^{b}=\Ibb_4$, which we refer to as the \textit{communication value} ($\CV$) game because the quantity $\max_{\Pbf\in\Qbb^{\Net}}\ip{\Vbf^{\CV}, \Pbf}$ is the communication value of a quantum channel \cite{chitambar_2023_cv_channel}.  In general, the CV game is expressed as $(d, \Vbf^{\CV}_{N\to N})$ where $\Vbf^{\CV}_{N \to N} = \Ibb_N$.  Note that this game was introduced in reference \cite{doolittle_2021_certify_classical_cost} as the maximum likelihood game and shown to be a facet inequality for all point-to-point network polytopes $\CSigd_{}$ where $1 < d < |\Xmc|=|\Ymc|$. 

\begin{figure}
    \centering
    \includegraphics[width=\columnwidth]{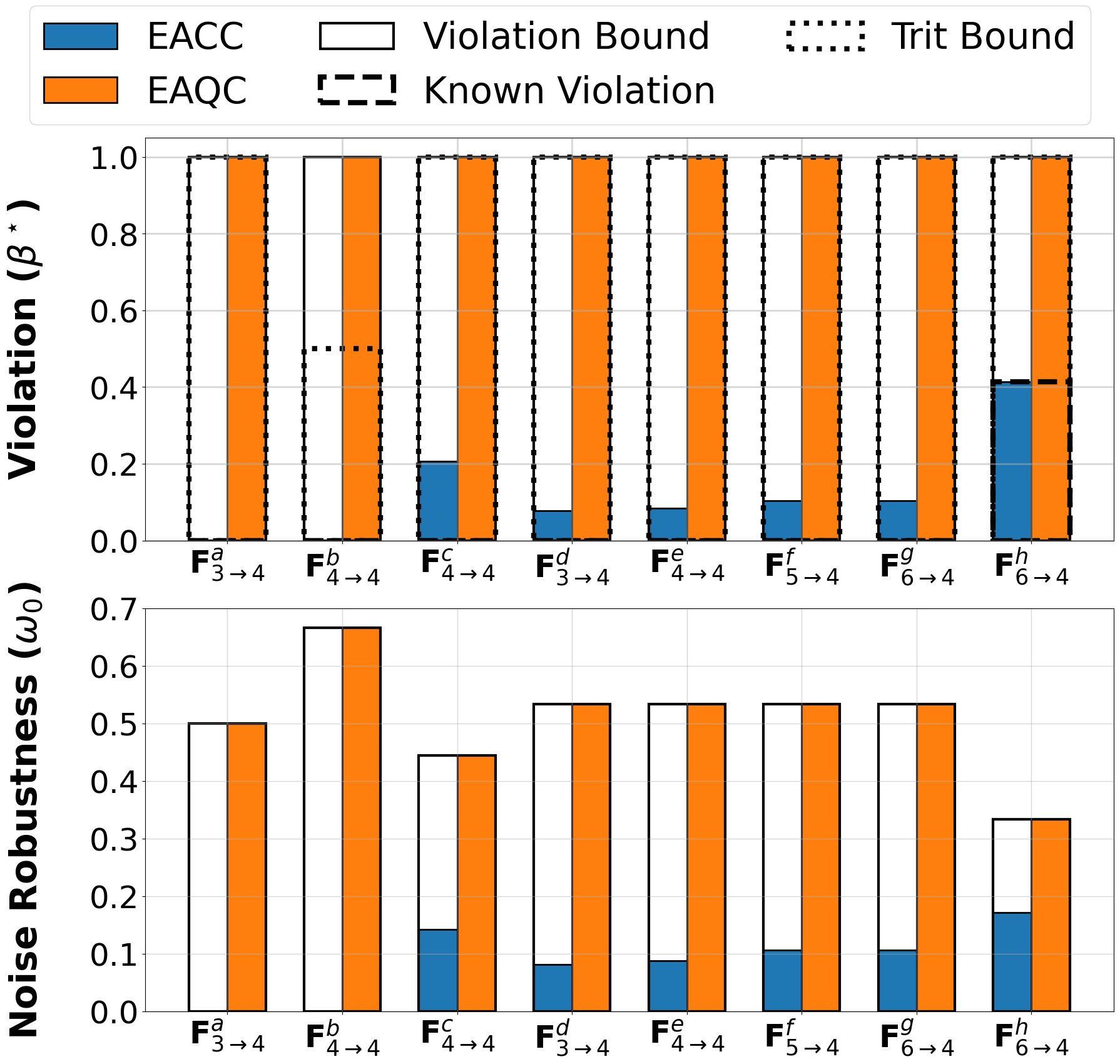}
    \caption{Point-to-point network violations (top) and noise robustness (bottom) using  entanglement-assisted quantum and  classical communication resources. The $x$-axis shows each linear nonclassicality witness in Table~\ref{Table:d2_signaling_dimension_witness} and the $y$-axis shows the maximal violation and noise robustness achieved by optimizing the variational ans\"{a}tze in Fig.~\ref{fig:point-to-point_circuit_ansatzes} (b) and (c). 
    The solid black outline shows the maximal possible violation for each nonclassicality witness, the dotted line shows the maximal violation for unassisted classical signaling of dimension $d=3$. The black dashed line shows the EACC violation of $\Fbf_{6\to 4}^{h}$ reported in Ref.~\cite{Frenkel2022ea_signaling_dim_violation}.}
    \label{fig:ea_violations_of_qubit_signaling_dimension}
\end{figure}

In the point-to-point signaling scenario, unassisted quantum communication does not demonstrate nonclassicality (i.e. $\Cbb^{\Xmc\text{\stackon[1pt]{$\to$}{$\scriptstyle d$}} \Ymc}=\Qbb^{\Xmc\text{\stackon[1pt]{$\to$}{$\scriptstyle d$}} \Ymc}$) \cite{Frenkel2015_classical_information_n-level_quantum_system}. However, nonclassical behaviors can be found using entanglement-assisted classical communication (EACC) or entanglement-assisted quantum communication (EAQC), $\Cbb^{\Xmc\text{\stackon[1pt]{$\to$}{$\scriptstyle d$}} \Ymc}_{\EA}$ and $\Qbb^{\Xmc\text{\stackon[1pt]{$\to$}{$\scriptstyle d$}} \Ymc}_{\EA}$, respectively.

For the case where $d=2$, we use variational optimization to maximize the violation of the signaling dimension witnesses in Table~\ref{Table:d2_signaling_dimension_witness}, plotting the results in Fig.~\ref{fig:ea_violations_of_qubit_signaling_dimension}.
The behaviors in the set $\Qbb^{\Xmc\text{\stackon[1pt]{$\to$}{$\scriptstyle 2$}} \Ymc}_{\EA}$  are able to achieve the maximal possible violation for each nonclassicality witness.
In general, EAQC resources can achieve the maximal score since entanglement plus one qubit communication allows for the transmission of two bits due to dense coding \cite{bennet1992_dense_coding}.
Interestingly, a trit  ($d=3$) of classical communication is sufficient to achieve the maximal possible violation for all facet inequalities in Table~\ref{Table:d2_signaling_dimension_witness} except $\Fbf_{4\to4}^b$. Thus, a sufficiently large violation of this inequality can witness EAQC of a qubit from a trit of classical communication ($d=3$). 

For EACC, we find no violations of the inequalities $\Fbf_{3\to 4}^a$ and $\Fbf_{4\to 4}^b$ because entanglement cannot improve the communication value of a classical channel \cite{chitambar_2023_cv_channel}. Nonetheless, EACC resources yield an operational advantage because all other nonclassicality witnesses can be violated. We find that these violations  require that the sender only use entanglement in the encoding for certain values of $x_1\in\Xmc_1$ while the entanglement is otherwise discarded.
For the game $\Fbf_{6\to 4}^h$, our methods successfully reproduce the maximal violation derived by Frenkel \textit{et al.} \cite{Frenkel2022ea_signaling_dim_violation}.

One application of these entanglement-assisted nonclassical behaviors is in semi-device-independent certification of entanglement-assisted communication channels. Suppose that it is known that a bit (or qubit) of communication is used between sender and receiver in a point-to-point network. Then, a violation of any $d=2$ inequality in Table~\ref{Table:d2_signaling_dimension_witness} witnesses the presence of entanglement between the two parties. Furthermore, entanglement-assisted quantum communication can be discerned from entanglement-assisted classical communication by optimizing the communication value. Indeed, such violations serve as a minimal example of a semi-device-independent test that  certifies LOCC and LOQC resources. 

Finally, our results demonstrate that dense-coding protocols can be optimized into a variational circuit ansatz that simulates entanglement-assisted quantum communication in a point-to-point network.   Since $\Vbf^{\CV}$ corresponds to a simulation game, it can be optimized using Algorithm~\ref{alg:vqo_simulation}. Upon completion of the algorithm, the behavior is optimized such that $\Pbf(\vec{\theta}^\star)$ minimizes the state discrimination error, or equivalently, maximizes the communication value of the channel. Remarkably, the minimized simulation error $\Delta(\Vbf^{\CV}, \Pbf(\vec{\theta}^\star)) = P_{\Error}$ precisely characterizes the quality of the established dense-coded channel. Thus, our framework could be used to automatically establish dense-coded channels of a desired quality.

\subsubsection{Prepare-and-Measure Networks}\label{section:prepare-and-measure_scenario}

Prepare-and-measure networks, $\PM(\sigarb{\Xmc_1,\Xmc_2}{d}{\Ymc_2})$, extend point-to-point networks by giving the receiver an independent input $x_2\in\Xmc_2$ (see Fig.~\ref{fig:bipartite_communication_scenarios}.b). This setting has been widely studied in literature, and it is known that QC, EACC, and EAQC can demonstrate nonclassicality. These nonclassical behaviors have known applications in quantum dimensionality witnessing \cite{gallego2010_dim_witnessing}, random access coding \cite{ambainis2008quantum}, and semi-device-independent protocols for key distribution \cite{pawlowski2011_sdi_qkd}, randomness generation \cite{li2011_sdi_qrng}, and quantum resource certification  \cite{moreno2021_sdi_certification_entanglement_sd_coding, tavaokoli2018_pm_self-test}.

The random access coding (RAC) task is a notable nonclassicality witness for prepare-and-measure networks \cite{ambainis2008quantum,Heinosaari2022}. In this task, sender $A$ is given an $n$-bit input $\xv_1=(b_i = \{0,1\})_{i=1}^n$, which is encoded into message and communicated to the receiver $B$ using a channel with signaling dimension $d=2$. Meanwhile, the receiver is given an input $x_2 \in \{0,\dots,n-1\}$, which conditions how the received message is decoded. The game's objective is for the receiver to output the value $y=b_{x_2}$ that corresponds to the $x_2^\text{th}$ bit of the sender's $n$-bit string $\xv_1$.

\begin{figure}[b!]
    \centering
    \includegraphics[width=\columnwidth]{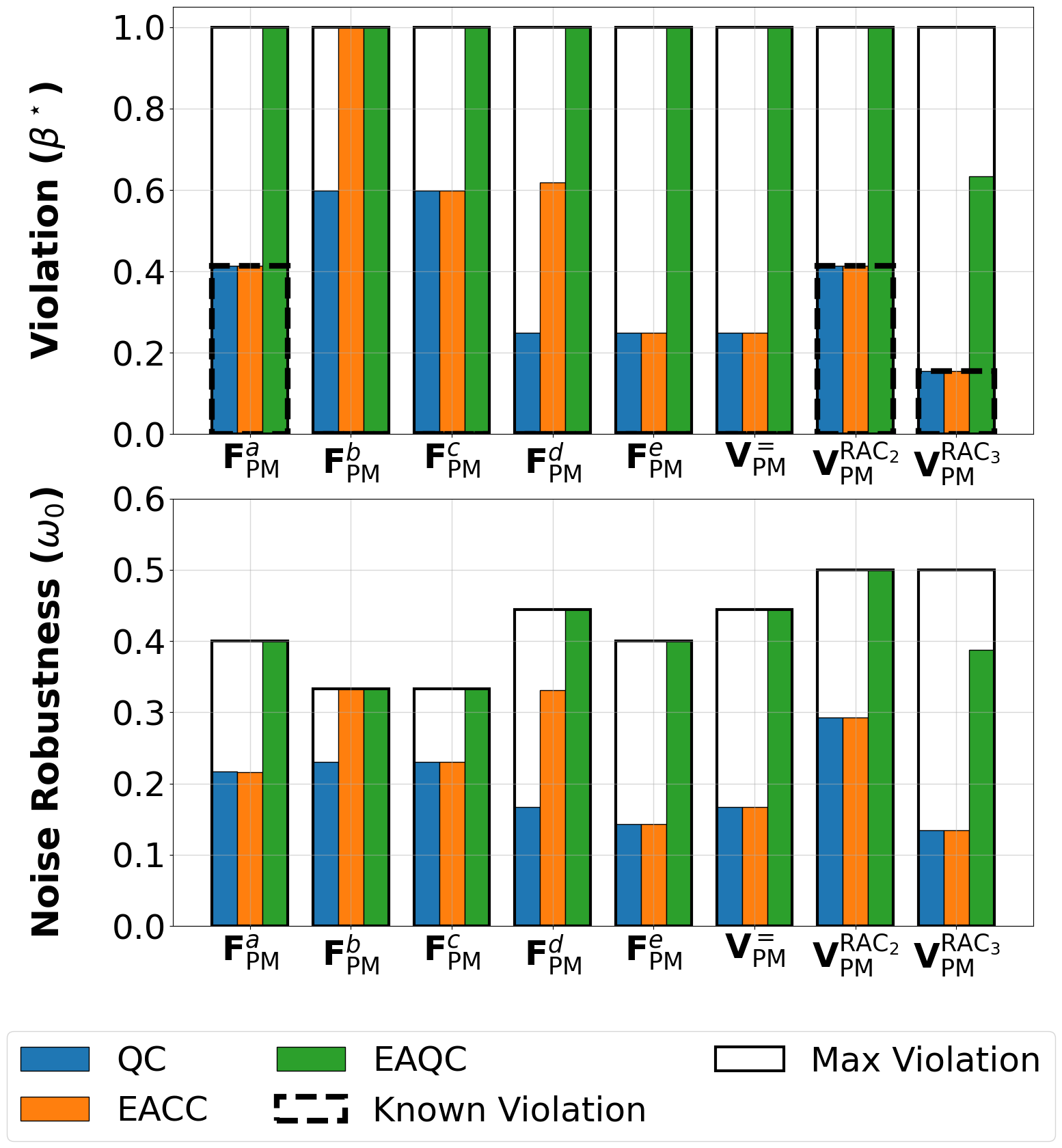}
    \caption{Prepare-and-measure network violations (top) and noise robustness (bottom) using QC, EACC, and EAQC resource configurations. The $x$-axis shows each nonclassicality witness in Table.~\ref{Table:prepare_and_measure-nonclassicality-witnesses} and the $y$-axis shows the maximal violation or noise robustness achieved by optimizing the variational ans\"{a}tze in Fig.~\ref{fig:point-to-point_circuit_ansatzes}. The solid black outlines show the maximal score possible from Eq.~\eqref{eq:maximal_black_box_game_score}.
    The dashed lines show the known violations for the QC case for $\Fbf_{\PM}^a$ \cite{gallego2010_dim_witnessing} and the RAC simulation games, $\Vbf_{\PM}^{\text{RAC}_2}$ and $\Vbf_{\PM}^{\text{RAC}_3}$, (see Eq.~\eqref{eq:quantum_rac_bound}). }
    \label{fig:quantum_violations_of_prepare_and_measure_nonclassicality}
\end{figure}

As a linear nonclassicality witness, the RAC task is expressed as the simulation game 
\begin{align}\label{eq:rac_linear_nonclassicality_witness}
    \gamma^{\text{RAC}_n} = |\XNet|P^{\text{RAC}_n}_{\text{Success}}, \quad V^{\text{RAC}_n}_{y|x_1,x_2} = \delta_{y,b_{x_2}}
\end{align}
where the maximal probability of winning the RAC game using one-way classical communication is \cite{ambainis2008quantum}
\begin{equation}\label{eq:rac_classical_bound}
    P^{\text{RAC}_n}_{\text{Success}} =\frac{1}{2} + \frac{1}{2^n}\binom{n-1}{\lfloor\frac{n-1}{2}\rfloor}.
\end{equation}
The linear nonclassicality witness in Eq.~\eqref{eq:rac_linear_nonclassicality_witness} can be violated using all quantum resource configurations shown in Fig.~\ref{fig:point-to-point_circuit_ansatzes}. 
For the unassisted QC case, the following success probability of the $n=2$ and $n=3$ RAC games can be achieved \cite{Tavakoli-2015a}
\begin{equation}\label{eq:quantum_rac_bound}
    P^{\text{RAC}_n}_{\text{Success}} = \frac{1}{2}\left(1 + \frac{1}{\sqrt{n}}\right).
\end{equation}
This violation of the classical bound in Eq.~\eqref{eq:rac_classical_bound} and demonstrates nonclassicality and an explicit communication advantage in the RAC task.

We list the nonclassicality witnesses  considered for the prepare-and-measure network in Table~\ref{Table:prepare_and_measure-nonclassicality-witnesses}. Three of the considered prepare-and-measure network nonclassicality witnesses have been previously studied. Facet $\Fbf_{\PM}^a$ corresponds to a well known dimensionality witness for the $\PM(\sigarb{3,2}{2}{2})$ network and $\Vbf_{\PM}^{\text{RAC}_n}$ corresponds to the $n$-bit random access simulation game, for which we consider the cases where $n=2$ and $n=3$. The remaining facet inequalities have not been reported to our knowledge where $\Fbf^b_{\PM}$, $\Fbf^c_{\PM}$, and $\Fbf^d_{\PM}$  bound the $\PM(\sigarb{3,3}{2}{2})$ network while $\Fbf^e_{\PM}$ bounds the $\PM(\sigarb{8,3}{2}{2})$.

For each prepare-and-measure nonclassicality witness in Table~\ref{Table:prepare_and_measure-nonclassicality-witnesses}, we optimize the variational ans\"{a}tze for the QC, EACC, and EAQC scenarios depicted in Fig. \ref{fig:point-to-point_circuit_ansatzes}. We plot the respective quantum violations and their noise robustness in Fig.~\ref{fig:quantum_violations_of_prepare_and_measure_nonclassicality}.
We observe in all cases that the maximum violation for EACC resources exceeds that of QC resources, suggesting that EACC is a stronger resource than QC such that
\rev{
\begin{equation}
    \Qbb^{\PM(\sigarb{\Xmc_1,\Xmc_2}{2}{\Ymc_2})} \subseteq \Cbb^{\PM(\sigarb{\Xmc_1,\Xmc_2}{2}{\Ymc_2})}_{\EA}.
\end{equation}
This hierarchy is shown to hold in References~\cite{pawlowski2012_eacc_qc_simulation,Tavakoli2022_ea_communications}, further validating our methods.}{Furthermore, References~\cite{pawlowski2012_eacc_qc_simulation,Tavakoli2022_ea_communications} show that $\Qbb^{\PM(\sigarb{\Xmc_1,\Xmc_2}{2}{\Ymc_2})} \subseteq \Cbb^{\PM(\sigarb{\Xmc_1,\Xmc_2}{2}{\Ymc_2})}_{\EA}$, meaning that one bit of entanglement-assisted classical communication can simulate one qubit of communication with zero error.}
We find in all but one case that EAQC resources are sufficient to achieve the maximal possible violation. The exception being  the three-bit RAC, $\Vbf_{\PM}^{\text{RAC}_3}$, because dense coding only boosts the classical communication to $d^2=4$, meaning that the sender's input alphabet of size $|\Xmc_1|=8$ cannot be fully communicated to the receiver, which prevents EAQC resources from achieving the maximal possible score.

\begin{figure}[b]
    \centering
    \includegraphics[width=\columnwidth]{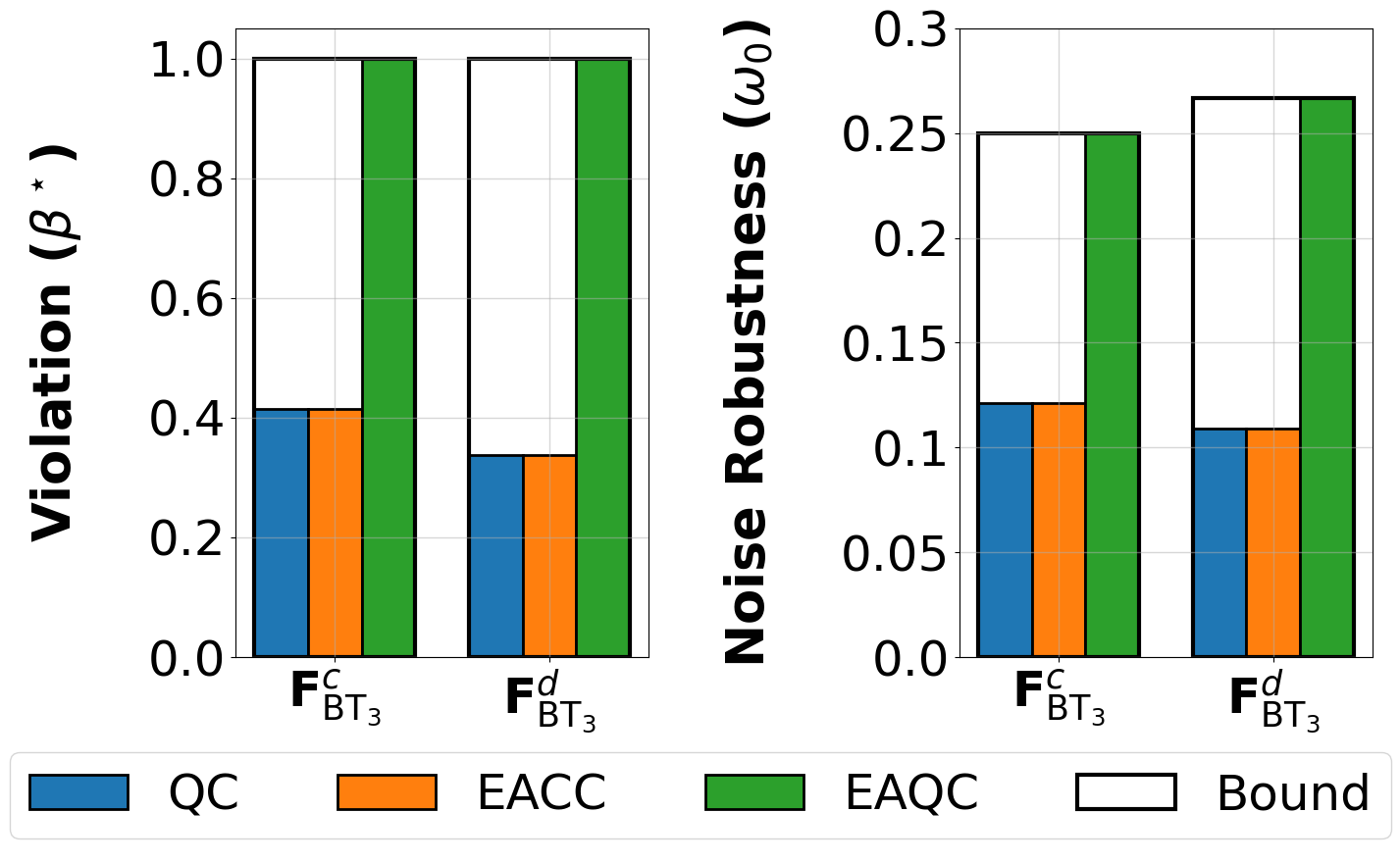}
    \caption{Bell scenarios with communication violations (left) and noise robustness (right). The $x$-axis shows each of the three-input facet inequality from Table~\ref{Table:Bell_inequalities_aux_communication} while the $y$-axis shows either the violation or the noise robustness of the facet inequality.}
    \label{fig:quantum_violations_Bell_sceanrios_aux_communication}
\end{figure}

\subsubsection{Bell Scenarios with Communication}\label{section:bell_scenario_aux_signaling}

In a Bell scenario with auxiliary communication there is one-way communication either from $A$ to $B$ or from $B$ to $A$. We refer to this scenario as the Bacon-Toner scenario \cite{bacon2003_bell_inequalities_aux_communication} and denote it as $\BT(\sigarb{\Xmc_1,\Xmc_2}{d}{\Ymc_1,\Ymc_2})$. The scenario with a  fixed direction of communication has also been studied \cite{Maxwell2014_bell_inequality_aux_comm}, however, if $\Xmc_1 = \Xmc_2$ and $\Ymc_1 = \Ymc_2$, then the device labels $A$ and $B$ can be swapped motivating our relaxation on the direction of communication.
When each device has two outputs, the facet inequalities of the Bacon-Toner scenario have been derived for the cases where each device has two inputs and three inputs, which are respectively denoted as $\BT 2$ and $\BT 3$ (see Table~\ref{Table:Bell_inequalities_aux_communication}).

In Fig.~\ref{fig:quantum_violations_Bell_sceanrios_aux_communication} we plot the violations and noise robustness of these facet inequalities. We only find violations of the inequalities for the $\BT 3$ case, suggesting that quantum resources provide no advantage over one-bit of classical signaling in the two-input case $\BT 2$. We also observe that the violations of unassisted QC and EACC resources achieve similar values whereas EAQC resources achieve the maximal possible violation of each facet inequality. Interestingly, when QC and EACC resources are used, the facet $\Fbb^c_{\BT 3}$ is more robust to noise, however, when EAQC is used the facet $\Fbb^d_{\BT 3}$ is more robust. Hence we find that the nonclassicality witness most robust to noise depends on the resource configuration.

\subsection{Nonclassicality in Multiaccess Networks}\label{section:nonclassicality_multiaccess_networks}

\begin{figure*}
    \centering
    \resizebox{0.7\textwidth}{!}{%
    \begin{tabular}{l l l l l}
        (a) CC & &  (b) ETx CC & &(c) GEA CC \\
        \begin{tikzpicture}
            \node[terminal] (x1) at (-0.15,1.25) {$x_1$};
            \node[terminal] (x2) at (-0.15,-1.25) {$x_2$};    
            \node[dev] (A1) at (1.4,1.25) {$P^{A_1}_{a_1|x_1\lambda}$};
            \node[dev] (A2) at (1.4,-1.25) {$P^{A_2}_{a_2|x_2\lambda}$};
            \node[dev] (B) at (2.8, 0) {$P^B_{y|a_1a_2\lambda}$};
            \node[terminal] (y) at (4.2, 0) {$y$};
        
            \path (x1) \cedge (A1);
            \path (x2) \cedge (A2);
            \path (A1) \cedge  node[el, above=3pt, xshift=4pt] {$a_1$} (B);
            \path (A2) \cedge node[el, below=3pt, xshift=4pt] {$a_2$} (B);
            \path (B) \cedge (y);
        \end{tikzpicture} & & \begin{tikzpicture}
            \node[terminal] (x1) at (0,1.25) {$x_1$};
            \node[terminal] (x2) at (0,-1.25) {$x_2$}; 
            \node[qsource] (lambda) at (0,0) {$\rho^{\Lambda}$};
            \node[meas_dev] (A1) at (1.4,1.25) {$\Pi^{A_1}_{a_1|x_1}$};
            \node[meas_dev] (A2) at (1.4,-1.25) {$\rho^{A_2}_{a_2|x_2}$};
            \node[dev] (B) at (2.8, 0) {$P^B_{y|a_1a_2}$};
            \node[terminal] (y) at (4.2, 0) {$y$};
        
            \path (lambda) \qedge node[el, above=-6pt, xshift=-8pt] {$\rho^{\Lambda}_1$} (A1);
            \path (lambda) \qedge node[el, above=3pt, xshift=1pt] {$\rho^{\Lambda}_2$} (A2);
            \path (x1) \cedge (A1);
            \path (x2) \cedge (A2);
            \path (A1) \cedge  node[el, above=4pt, xshift=4pt] {$a_1$} (B);
            \path (A2) \cedge node[el, below=4pt, xshift=4pt] {$a_2$} (B);
            \path (B) \cedge (y);
        \end{tikzpicture} & & \begin{tikzpicture}
            \node[terminal] (x1) at (0,1.25) {$x_1$};
            \node[terminal] (x2) at (0,-1.25) {$x_2$}; 
            \node[qsource] (lambda) at (0,0) {$\rho^{\Lambda}$};
            \node[meas_dev] (A1) at (1.4,1.25) {$\Pi^{A_1}_{a_1|x_1}$};
            \node[meas_dev] (A2) at (1.4,-1.25) {$\rho^{A_2}_{a_2|x_2}$};
            \node[meas_dev] (B) at (2.8, 0) {$\Pi^B_{y|a_1a_2}$};
            \node[terminal] (y) at (4.2, 0) {$y$};
        
            \path (lambda) \qedge node[el, above=-6pt, xshift=-8pt] {$\rho^{\Lambda}_1$} (A1);
            \path (lambda) \qedge node[el, below=-6pt, xshift=-12pt] {$\rho^{\Lambda}_3$} (A2);
            \path (lambda) \qedge node[el, above=3pt, xshift=4pt] {$\rho^{\Lambda}_2$} (B);
            \path (x1) \cedge (A1);
            \path (x2) \cedge (A2);
            \path (A1) \cedge  node[el, above=4pt, xshift=4pt] {$a_1$} (B);
            \path (A2) \cedge node[el, below=4pt, xshift=4pt] {$a_2$} (B);
            \path (B) \cedge (y);
        \end{tikzpicture}\\
        \hfill \\
        (d) QC & & (e) ETx QC  & & (f) GEA QC \\
        \begin{tikzpicture}
            \node[terminal] (x1) at (0,1.25) {$x_1$};
            \node[terminal] (x2) at (0,-1.25) {$x_2$};    
            \node[prep_dev] (A1) at (1.4,1.25) {$\rho^{A_1}_{x_1}$};
            \node[prep_dev] (A2) at (1.4,-1.25) {$\rho_{x_2}^{A_2}$};
            \node[meas_dev] (B) at (2.8, 0) {$\Pi^B_{y}$};
            \node[terminal] (y) at (4.2, 0) {$y$};
        
            \path (x1) \cedge (A1);
            \path (x2) \cedge (A2);
            \path (A1) \qedge  node[el, above=4pt, xshift=4pt] {$\rho^{A_1}_{x_1}$} (B);
            \path (A2) \qedge node[el, below=2pt, xshift=4pt] {$\rho^{A_2}_{x_2}$} (B);
            \path (B) \cedge (y);
        \end{tikzpicture} & & \begin{tikzpicture}
            \node[terminal] (x1) at (0,1.25) {$x_1$};
            \node[qsource] (lambda) at (0,0) {$\rho^{\Lambda}$};
            \node[terminal] (x2) at (0,-1.25) {$x_2$};    
            \node[proc_dev] (A1) at (1.4,1.25) {$\Emc_{x_1}^{A_1}$};
            \node[proc_dev] (A2) at (1.4,-1.25) {$\Emc_{x_2}^{A_2}$};
            \node[meas_dev] (B) at (2.8, 0) {$\Pi^B_{y}$};
            \node[terminal] (y) at (4.2, 0) {$y$};
        
            \path (lambda) \qedge node[el, above=-6pt, xshift=-8pt] {$\rho^{\Lambda}_1$} (A1);
            \path (lambda) \qedge node[el, above=3pt, xshift=1pt] {$\rho^{\Lambda}_2$} (A2);
            \path (x1) \cedge (A1);
            \path (x2) \cedge (A2);
            \path (A1) \qedge  node[el, above=4pt, xshift=4pt] {$\rho^{A_1}_{x_1}$} (B);
            \path (A2) \qedge node[el, below=2pt, xshift=4pt] {$\rho^{A_2}_{x_2}$} (B);
            \path (B) \cedge (y);
        \end{tikzpicture} & & \begin{tikzpicture}
            \node[terminal] (x1) at (0,1.25) {$x_1$};
            \node[terminal] (x2) at (0,-1.25) {$x_2$}; 
            \node[qsource] (lambda) at (0,0) {$\rho^{\Lambda}$};
            \node[proc_dev] (A1) at (1.4,1.25) {$\Emc^{A_1}_{x_1}$};
            \node[proc_dev] (A2) at (1.4,-1.25) {$\Emc^{A_2}_{x_2}$};
            \node[meas_dev] (B) at (2.8, 0) {$\Pi^B_{y}$};
            \node[terminal] (y) at (4.2, 0) {$y$};
        
            \path (lambda) \qedge node[el, above=-6pt, xshift=-8pt] {$\rho^{\Lambda}_1$} (A1);
            \path (lambda) \qedge node[el, below=-6pt, xshift=-12pt] {$\rho^{\Lambda}_3$} (A2);
            \path (lambda) \qedge node[el, above=3pt, xshift=1pt] {$\rho^{\Lambda}_2$} (B);
            \path (x1) \cedge (A1);
            \path (x2) \cedge (A2);
            \path (A1) \qedge  node[el, above=4pt, xshift=4pt] {$\rho^{A_1}_{x_1}$} (B);
            \path (A2) \qedge node[el, below=2pt, xshift=4pt] {$\rho^{A_1}_{x_1}$} (B);
            \path (B) \cedge (y);
        \end{tikzpicture} \\
    \end{tabular}
    }
    \caption{Multiaccess network DAGs. a) Classical communication. b) Entanglement-assisted senders using classical communication. c) Classical communication using global entanglement assistance. d) Quantum communication. e) Entanglement-assisted senders using quantum communication. f) Quantum communication using global entanglement assistance.}
    \label{fig:multiaccess_network_dags}
\end{figure*}

A multiaccess network $\MA(\sigarb{\XNet}{\vec{d}}{\Ymc})$ has multiple independent senders $\Av = \{A_i\}_{i=1}^n$ and one receiver $B$. Each sender is given the classical input $x_i\in\Xmc_i$ where the network's total input alphabet is $\XNet = \Xmc_1\times\cdots\times\Xmc_n$. A noiseless communication channel $\id^{A_i \to B}_{d_i}$ having signaling dimension  $d_i$ connects sender $A_i$ to the receiver where $\vec{d} =(d_1, \dots, d_n)$. The receiver $B$ jointly processes the messages from all senders to produce the value $y\in\Ymc$. 

\begin{table}[b!]
    \centering
    \resizebox{\columnwidth}{!}{%
    \begin{tabular}{| c | c |}
        \hline
        \vspace{-6pt}&\\
        \textbf{Set} & \textbf{Behavior Decomposition}  \\
        \vspace{-6pt}&\\
        \hhline{|=|=|}
        \vspace{-6pt}&\\
        a) $\Cbb^{\MA}_{}$ & $P_{y|\xv} = \prod_{\av\in\ANet} P^B_{y|\av} \prod_{i=1}^n P^{A_i}_{a_i|x_i}$\\
        \vspace{-6pt}&\\
        \hline
        \vspace{-6pt}&\\
        b) $\Cbb^{\MA}_{\ETx}$ & $P_{y|\xv} = \sum\limits_{\av\in\ANet}P^B_{y|\av}\tr{\Big(\textstyle \bigotimes\limits_{i=1}^n \Pi^{A_i}_{a_i|x_i}\Big) \rho^{\Lambda}}$ \\
        \vspace{-6pt}&\\
        \hline
        \vspace{-6pt}&\\
        c) $\Cbb^{\MA}_{\GEA}$ & $P_{y|\xv} = \sum\limits_{\av\in\ANet}\tr{\Big(\Pi^B_{y|\av}\otimes \textstyle\bigotimes\limits_{i=1}^n \Pi^{A_i}_{a_i|x_i} \Big) \rho^{\Lambda}} $\\
        \vspace{-6pt}&\\
        \hline
        \vspace{-6pt}&\\
        d) $\Qbb^{\MA}$ & $P_{y|\xv} = \tr{\Pi^B_{y} \bigotimes_{i=1}^n \rho^{A_i}_{x_i}}$\\
        \vspace{-6pt}&\\
        \hline
        \vspace{-6pt}&\\
        e) $\Qbb^{\MA}_{\ETx} $ & $P_{y|\xv} = \tr{\Pi^B_y \Big(\textstyle\bigotimes_{i=1}^n \Emc^{A_i}_{x_i}\Big)(\rho^{\Lambda}) }$ \\
        \vspace{-6pt}&\\
        \hline
        \vspace{-6pt}&\\
        f) $ \Qbb^{\MA}_{\GEA}$ & $P_{y|\xv = }\tr{\Pi^B_y \Big(\id^{\Lambda_0\to B}\otimes\textstyle\bigotimes_{i=1}^n \Emc^{A_i}_{x_i}\Big)(\rho^{\Lambda}) } $ \\
        &\\
        \hline
    \end{tabular}
    }
    \caption{Sets of behaviors for multiaccess network resource configurations. a) Classical communication, b) classical communication using entanglement-assisted senders, c) classical communication using global entanglement assistance, d) quantum communication, e) quantum communication using entanglement-assisted senders, f) quantum communication using global entanglement assistance. The respective DAGs for each of these sets is shown for the two-sender case in Fig.~\ref{fig:multiaccess_network_dags}.  }
    \label{table:multiaccess_network_behavior_sets}
\end{table}

We focus on the bipartite multiaccess network having two senders $A_1$ and $A_2$ where $d_i < |\Xmc_i|$ such that the communication is restricted. This case generalizes the bipartite \textit{prepare-and-measure} scenario discussed in Section~\ref{section:prepare-and-measure_scenario} where 
$\MA(\sigarb{\Xmc_1,\Xmc_2}{d_1,|\Xmc_2|}{\Ymc}) = \PM(\sigarb{\Xmc_1,\Xmc_2}{d_1}{\Ymc_2})$. \rev{The multiaccess network DAGs are shown in Fig.~\ref{fig:multiaccess_network_dags} and the corresponding sets of behaviors are shown in Table~\ref{table:multiaccess_network_behavior_sets}.}{}

\rev{Overall, we show that quantum communication or entanglement between two or more devices are each sufficient resources to achieve nonclassicality in multiaccess networks. In Section~\ref{section:multiaccess_nonclassicality_witnesses}, we construct a broad set of facet inequalities and simulation games for multiaccess networks. In Section~\ref{section:multiaccess_numerical_results}, we present our numerical results showing the maximum quantum violation of each classical bound and its noise robustness. Finally, in Section~\ref{section:multiaccess_protocols}, we give explicit protocols for the observed communication advantages in multiaccess networks with entanglement-assisted senders. These protocols include a dense bitwise XOR protocol for quantum multiaccess networks (Protocol~\ref{protocol:bitwise_xor_behavior}), and a trit equality protocol for classical multiaccess networks (Protocol~\ref{protocol:multiaccess_cmac_etx}).}{We obtain nonclassicality witnesses for bipartite multiaccess networks having up to four inputs and outputs and one bit of communication from each sender to the receiver. 
These nonclassicality witnesses are listed in Table~\ref{Table:ma-33-dd-2-nonclassicality-witnesses}, which includes both multiaccess network facet inequalities and simulation games.
We then apply variational optimization to obtain examples of quantum nonclassicality. Finally, we investigate the nonclassical quantum strategies that lead to advantages in simulation games corresponding to tasks including bitwise XOR operations, calculating the distance between two inputs, and comparing two inputs.}

\subsubsection{Multiaccess Network Nonclassicality Witnesses}\label{section:multiaccess_nonclassicality_witnesses}

A multiaccess network with $n$ senders can simulate any behavior in $\Pbb_{\YNet|\XNet}$ if $d_i = |\mc{X}_i|$ for all $i\in \{1, \dots, n\}$. It follows that nonclassicality in the multiaccess network requires at least one sender to have a signaling dimension $d_i < |\Xmc_i|$ and $2 \leq |\Ymc|$. We consider the multiaccess network with two senders as the simplest nontrivial case, focusing on $\MA(\sigarb{\Xmc_1,\Xmc_2}{2,2}{\Ymc})$ with input and output alphabets of size 4 or less. We first reproduce previous nonclassicality results regarding the prepare-and-measure scenario. Then we investigate examples of nonclassicality in multiaccess networks.

We can use PoRTA \cite{PORTA} to compute the full multiaccess network polytope $\Cbb^{\MA(\sigarb{3,3}{d_1,d_2}{2})}_{}$ for $d_1,d_2\in\{2,3\}$ (see upper section in  Table~\ref{Table:ma-33-dd-2-nonclassicality-witnesses}). Additionally, we compute the joint multiaccess network polytope when either $d_1$ or $d_2$ equals three while the other equals two,
\begin{equation}\label{eq:ma_23/32_polytope}
    \scriptstyle \Cbb^{\MA(\sigarb{3,3}{\{2,3\}}{2})} = \conv{\Cbb^{\MA(\sigarb{3,3}{2,3}{2})}_{}\cup \Cbb^{\MA(\sigarb{3,3}{3,2}{2})}_{}}
\end{equation}
where we use the curly braces in $\sigarb{3,3}{\{2,3\}}{2}$ to denote that either channel could send a trit while the other sends a bit (see lower section of Table~\ref{Table:ma-33-dd-2-nonclassicality-witnesses}). 
Note that for a behavior $\Pbf\not\in\Cbb^{\MA(\sigarb{3,3}{\{2,3\}}{2})}_{}$, each sender must use a trit of communication to simulate the behavior.

We also investigate a handful of simulation games $(\gamma, \Vbf)$ where $\Vbf$ is a deterministic nonclassical behavior. Since the multiaccess networks accepts two inputs, $x_1$ and $x_2$, and maps them to a single output $y$, it is natural to consider arithmetic operations as deterministic communication tasks (see Table~\ref{table:multiaccess_simulation_games}). Later, in Section~\ref{section:nonclassicality_multipoint_networks}, we consider multiaccess networks having up to $|\Ymc| = 9$ outputs, allowing the consideration of a broader set of communication tasks including multiplication and addition.

\begin{table}[t!]
    \centering
    \resizebox{\columnwidth}{!}{%
    \begin{tabular}{| c | c | c |}
        \hline
        \vspace{-6pt}&&\\
        \textbf{Task} & \textbf{Symbol} & \textbf{Definition} \\
        \vspace{-6pt}&&\\
        \hhline{|=|=|=|}
        \vspace{-6pt}&&\\
        Distance & $\Vbf^{-}_{\Xmc_1\Xmc_2\to \Ymc}$ & $V^-_{y,x_1,x_2} = \delta_{y,|x_1 - x_2|}$\\
        \vspace{-6pt}&&\\
        \hline
        \vspace{-6pt}&&\\
         \shortstack{Bitwise \\ XOR} & $\Vbf^{\oplus}_{4,4\to 4}$ & $V^{\oplus}_{\yv,\xv_{1},\xv_2} = \left\{  \begin{matrix} 
            1 & \text{if} \; {\genfrac{}{}{0pt}{2}{y_j=x_{1j}\oplus x_{2j}}{\forall \; j\in\{0,1\}} }\\
            0 & \text{otherwise}
        \end{matrix}\right. $\\
        \vspace{-6pt}&&\\
        \hline
        \vspace{-6pt}&&\\
        Compare & $\Vbf^{\gtrless}_{\Xmc_1\Xmc_2\to 3}$ & $V^{\gtrless}_{y,x_1,x_2} = \left\{  \begin{matrix} 
            1 & {\begin{matrix}\text{if} \;  y = 0 \; \text{and} \; x_1 = x_2 \\ \text{or} \; y=1 \; \text{and}\; x_1 > x_2 \\ y=2 \; \text{and} \; x_1 < x_2  \end{matrix} }\\
            0 & \text{otherwise}
        \end{matrix}\right.$\\
        \vspace{-6pt}&&\\
        \hline
        \vspace{-6pt}&&\\
        Equals & $\Vbf^{=}_{\Xmc_1\Xmc_2\to 2}$ & $V^{=}_{y,x_1,x_2} = \left\{  \begin{matrix} 
            1 &  { \begin{matrix}\text{if} \; y = 0 \; \text{and} \; x_1 = x_2 \\ \text{or} \; y=1 \; \text{and}\; x_1 \neq x_2 \end{matrix}} \\
            0 & \text{otherwise}
        \end{matrix}\right.$\\
        &&\\
        \hline
    \end{tabular}%
    }
    \caption{Simulation games for multiaccess networks. The classical bound is computed for one-bit of communication from each sender.}
    \label{table:multiaccess_simulation_games}
\end{table}

\subsubsection{Numerical Quantum Violations of Classical Bounds in Multiaccess Networks}\label{section:multiaccess_numerical_results}

\begin{figure*}
    \centering
    \begin{tabular}{c}
        \includegraphics[width=0.9\textwidth]{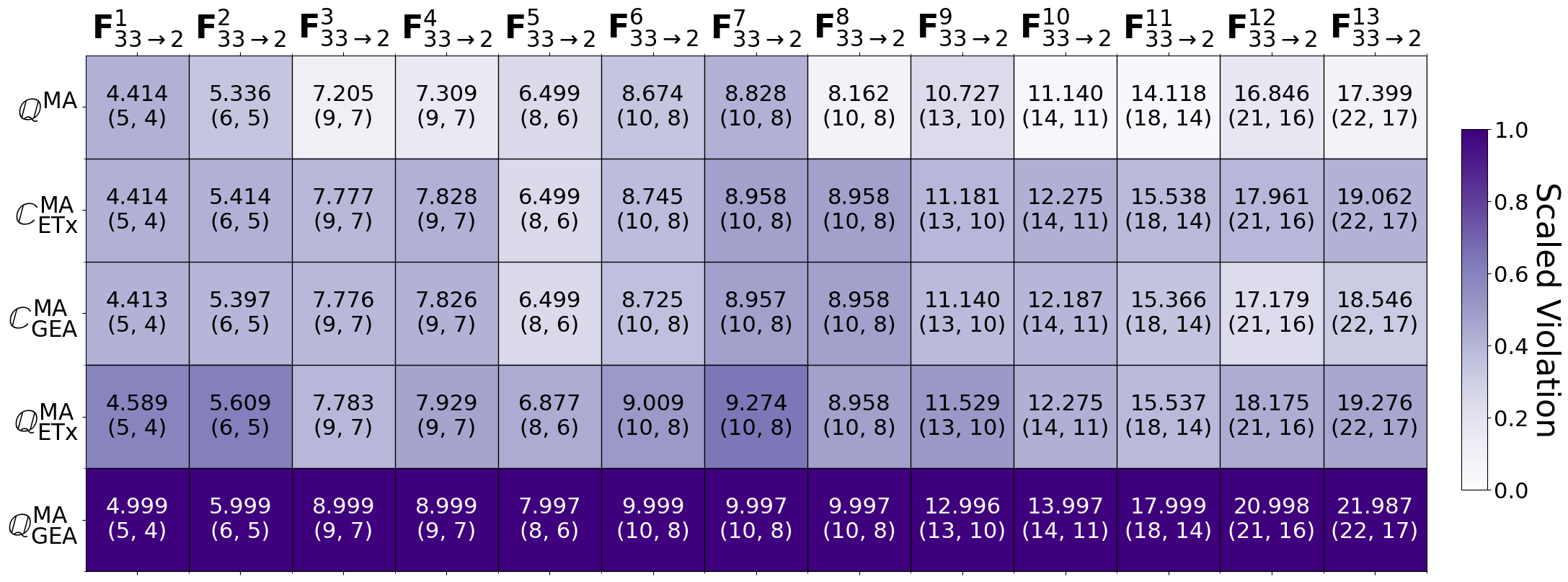}  \\
        \includegraphics[width=0.9\textwidth]{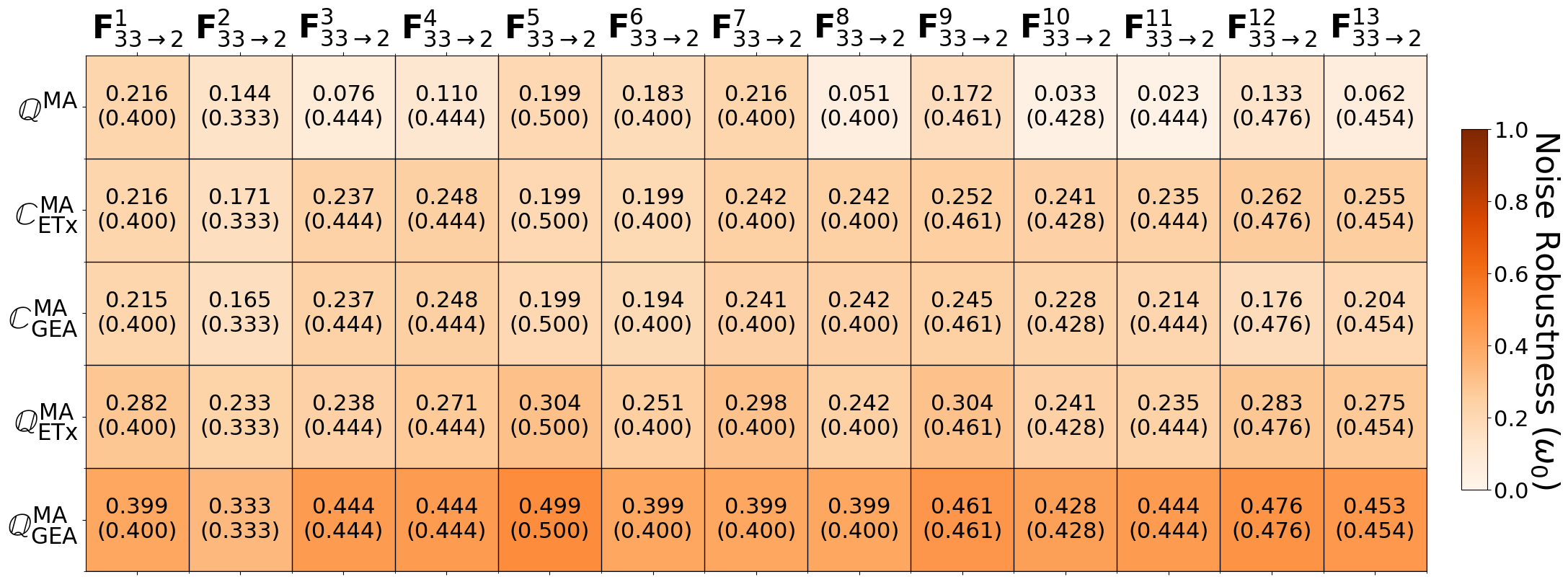} 
    \end{tabular}
    \caption{Max violations (top) and noise robustness (bottom) for $\MA(\sigarb{3,3}{2,2}{2})$ multiaccess network. Each column corresponds to a facet inequality listed in Table~\ref{Table:ma-33-dd-2-nonclassicality-witnesses}. Each row corresponds to a quantum resource configuration in figure \ref{fig:multiaccess_network_dags}. The top number in each cell shows the largest numerical violation obtained via variational optimization. The lower tuple, $(\hat{\gamma}, \gamma)$, shows the largest possible score $\hat{\gamma}$ and the classical bound $\gamma$ for each linear black box game where shading is used to show the magnitude of violation. In the noise robustness plot, the top number of each cell shows the noise robustness while the bottom number shows the noise robustness for the maximal possible violation.
   }
    \label{fig:quantum_violations_33-22-2_ma_network}
\end{figure*}

\begin{figure*}
    \centering
    \begin{tabular}{cc}
        \includegraphics[width=0.99\columnwidth]{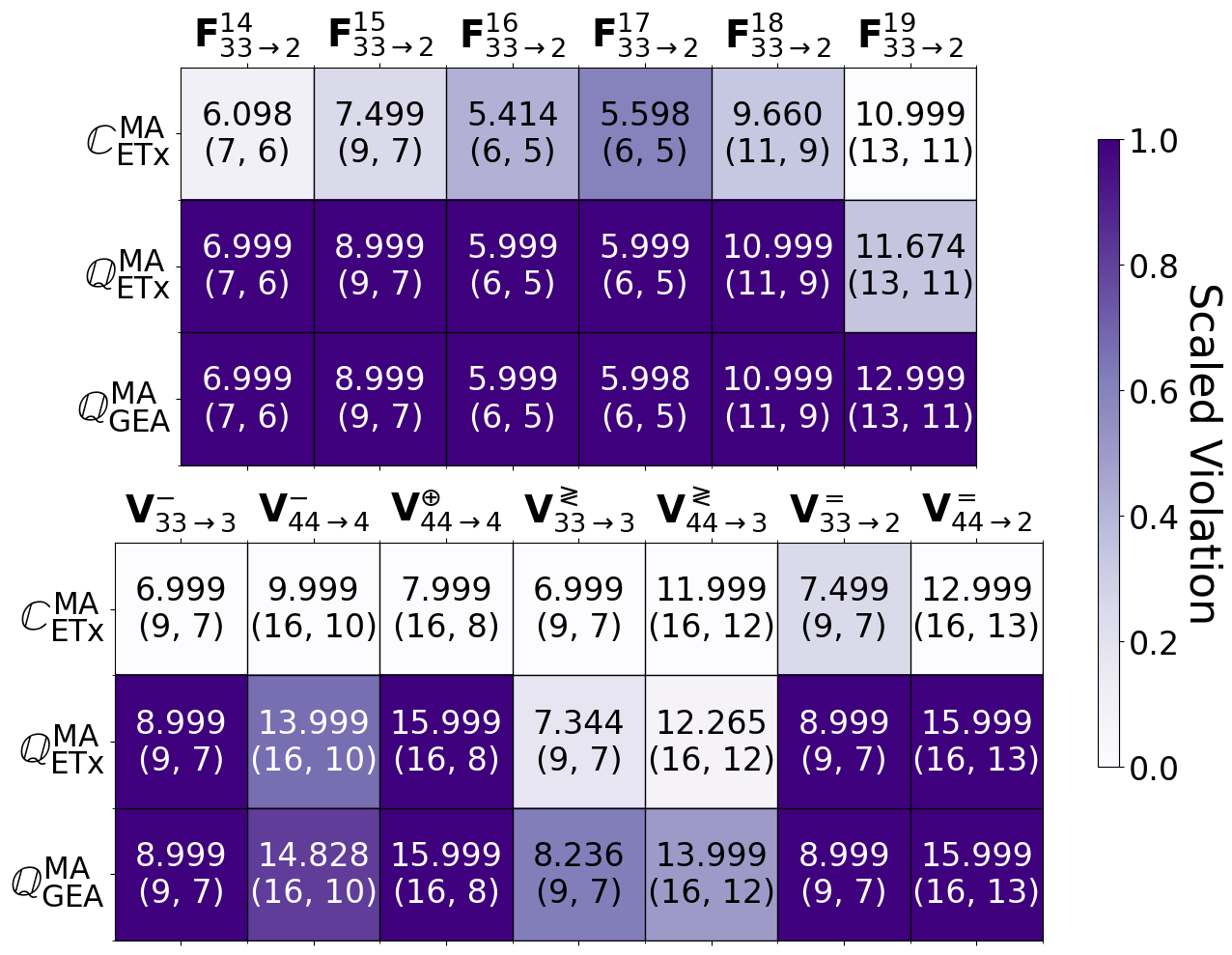} & \includegraphics[width=0.99\columnwidth]{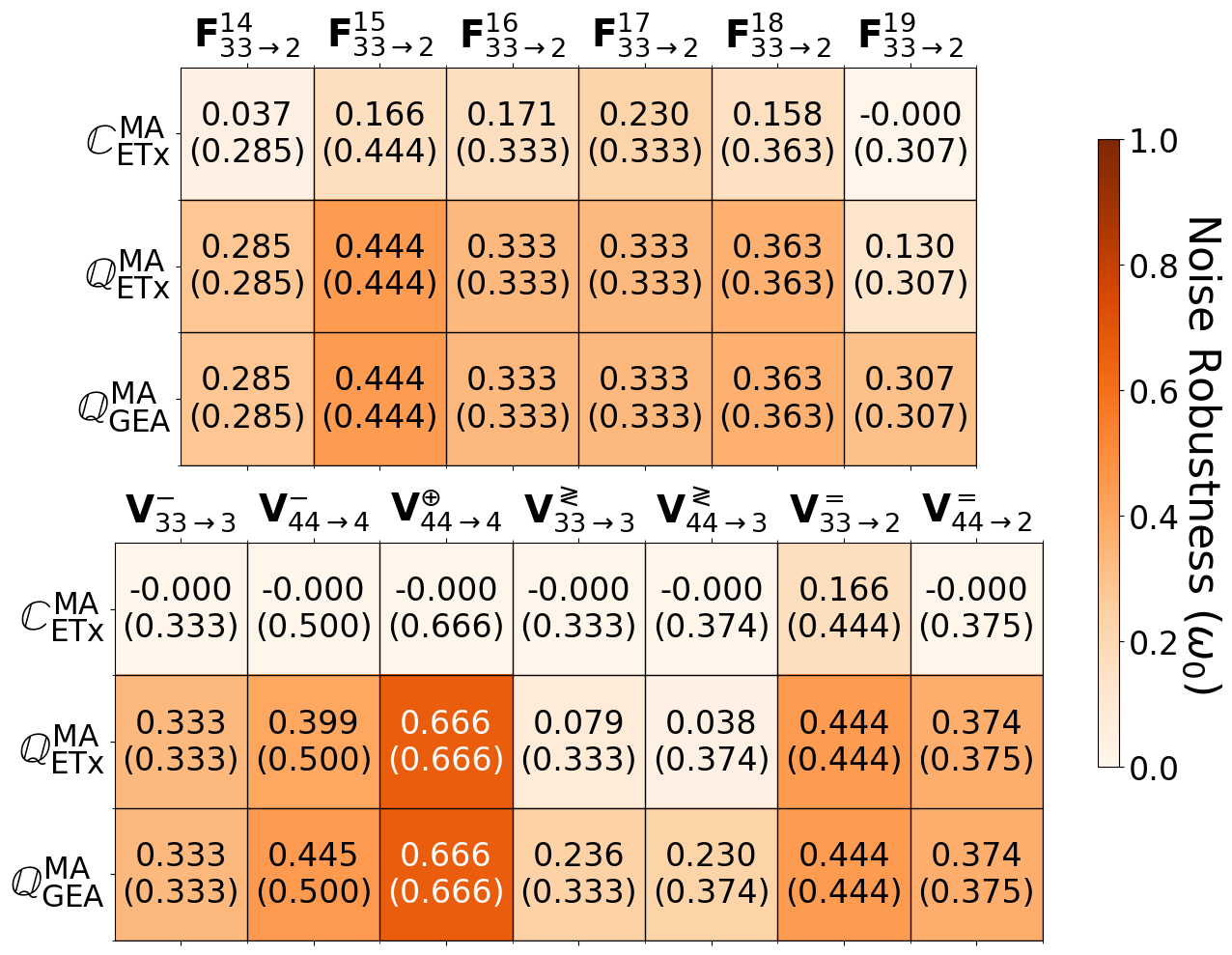} \\
    \end{tabular}
    
    \caption{Multiaccess network violations (left) and noise robustness (right). The upper plot on each side shows the violations or noise robustness of classicality in $\MA(\sigarb{3,3}{\{2,3\}}{2})$ multiaccess networks shown in Table~\ref{Table:ma-33-dd-2-nonclassicality-witnesses}. The lower plots show the violations and noise robustness of each simulation game in  Table~\ref{table:multiaccess_simulation_games}. The column of each plot corresponds to a different nonclassicality witness while each row corresponds to a different resource configuration. The top number in each cell shows the largest numerical violation obtained or its corresponding noise robustness. The lower tuple, $(\hat{\gamma}, \gamma)$, shows the largest possible score $\hat{\gamma}$ and the classical bound $\gamma$, or the noise robustness of the largest possible violation.}
    \label{fig:ma_simulation_game_numerics}
\end{figure*}

In Figure~\ref{fig:quantum_violations_33-22-2_ma_network}, we plot the violations for each facet inequality bounding the classical network polytope $\Cbb^{\MA(\sigarb{\Xmc_1,\Xmc_2}{2,2}{2})}_{}$ (see top section of Table~\ref{Table:ma-33-dd-2-nonclassicality-witnesses}). Remarkably, all considered quantum resource configurations can produce nonclassical behaviors. In Figure~\ref{fig:ma_simulation_game_numerics}, we show the violations of the facet inequalities of the  multiaccess network polytope facet inequalities $\Cbb^{\MA(\sigarb{3,3}{\{2,3\}}{2})}_{}$ from the bottom section of Table~\ref{Table:ma-33-dd-2-nonclassicality-witnesses}. We find that unassisted qubit communication  $\Qbb^{\MA}$ is unable to violate these  bounds, suggesting that qubit communication is classically simulable by the multiaccess network using a bit and a trit of communication, $\Qbb^{\MA}\subseteq \Cbb^{\MA(\sigarb{3,3}{\{2,3\}}{2})}_{}$. On the other hand, we find that entanglement-assisted senders are able to still violate these classical bounds, indicating that entanglement-assisted senders require at least two-trits of classical communication to simulate.

In the bottom plot of Figure~\ref{fig:ma_simulation_game_numerics}, we show the violations of the classical bound for each of the simulation games in Table~\ref{table:multiaccess_simulation_games}. We find that entanglement-assisted senders using classical communication, $\Cbb^{\MA}_{\ETx}$, are able to demonstrate an advantage in the trit equality game $\Vbf^=_{3,3\to 2}$. We find that entanglement-assisted senders using quantum communication $\Qbb^{\MA}_{\ETx}$ are able to achieve the maximal possible score for the games $\Vbf^-_{3,3\to 3}$, $\Vbf^{\oplus}_{4,4\to 4}$, $\Vbf^=_{3,3\to 2}$, and $\Vbf^=_{4,4 \to 2}$. The strongest violations are achieved by the strongest resource configuration $\Qbb^{\MA}_{\GEA}$ where entanglement is shared globally across all three parties. Moreover, the variational ansatz circuit for the sets $\Cbb^{\MA}_{\GEA}$ and $\Qbb^{\MA}_{\GEA}$ parameterize general three-qubit entangled states. Note that the global entanglement-assisted quantum signaling setting does not admit the maximal possible scores for all games. In the games where the maximal possible score can be obtained as $\gamma^\star = \ip{\Vbf, \Pbf^{\Net}(\vec{\theta}^\star)}$, Eq.~\ref{eq:deterministic_behavior_simulation_error} shows that $\Vbf \approx \Pbf^{\Net}(\vec{\theta})$. As a result,   Algorithm~\eqref{alg:vqo_simulation} can be used to automatically establish these deterministic tasks in a multiaccess network.

In many cases, quantum communication with entanglement-assisted senders or global entanglement assistance is able to achieve the maximal possible score, however, the noise robustness is not the same in each of these cases. Indeed, the nonclassicality witness most robust to noise is found to be the bitwise XOR simulation game $\Vbf^{\oplus}_{44\to 4}$, which can demonstrate nonclassicality with up to 2/3 mixture of white noise. 

\rev{When the max violation of each nonclassicality witness is compared across all multiaccess network resource configurations, we observe the hierarchy
\begin{equation}\label{eq:multiaccess_resource_hierarchy}
    \scriptstyle \Cbb^{\MA}_{}\; \subseteq \;\Qbb^{\MA}\;  \overset{?}{\subseteq}\; \Cbb^{\MA}_{\ETx}\;\overset{?}{\subseteq}\;\Cbb^{\MA}_{\GEA}\; \overset{?}{\subseteq}\; \Qbb^{\MA}_{\ETx}\;\subseteq\; \Qbb^{\MA}_{\GEA}
\end{equation}
where the maximum violations of each resource configuration are nondecreasing from left to right. Note that all subset relations annotated with question marks in Eq.~\eqref{eq:multiaccess_resource_hierarchy} are conjectured and left as an open problem. The remaining relations hold trivially because quantum communication can simulate classical communication (see Eq.~\eqref{eq:trivial_classical_quantum_hierarchy}). }{We leave it as an open problem whether the conjectured resource hierarchy in Eq.~\eqref{eq:multiaccess_resource_hierarchy} generally holds for multiaccess networks.}

In some cases, we find that certain resource configurations do not yield violations while others do.
For instance, we deduce from Fig.~\ref{fig:ma_simulation_game_numerics} that entanglement-assisted classical senders $\Cbb^{\MA}_{\ETx}$  do not violate many of the considered simulation games, while quantum multiaccess networks with entanglement-assisted senders $\Qbb^{\MA}_{\ETx}$ do yield violations. These nonclassicality witnesses that require certain resource configurations to violate are valuable for quantum resource certification. It is important to investigate, classify, and characterize such examples.

\subsubsection{Protocols for Nonclassicality in Multiaccess Networks with Entanglement-Assisted Senders}\label{section:multiaccess_protocols}

We now present protocols for achieving nonclassicality in multiaccess networks with entanglement-assisted senders. These protocols correspond to deterministic information processing tasks that can be implemented using quantum resources, but not classical resources, for a given signaling dimension. Notably, the following example shows a form of dense information processing for the bitwise XOR behavior $\Vbf^{\oplus}_{4,4\to 4}$. 

\begin{protocol}\label{protocol:bitwise_xor_behavior}
    Achieve a zero-error simulation of the bitwise XOR behavior $\Vbf^{\oplus}_{4,4\to 4}$ using a multiaccess network $\Qbb^{\MA(\sigarb{4,4}{2,2}{4})}_{\ETx}$.
    \begin{enumerate}
        \item The source $\Lambda$ prepares the maximally entangled state $\ket{\Phi^+} = \frac{1}{\sqrt{2}}(\ket{00} + \ket{11})$ and distributes it between two senders $A_1$ and $A_2$.
        \item Each sender applies a unitary
        \begin{equation}
            U^{A_i}_{\xv_i} \in (\Ibb_2, \sigma_z^{A_i}, \sigma_x^{A_i}, \sigma_y^{A_i})
        \end{equation}
         conditioned on the two-bit input $\xv_i\in \Xmc_i = \{0,1\}^2$. The resulting quantum state is then
         \begin{align}
             \ket{\psi_{\xv_1\xv_2}} &= U^{A_1}_{\xv_1}\otimes U^{A_2}_{\xv_2}\ket{\Phi^+} \\ 
             &= U^{A_1}_{\xv_1}(U^{A_2}_{\xv_2})^T\otimes \Ibb_{2} \ket{\Phi^+},
         \end{align}
         for which we verify the following cases:
        \begin{align}
            \text{when} \quad & 00 = \xv_1\oplus\xv_2, & \ket{\psi_{\xv_1\xv_2}} = \nu \ket{\Phi^+} \\
            \text{when} \quad & 01 =\xv_1\oplus \xv_2, & \ket{\psi_{\xv_1\xv_2}} = \nu\ket{\Phi^-}\\
            \text{when}\quad & 10 =\xv_1\oplus \xv_2 & \ket{\psi_{\xv_1\xv_2}} = \nu \ket{\Psi^+} \\
            \text{when}\quad & 11 =\xv_1\oplus \xv_2 & \ket{\psi_{\xv_1\xv_2}} = \nu \ket{\Psi^-}
        \end{align}
        Note that $\nu=\pm 1$ represents a global phase factor dependent on both $\xv_1$ and $\xv_2$.
        \item The receiver $B$ jointly measures the two-qubits in the Bell basis $\{\ket{\Phi^+},\ket{\Phi^-},\ket{\Psi^+},\ket{\Psi^-}\}$ to obtain the output $y= \xv_1 \oplus \xv_2$ with zero error.
    \end{enumerate}
    
    \remark The classical bound for one bit of signaling is $\Delta(\Vbf^{\oplus}_{4,4\to 4}, \Pbf) = P_{\Error} = \frac{1}{2}$, whereas two bits of signaling from each sender are needed to achieve the simulation error $P_{\Error} = 0$.
\end{protocol}

Protocol~\ref{protocol:bitwise_xor_behavior} describes a quantum advantage where the bitwise XOR is performed using two bits fewer than necessary in a classical setting. That is, if $N$ pairs of entangled states are shared between the two senders, then the XOR of two $2N$-bit strings can be evaluated by the multiaccess network using only $2N$ qubits of communication where $4N$ classical bits are required in the classical case. Hence this quantum communication advantage demonstrates a form of dense computation similar to dense coding \cite{bennet1992_dense_coding}.

We also find an interesting simulation advantage for entanglement-assisted senders using classical signaling. In the following protocol, we outline the quantum strategy that achieves a violation of the nonclassicality witness $(7, \Vbf^=_{3,3\to 2})$ (see facet inequality $\Fbf^{15}_{33\to 2}$) in Table~\ref{Table:ma-33-dd-2-nonclassicality-witnesses}). One application of this violation is demonstrating the use of entanglement between two senders.

\begin{protocol}\label{protocol:multiaccess_cmac_etx}
    Achieve the simulation game violation $7.5 = \ip{\Vbf^{=}_{3,3\to 2}, \Pbf} > 7$ in the classical multiaccess network with entanglement-assisted senders $\Cbb^{\MA(\sigarb{3,3}{2,2}{2})}_{\ETx}$.
    \begin{enumerate}
        \item The source prepares the maximally entangled state $\ket{\Phi^+} = \frac{1}{\sqrt{2}}(\ket{00}+\ket{11})$ and distributes it to the senders $A_1$ and $A_2$.
        \item Each sender measures their local qubit to obtain a one-bit outcome, with measurement bases $\scriptstyle \left\{\ket{\phi^{A_i}_{a_i|x_i}}\right\}_{a_i\in\{0,1\}}$ given as   
        \begin{align}
            \scriptstyle
            \Big\{\ket{\phi^{A_i}_{0|0}} & \scriptstyle = \ket{0},\; \;\ket{\phi^{A_i}_{1|0}}=\ket{1}\Big\}, \label{eq:fp-basis-0}\\
            \scriptstyle \Big\{\ket{\phi^{A_i}_{0|1}} &\;\;\scriptstyle= \frac{1}{2}(\ket{0} + \sqrt{3}\ket{1}),\ket{\phi^{A_i}_{1|1}} = \frac{1}{2}(\sqrt{3}\ket{0} - \ket{1}) \Big\},\label{eq:fp-basis-1} \\
            \scriptstyle\Big\{\ket{\phi^{A_i}_{0|2}} &\scriptstyle= \frac{1}{2}(\ket{0} - \sqrt{3}\ket{1}), \;\;\ket{\phi^{A_i}_{1|2}} = \frac{1}{2}(\sqrt{3}\ket{0} + \ket{1})\Big\}\label{eq:fp-basis-2}.
        \end{align}
        Note that When $x_1=x_2$, the senders $A_1$ and $A_2$ measure $\ket{\Phi^+}$ in the same basis resulting in outcomes that have even parity, $a_1\oplus a_2 = 0$. Otherwise, if $x_1\neq x_2$, then we obtain even and odd parity results with probabilities
        \begin{align}
            &P(a_1\oplus a_2 = 0|x_1,x_2) = 0.25, \\
            &P(a_1\oplus a_2 = 1|x_1,x_2) = 0.75.
        \end{align}
                
        \item Each sender transmits their one-bit measurement result $a_i$ via a classical channel to the receiver $B$ who outputs the value $y = a_1 \oplus a_2$. The resulting behavior is then
        \begin{align}\label{eq:optimal-33-22-2_finger_printing_behavior}
            \mbf{P}^\star = \frac{1}{4}\begin{bmatrix}
                4 & 1 & 1 & 1 & 4 & 1 & 1 & 1 & 4 \\
                0 & 3 & 3 & 3 & 0 & 3 & 3 & 3 & 0
            \end{bmatrix}\notag,
        \end{align}
        which achieves the score $\ip{\Vbf^{=}_{3,3\to 2},\Pbf^{\star}} = 7.5$.
    \end{enumerate}
\end{protocol}

\subsection{Nonclassicality in Broadcast Networks}\label{section:nonclassicality_broadcast_networks}

\begin{figure*}
    \centering
    \includegraphics[width=\textwidth]{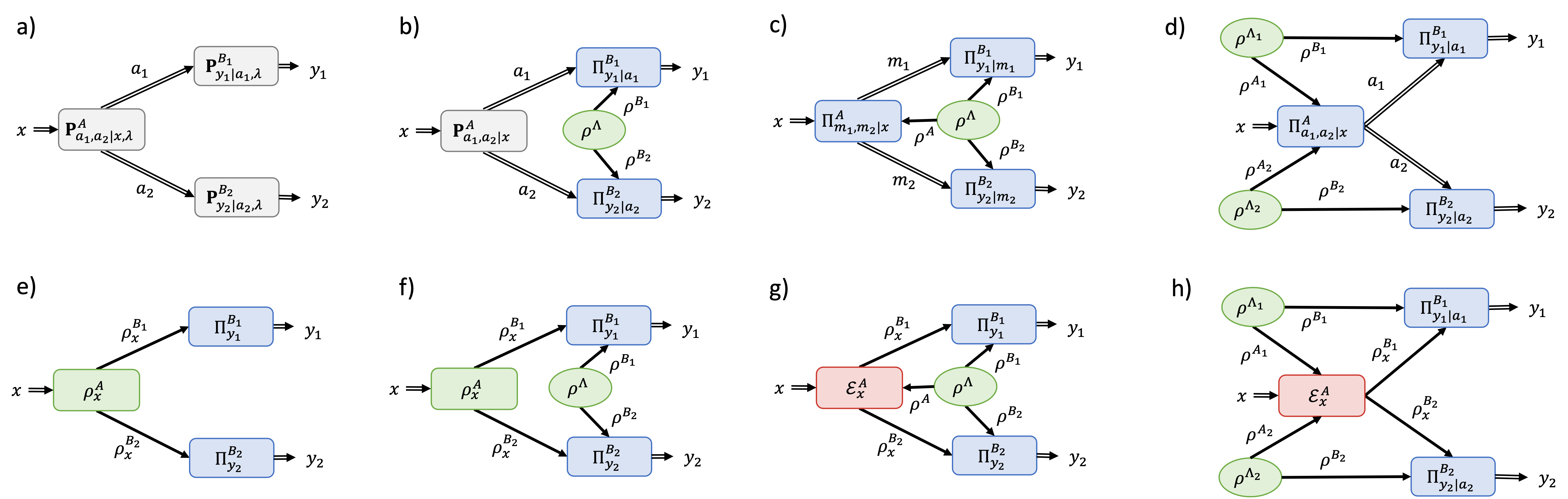}
    \caption{Broadcast network DAGs. a) Classical communication $\Cbb^{\BC}$. b) Classical communication using entanglement-assisted receivers $\Cbb^{\BC}_{\ERx}$. c) Classical communication with global entanglement assistance $\Cbb^{\BC}_{\GEA}$. d) Entanglement-assisted classical communication $\Cbb^{\BC}_{\EA}$. e) Quantum communication $\Qbb^{\BC}$. f) Quantum communication using entanglement-assisted receivers $\Qbb^{\BC}_{\ERx}$. g) Quantum communication using global entanglement-assistance $\Qbb^{\BC}_{\GEA}$. h) Entanglement-assisted quantum communication $\Qbb^{\BC}_{\EA}$.}
    \label{fig:broadcast_network_dags}
\end{figure*}

A broadcast network $\BC(\sigarb{\Xmc}{\dv}{\YNet})$ consists of one sender $A$ and multiple receivers $\Bv = (B_1,\dots, B_n)$ where a noiseless channel $\id^{A\to B_i}_{d_i}$ with signaling dimension $d_i$ connects the sender to receiver $B_i$. The broadcast network as a single input $x\in\Xmc$ that is given to the sender while each receiver outputs $y_i\in \Ymc_i$, hence broadcast network behaviors belong to the probability polytope $\Pbb_{\YNet|\Xmc}$ where $\YNet \equiv \Ymc_1\times\dots\times\Ymc_{n}$.
We primarily focus on broadcast networks with two receivers where DAGs depicting the considered communication resource configurations are shown in Fig.~\ref{fig:broadcast_network_dags} and the behavior sets for each configuration are given in Table~\ref{table:broadcast_behavior_decompositions}.

Overall, we show that entanglement is necessary for nonclassicality in broadcast networks.
\rev{In Section~\ref{section:broadcast_nonclassicality_witnesses}, we derive many examples of broadcast network facet inequalities and simulation games. In Section~\ref{section:broadcast_classical_sim}, we present Theorem~\ref{thm:broadcast_classicality}, which shows that all quantum broadcast networks (see Fig.~\ref{fig:broadcast_network_dags}.e) are classically simulable. In Section~\ref{section:broadcast_numerics}, we present our numerical results that demonstrate nonclassicality in all cases where communication-constrained parties share entanglement, paving the way for new entanglement certification approaches in broadcast networks.}{}

\begin{table}[t!]
    \centering
    \resizebox{\columnwidth}{!}{%
    \begin{tabular}{| c | c |}
        \hline
        \vspace{-6pt}&\\
        \textbf{Set} & \textbf{Behavior Decomposition}  \\
        \vspace{-6pt}&\\
        \hhline{|=|=|}
        \vspace{-6pt}&\\
        a) $\Cbb^{\BC}_{}$ & $P_{\yv|x} = \sum_{\av\in\ANet}\big(\prod_{i=1}^{|\ANet|}P^{B_i}_{y_i|a_i}\big)P^A_{\av|x}$ \\
        \vspace{-6pt}&\\
        \hline
        \vspace{-6pt}&\\
        b) $\Cbb^{\BC}_{\ERx}$ & $P_{\yv|x} = \sum\limits_{\av\in\ANet}\tr{\Pi^{B_1}_{y_1|a_1}\otimes\Pi^{B_2}_{y_2|a_2}\rho^{\Lambda}}P^{A}_{\av|x}$ \\
       \vspace{-6pt} &\\
        \hline
        \vspace{-6pt}&\\
        c) $\Cbb^{\BC}_{\GEA}$ & $P_{\yv|x} = \sum\limits_{\av\in\ANet}\tr{\bigotimes\limits_{i=1}^n \Pi^{B_i}_{y_i|a_i}\ptr{A}{\Pi^{A}_{\av|x}\otimes\id^{\Lambda\to\Bv}\rho^{\Lambda}}}$ \\
        \vspace{-6pt}&\\
        \hline
        \vspace{-6pt}&\\
        d) $\Cbb^{\BC}_{\EA}$ & $P_{\yv|x} =  \tr{\sum\limits_{\av\in\ANet} \bigotimes\limits_{i=1}^n\Pi^{B_i}_{y_i|a_i} \ptr{A}{\Pi^{A}_{\av|x}\otimes\id^{\Lambda\to\Bv}\bigotimes\limits_{j=1}^n\rho^{\Lambda_j}}}$ \\
        \vspace{-6pt}&\\
        \hline
        \vspace{-6pt}&\\
        e) $\Qbb^{\BC}$ & $P_{\yv|x} = \tr{\bigotimes_{i=1}^n\Pi^{B_i}_{y_i}\rho^A_x}$ \\
        \vspace{-6pt}&\\
        \hline
        \vspace{-6pt}&\\
        f) $\Qbb^{\BC}_{\ERx}$ & $P_{\yv|x} = \tr{\bigotimes\limits_{i=1}^n\Pi^{B_i}_{y_i} \id^{A,\Lambda\to\Bv}(\rho^{A}_x \otimes \rho^{\Lambda})}$ \\
        \vspace{-6pt}&\\
        \hline
        \vspace{-6pt}&\\
        g) $\Qbb^{\BC}_{\GEA}$ & $P_{\yv|x} = \tr{\bigotimes\limits_{i=1}^n \Pi^{B_i}_{y_i}\left(\Emc^A_{x} \otimes \id^{\Bv}(\rho^{\Lambda})\right)}$ \\
        \vspace{-6pt}&\\
        \hline
        \vspace{-6pt}&\\
        h) $\Qbb^{\BC}_{\EA}$ & $\begin{matrix}P_{\yv|x} = \tr{\bigotimes\limits_{i=1}^n \Pi^{B_i}_{y_i} \Emc^{A}_x\otimes\id^{\Lambda\to\Bv} \left(\bigotimes\limits_{j=1}^n\rho^{\Lambda_j}\right)} \\ \vspace{-9pt}\end{matrix}$\\
        \hline
    \end{tabular}%
    }
    \caption{Sets of behaviors for broadcast networks. a) Classical communication. b) Classical communication with entanglement-assisted receivers. c) Classical communication assisted by global entanglement. d)  Entanglement-assisted classical communication. e) Quantum communication. f) Quantum communication with entanglement-assisted receivers. g) Quantum communication assisted by global entanglement. h) Entanglement-assisted quantum communication. }
    \label{table:broadcast_behavior_decompositions}
\end{table}

\subsubsection{Broadcast Network Nonclassicality Witnesses}\label{section:broadcast_nonclassicality_witnesses}

A classical broadcast network with $n$ receivers can simulate any behavior in $\Pbb_{\YNet|\Xmc}$ if either $|\Ymc_i| \leq d_i$ for all $i\in\{1,\dots,n\}$ or $|\Xmc| \leq \min\{d_i\}_{i=1}^n$. We therefore consider two simple nontrivial cases of broadcast networks,  $\BC(\sigarb{3}{2,2}{3,3}) = \BC 3$ and $\BC(\sigarb{4}{2,2}{4,4}) = \BC 4$.

The facet inequalities of the broadcast network $\BC 3$ are shown in Table~\ref{Table:broadcast_X_22_YY_polytope}.
We also compute the facet inequalities in the case where one trit and one bit of communication is used in the network.
\begin{equation}
    \scriptstyle \Cbb^{\BC(\sigarb{3}{\{2,3\}}{33})}=\conv{\Cbb^{\BC(\sigarb{3}{2,3}{33})}\cup \Cbb^{\BC(\sigarb{3}{3,2}{33})}}.
\end{equation}
The resulting facet inequality corresponds to an interesting simulation game, which we refer to as the broadcast communication value (BCV).
The simulation game is denoted $(\gamma^{\text{BCV}}, \Vbf^{\text{BCV}})$ where $\gamma^{\text{BCV}} = \min\{d_1,d_2\}$ and
\begin{equation}\label{eq:game_bcv}
    V^{\text{BCV}}_{y_1,y_2,x} = \left\{ \begin{matrix} 1& \quad \text{if} \;\; y_1=y_2=x \\
    0& \quad \text{otherwise.} \end{matrix} \right.
\end{equation}

In the case of $\BC4$, we apply broadcast communication value simulation game $(2,\Vbf^{\text{BCV}}_{\BC4})$, and we apply the linear program in Eq.~\eqref{eq:facet_linear_program} to obtain facet inequalities, $\Fbf^a_{\BC4}$ and $\Fbf^b_{\BC4}$, of the $\Cbb^{\BC4}$ classical network polytope (see Table~\ref{Table:broadcast_X_22_YY_polytope}). Notably, the facet inequality $(8, \Fbf^b_{\BC4}$) was obtained from the linear program in Eq.~\eqref{eq:facet_linear_program} by inputting the nonclassical behavior that simulates a Popescu-Rohrlich (PR) box \cite{Popescu1994} used by the receivers to achieve the maximal possible violation of the Clauser, Horne, Shimony, and Holt (CHSH) inequality \cite{chsh1969}. Since $|\Ymc_1| = |\Ymc_2| = 4$, each receiver outputs a two-bit value containing their measurement result and their measurement basis.

\rev{
\subsubsection{Classical Simulability of Quantum Broadcasts}\label{section:broadcast_classical_sim}

In this section we prove that unassisted quantum broadcast networks do not exhibit nonclassicality, and therefore, do not provide any observable communication advantage. It follows that entanglement is necessary for nonclassicality in broadcast networks.

\begin{theorem} \label{thm:broadcast_classicality}
Given any broadcast network $\BC(\sigarb{\Xmc}{\dv}{\vec{\Ymc}})$ with $n$ receivers, $\vec{B}=(B_1,\dots,B_n)$, and signaling dimensions, $\vec{d}=(d_1,\dots,d_n)$, the set of quantum broadcast behaviors is equivalent to the set of classical broadcast behaviors,
\begin{equation}
    \Cbb^{\BC(\sigarb{\Xmc}{\dv}{\vec{\Ymc}})} = \Qbb^{\BC(\sigarb{\Xmc}{\dv}{\vec{\Ymc}})}.
\end{equation}
Thus, nonclassicality cannot be exhibited by unassisted quantum broadcast networks (see Fig.~\ref{fig:broadcast_network_dags}.e).

\begin{proof}

From Eq.~\eqref{eq:trivial_classical_quantum_hierarchy}, the inclusion $\Cbb^{\BC(\sigarb{\Xmc}{\dv}{\vec{\Ymc}})} \subseteq \Qbb^{\BC(\sigarb{\Xmc}{\dv}{\vec{\Ymc}})}$ is trivial. Therefore, it is sufficient to show that each $\Pbf\in\Qbb^{\BC(\sigarb{\Xmc}{\dv}{\vec{\Ymc}})}$ is classically simulable such that $\Pbf \in \Cbb^{\BC(\sigarb{\Xmc}{\dv}{\vec{\Ymc}})}$.
We first show that, given any classical broadcast network behavior $\Pbf \in \Cbb^{\BC(\sigarb{\Xmc}{\dv}{\vec{\Ymc}})}$, there exists a valid communication strategy in which sender $A$ has complete knowledge of the network's output, $\yv = (y_1, \dots, y_n)$. From the decomposition of a classical broadcast network behavior in Table~\ref{table:broadcast_behavior_decompositions}.a, the transition probabilities decompose as
\begin{equation}
    P_{\yv|x} = \sum_{\lambda} P_\lambda \sum_{\av} \left(\textstyle \prod_{i=1}^n P^{B_i}_{y_i|a_i,\lambda}\right)P^A_{\av|x,\lambda}
\end{equation}
where the sum over $\av=(a_1,\dots,a_n)$ accounts for all message encodings and the GSR is modeled by the shared random variable $\lambda$ and its prior $P_{\lambda}$.
Since an unlimited amount of GSR is assumed, any local stochastic behavior can be decomposed into a convex combination of deterministic behaviors
\begin{align}
    P^A_{\av|x,\lambda} &= \textstyle\sum_{\mu_0} P_{\mu_0} V^{A}_{\av|x,\lambda,\mu_0} \label{eq:broadcast_determinism_a}\\ 
    P^{B_i}_{y_i|a_i,\lambda} &= \textstyle\sum_{\mu_i} P_{\mu_i} V^{B_i}_{y_i|a_i,\lambda,\mu_i}. \label{eq:broadcast_determinism_b}
\end{align}
Without loss of generality, each party's local random variable $\mu_i$ can be incorporated into the GSR as public knowledge, such that all parties share the random variables $\vec{\lambda}=(\lambda,\mu_0,\dots,\mu_{n})$. If we assume that each party's deterministic behavior $\Vbf^{B_i}_{\lambda,\mu_i}$  is known by the sender $A$, then the sender can predict the output $\yv$ with zero-error given $x\in\Xmc$.

Now consider an unassisted quantum broadcast network behavior from Table~\ref{table:broadcast_behavior_decompositions}.e,
\begin{align}\label{eq:broadcast_network_behavior_proof}
    P_{\yv|x} &= \tr{\left(\textstyle\bigotimes_{i=1}^n\Pi^{B_i}_{y_i}\right) \rho^{A}_x},
\end{align}
where $\rho^{A}_x \in D(\Hmc_d^n)$ is an $n$-qudit density operator and $\Pi^{B_i}_{y_i}$ is a POVM element.
Our goal is to decompose Eq.~\eqref{eq:broadcast_network_behavior_proof} into $n$ parallel single-qudit channels that can be classically simulated given a classical sender's  privileged knowledge of   the network's output $\yv$.

Suppose that receiver $B_n$ measures their qudit by taking the partial trace over the $n^{th}$ qudit, then
\begin{align}
    P_{\yv|x} &= \tr{\textstyle  \bigotimes_{i=1}^{n-1}\Pi^{B_i}_{y_i} \ptr{n}{ \Ibb_{d^{n-1}}\otimes \Pi^{B_n}_{y_n} \rho^{A}_x}} \\
    &= \tr{\textstyle \bigotimes_{i=1}^{n-1}\Pi^{B_i}_{y_i}\widetilde{\Pi}^{A}_{y_n|x}} \\
    &= \tr{\textstyle \bigotimes_{i=1}^{n-1}\Pi^{B_i}_{y_i}\rho^{A}_{y_n|x}}P_{y_n|x} \label{eq:broadcast_post_meas_state}
\end{align}
where $\widetilde{\Pi}^{A}_{y_n|x}$ is the unnormalized reduced density matrix of the $(n-1)$-qudit post-measurement state $\rho^{A}_{y_n|x} = \widetilde{\Pi}^{A}_{y_n|x}\frac{1}{P_{y_n|x}}$. Since $\Pi^{B_n}_{y_n}$ is a POVM, the normalization factor is the conditional probability
\begin{equation}\label{eq:broadcast_post_meas_conditional_prob}
    P_{y_n|x} = \tr{\Pi^{B_n}_{y_n} \ptr{(1,\dots,n-1)}{\rho^{A}_x}}.
\end{equation}
Next, we substitute Eq.~\eqref{eq:broadcast_post_meas_conditional_prob} into Eq.~\eqref{eq:broadcast_post_meas_state} to obtain
\begin{align}\label{eq:quantum_broadcast_simulable_decomposition}
    P_{\yv|x} = \tr{\textstyle \bigotimes_{i=1}^{n-1}\Pi^{B_i}_{y_i} \rho^{A}_{y_n|x} }\tr{\Pi^{B_n}_{y_n} \rho_x^{A_n}},
\end{align}
in which party $B_n$ is separated from the receivers, $B_1, \dots, B_{n-1}$, such that $y_n$ only depends on the marginal qudit state $\rho_x^{A_n} = \ptr{(1,\dots,n-1)}{\rho^{A}_x}$.

We can now prove that a quantum broadcast network with $n$ receivers is equivalent to a quantum broadcast network with $(n-1)$ receivers and a classical point-to-point channel.
First, we apply the point-to-point simulability result of Frenkel and Weiner \cite{Frenkel2015_classical_information_n-level_quantum_system} to show that the transition probabilities of the qudit channel in Eq.~\eqref{eq:quantum_broadcast_simulable_decomposition}, $P_{y_n|x}=\tr{\Pi^{B_n}_{y_n} \rho_x^{A_n}}$, can be simulated with zero-error using GSR and a classical channel of signaling dimension $d$.
Next, Eqs.~\eqref{eq:broadcast_determinism_a} and~\eqref{eq:broadcast_determinism_b} show that, given the input $x$ and the GSR value $\vec{\lambda}$, the sender $A$ can determine the output $y_n$ of receiver $B_n$, provided that each party behaves deterministically according to the GSR.
Finally, the sender $A$ uses the value $y_n$ to prepare the $(n-1)$-qudit state $\rho^{A}_{y_n|x}$ and broadcast it to receivers $B_1,\dots,B_{n-1}$.

This base case can be extended to the remaining $(n-1)$-qudit system in Eq.~\eqref{eq:quantum_broadcast_simulable_decomposition}, separating each qudit channel to obtain a product of $n$ quantum channels
\begin{align}\label{eq:fully_separated_quantum_broadcast}
    P_{\yv|x} &= \tr{\Pi^{B_1}_{y_1}\rho^{A_1}_{y_2,\dots,y_n|x}}\times\dots \\ \notag
    &\dots\times\tr{\Pi^{B_{n-1}}_{y_{n-1}} \rho_{y_n|x}^{A_{n-1}}}\tr{\Pi^{B_n}_{y_n} \rho_x^{A_n}}.
\end{align}
If all parties operate deterministically according to the GSR, then we can apply the same argument as the base case to assert that each of the $n$ quantum channels in Eq.~\eqref{eq:fully_separated_quantum_broadcast} is classically simulable.
Thus, any quantum broadcast network with $n$ receivers is simulable using GSR and a classical broadcast network with $n$ receivers and signaling dimension $\vec{d}$.

\end{proof}

\end{theorem}

}{}

\subsubsection{Numerical Quantum Violations of Classical Bounds in Broadcast Networks}\label{section:broadcast_numerics}

\begin{figure}[t!]
    \centering
    \begin{tabular}{c}
        \includegraphics[width=0.99\columnwidth]{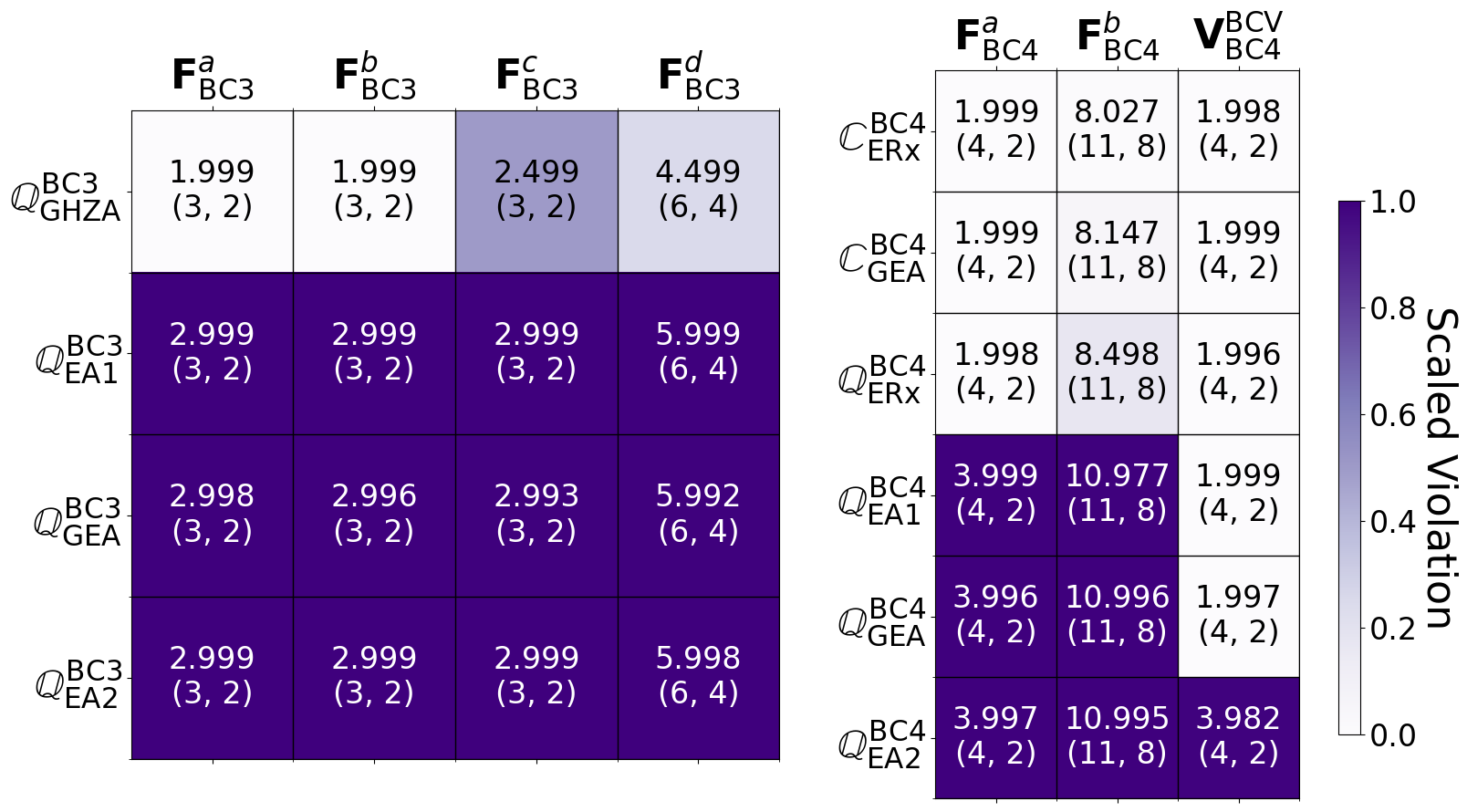} \\
        \includegraphics[width=0.99\columnwidth]{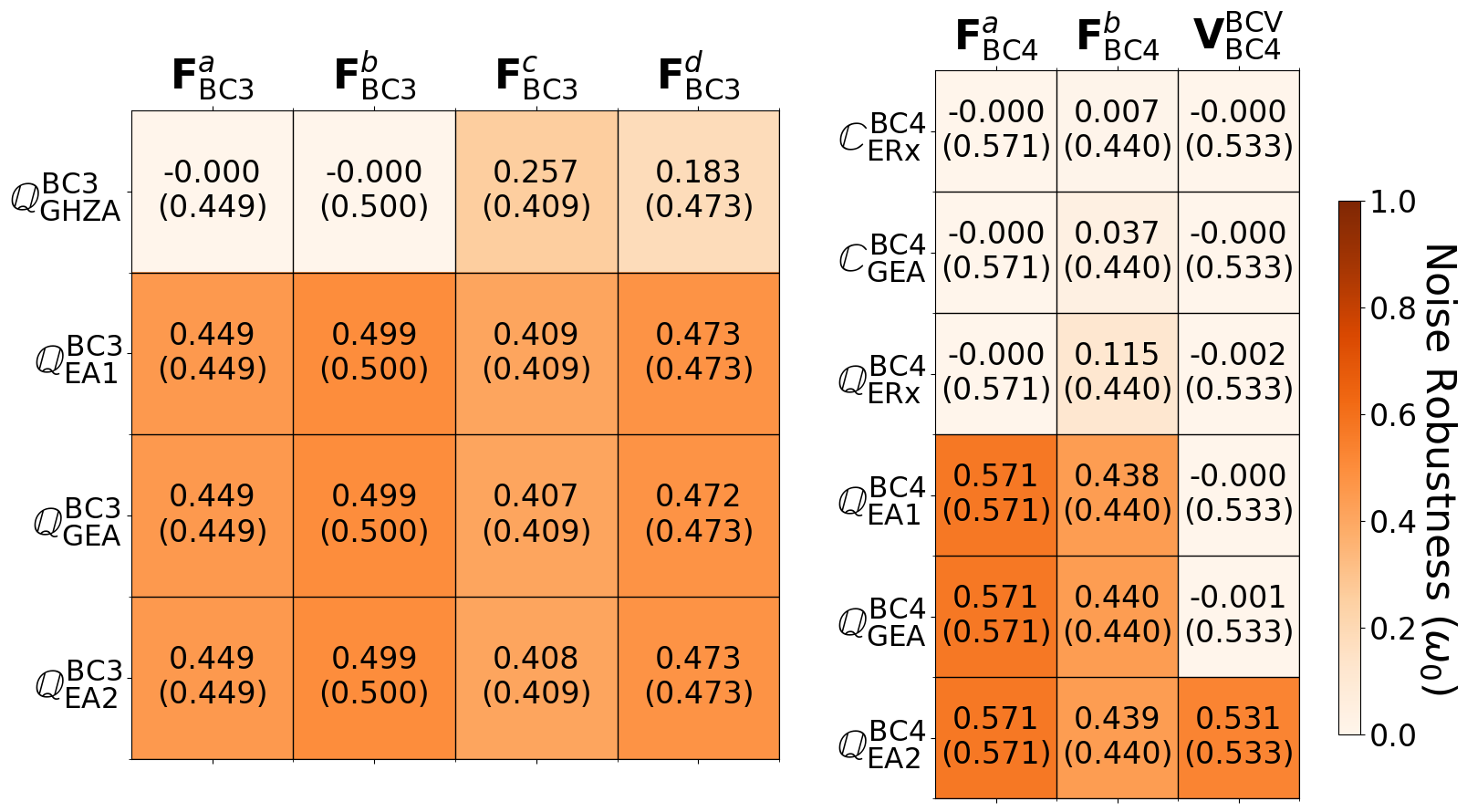}
    \end{tabular}
    
    \caption{Nonclassicality in entanglement-assisted broadcast network violations (top) and noise robustness (bottom). (Left) The max violations and noise robustness  of the facet inequalities listed in Table~\ref{Table:broadcast_X_22_YY_polytope} for the broadcast network  $\BC 3$. (Right) The max violations and noise robustness of the $\BC 4$ facet inequalities in Table~\ref{Table:broadcast_X_22_YY_polytope} and the $\Vbf^{\text{BCV}}$ simulation game from Eq.~\eqref{eq:game_bcv}. Each column  corresponds to a different nonclassicality witness while each row corresponds to a different resource configuration. The top number in each cell shows the largest numerical violation obtained via variational optimization or its corresponding noise robustness. The lower tuple, $(\hat{\gamma}, \gamma)$, shows the largest possible score $\hat{\gamma}$ and the classical bound $\gamma$ for each linear black box game or the noise robustness of the maximal possible violation. Note that EA1 corresponds to entanglement-assistance on one channel, while EA2 corresponds to entanglement-assistance on both channels.}
    \label{fig:violations_of_classicality_broadcast_networks}
\end{figure}

We investigate the nonclassical behaviors that can be generated in the broadcast networks $\BC3$ and $\BC4$ (see Fig.~\ref{fig:broadcast_network_dags}). 
Using variational optimization, we maximize the violation of each nonclassicality witness in Table~\ref{Table:broadcast_X_22_YY_polytope} for each quantum resource configuration shown in Fig.~\ref{fig:broadcast_network_dags}. We plot the max violation and noise robustness of each nonclassicality witnesses in Fig.~\ref{fig:violations_of_classicality_broadcast_networks}. \rev{Our numerical results support Theorem~\ref{thm:broadcast_classicality}, which proves that unassisted quantum communication $\Qbb^{\BC}$ cannot violate any nonclassicality witness.}{Therefore we conjecture that $\Qbb^{\BC} = \Cbb^{\BC}$ for all broadcast networks. Such a result would be similar to the bound on quantum point-to-point networks derived by Frenkel and Weiner \cite{Frenkel2015_classical_information_n-level_quantum_system}.} \rev{ This limitation distinguishes broadcast networks from multiaccess networks, which can demonstrate nonclassicality using unassisted quantum communication. }{}

We find the strongest examples of nonclassicality when entanglement-assisted communication channels are used. We consider the basic example of a broadcast network with entanglement-assisted quantum signaling, $\Qbb^{\BC}_{\EA2}$, as shown in Fig.~\ref{fig:broadcast_network_dags}.h. We also consider the case where entanglement assists only one communication channel $\Qbb^{\BC}_{\EA 1}$. Entanglement-assisted quantum communication enables dense coding communication \cite{bennet1992_dense_coding}, allowing two bits to be communicated. Since we consider input and output alphabets of size no greater than four it holds that $\Qbb^{\BC4}_{\EA2}  = \Pbb_{\YNet|\Xmc}$, implying that all broadcast behaviors having $|\Xmc|=|\Ymc_1|=|\Ymc_2|\leq 4$ can be simulated.
Therefore we demonstrate our optimization framework's ability to learn dense coded communication protocols.

In particular, we find that $\Qbb^{\BC4}_{\EA2}$ is the only considered resource configuration that can violate the classical bound of the broadcast communication value game $\Vbf^{\text{BCV}}_{\BC4}$ defined in Eq.~\eqref{eq:game_bcv}, which requires  a signaling dimension of $d_1=d_2=4$. Therefore, the simulation game $(\gamma^{\text{BCV}}, \Vbf^{\text{BCV}})$ could serve as a useful nonclassicality witness for establishing dense coded channels using Algorithm~\ref{alg:vqo_simulation} or certifying entanglement-assisted quantum channels.  

We find a more nuanced example of nonclassicality when a 3-qubit entangled state is distributed across all parties, $\Cbb^{\BC}_{\GEA}$ and $\Qbb^{\BC}_{\GEA}$ as shown in Fig.~\ref{fig:broadcast_network_dags}.c,g. Although our variational ansatz considers arbitrary three qubit state preparations, in the $\Qbb^{\BC3}_{\GEA}$ resource configuration, we also consider GHZ state preparations, $\Qbb^{\BC3}_{\text{GHZ}}$. We find that the GHZ state only shows violations for $\Fbf^c_{\BC3}$ and $\Fbf^d_{\BC3}$, whereas the general state preparation $\Qbb^{\BC3}_{\GEA}$ achieves the maximal score by using the entanglement along one channel to perform dense coding. As shown in Fig.~\ref{fig:violations_of_classicality_broadcast_networks}, the violations of $\Qbb^{\BC}_{\GEA}$ match $\Qbb^{\BC}_{\EA 1}$, raising the question of whether global entanglement-assistance in the broadcast network can provide an advantage over entanglement-assisted quantum communication along one edge of the broadcast network.

Remarkably, nonclassicality is witnessed in classical and quantum broadcast networks with entanglement-assisted receivers, $\Cbb^{\BC}_{\ERx}$ and $\Qbb^{\BC}_{\ERx}$ (see Fig.~\ref{fig:broadcast_network_dags}.b,f). We find the strongest violation to the facet inequality $(8, \Fbf_{\BC 4}^b)$ Table~\ref{Table:broadcast_X_22_YY_polytope} is $8.5 = \max_{\Pbf \in \Qbb^{\BC}_{\ERx}} \ip{\Fbf_{\BC 4}^b, \Pbf} \geq \gamma = 8$, which we detail in the following protocol. 

\rev{

\begin{protocol}\label{protocol:earx_broadcast_strategy}
    Achieve the facet inequality violation $8.5 \approx \max_{\Pbf \in \Qbb^{\BC}_{\ERx}} \ip{\Fbf_{\BC 4}^b, \Pbf} \geq 8$ in the quantum broadcast network with entanglement-assisted receivers $\Qbb^{\BC}_{\ERx}$ as shown in Fig.~\ref{fig:broadcast_network_dags}.f.
    
    \begin{enumerate}
        \item Let the broadcast network be modeled by the variational ansatz shown in Fig.~\ref{fig:earx_quantum_broadcast_nonclassicality_ansatz} where the optimal settings $\vec{\theta}^\star$ for each unitary are as follows:
        \begin{align}
            \vec{\theta}^A_{\xv} &= \text{\footnotesize $\left((0, \frac{\pi}{2}),\;(0, \frac{3\pi}{2}), \;(\pi, \frac{3\pi}{2}), \;(\pi, \frac{\pi}{2}) \right)$}, \\
            \vec{\theta}^{B_1} &= \text{\footnotesize $\left(\frac{3\pi}{2}, \frac{\pi}{2}, \frac{\pi}{2}, \frac{\pi}{4}\right)$}, \; \vec{\theta}^{B_2} = \text{\footnotesize $\left(\theta^{B_2}_{0}, \pi, \frac{\pi}{4}, \frac{\pi}{2}\right)$},
        \end{align}
        where we numerically find $\theta^{B_2}_{0} \approx -2.498091860$.
        \item Upon measurement the following behavior is achieved, rounded to the eighth decimal place:
        \begin{align}
            &\Pbf^{\BC}(\vec{\theta}^\star) = \label{eq:earx_opt_behavior}\\
            &\text{\scriptsize $\begin{bmatrix}
                 0.45000005 & 0.         & 0.         & 0.         \\
                 0.         & 0.45000005 & 0.         & 0.         \\
                 0.04999995 & 0.         & 0.         & 0.         \\
                 0.         & 0.04999995 & 0.         & 0.         \\
                 0.04999995 & 0.         & 0.         & 0.         \\
                 0.         & 0.04999995 & 0.         & 0.         \\
                 0.45000005 & 0.         & 0.         & 0.         \\
                 0.         & 0.45000005 & 0.         & 0.         \\
                 0.         & 0.         & 0.         & 0.10000006 \\
                 0.         & 0.         & 0.39999994 & 0.         \\
                 0.         & 0.         & 0.         & 0.39999994 \\
                 0.         & 0.         & 0.10000006 & 0.         \\
                 0.         & 0.         & 0.         & 0.39999994 \\
                 0.         & 0.         & 0.10000006 & 0.         \\
                 0.         & 0.         & 0.         & 0.10000006 \\
                 0.         & 0.         & 0.39999994 & 0.         \\
            \end{bmatrix}$}.\notag
        \end{align}
            \item The score $8.5 = \ip{\Fbf^b_{\BC 4},\Pbf^{\BC}(\vec{\theta}^\star)} + \epsilon$ where $0\leq \epsilon \approx 10^{-12}$ is achieved numerically, while the rounded behavior in Eq.~\eqref{eq:earx_opt_behavior} achieves $\epsilon \approx 10^{-7}$.
        \end{enumerate}
        \end{protocol}     
}{}

\begin{figure}[b!]
    \centering
    \resizebox{0.99\columnwidth}{!}{%
    \begin{tikzcd}[column sep=0.5cm]
            \gategroup[3,style={prep_gate_group, inner xsep=22pt, inner ysep=12pt, xshift=32pt}, label style={yshift=2pt}]{Source $\Lambda$} 
            \gategroup[2,style={meas_gate_group, inner xsep=92pt, inner ysep=12pt, xshift=219pt}, label style={yshift=2pt}]{Receiver $B_1$} 
            \lstick{$\ket{0}$} & \gate[style={prep_gate}]{H} & \ctrl{2}  & \qw & \ctrl{1} & \gate[style=meas_gate]{R_y(\vec{\theta}^{B_1}_{0})} & \ctrl{1} & \gate[style=meas_gate]{R_y(\vec{\theta}^{B_1}_{2})} & \meter[style={meas_gate}]{} & \rstick[2]{$y_1$}\cw \\
            \gategroup[3,style={prep_gate_group, inner xsep=25pt, inner ysep=12pt, xshift=90pt}, label style={yshift=-116pt}]{Sender $A$} 
            \lstick{$\ket{0}$} & \qw & \qw & \gate[style=prep_gate]{R_y(\vec{\theta}^A_{x,0})} & \targ{} & \gate[style=meas_gate]{R_y(\vec{\theta}^{B_1}_{1})} & \targ{} & \gate[style=meas_gate]{R_y(\vec{\theta}^{B_1}_{3})} & \meter[style={meas_gate}]{} & \cw \\ 
            \gategroup[2,style={meas_gate_group, inner xsep=92pt, inner ysep=12pt, xshift=219pt}, label style={yshift=-80pt}]{Receiver $B_2$} 
            \lstick{$\ket{0}$} & \qw & \targ{} & \qw & \ctrl{1} & \gate[style=meas_gate]{R_y(\vec{\theta}^{B_2}_{0})} & \ctrl{1} & \gate[style=meas_gate]{R_y(\vec{\theta}^{B_2}_{2})} & \meter[style={meas_gate}]{} & \rstick[2]{$y_2$}\cw \\
            \lstick{$\ket{0}$} & \qw & \qw  &  \gate[style=prep_gate]{R_y(\vec{\theta}^A_{x,1})} & \targ{} & \gate[style=meas_gate]{R_y(\vec{\theta}^{B_2}_{1})} & \targ{} & \gate[style=meas_gate]{R_y(\vec{\theta}^{B_2}_{3})} & \meter[style={meas_gate}]{} & \cw \\
        \end{tikzcd}
    }
    \caption{Minimal variational ansatz for maximal violation of the $\Fbb^b_{\BC 4}$ facet inequality by the entanglement-assisted broadcast network $\Qbb^{\BC4}_{\ERx}$ where $R_y(\vec{\theta}) = e^{-i \vec{\theta} \sigma_y/2}$.}
    \label{fig:earx_quantum_broadcast_nonclassicality_ansatz}
\end{figure}

One application of these broadcast nonclassicality witnesses is to apply them to certify the presence of entanglement-assisted resources, such as entanglement entanglement-assisted two-party measurements. For instance, \rev{if unassisted quantum communication is used, then Theorem~\ref{thm:broadcast_classicality} implies that a violation of a broadcast network nonclassicality witness requires}{} that the receivers share a resource stronger than GSR, such as entanglement. We expect that future investigations of broadcast network nonclassicality witnesses and their associated violations will uncover interesting tests for verifying entangled measurement devices and entanglement-assisted communication in communication networks.

\subsection{Nonclassicality in Multipoint Networks}\label{section:nonclassicality_multipoint_networks}

\rev{In this section, we apply our operational framework to obtain novel nonclassical behaviors in multipoint communication networks.}{} 
A multipoint network consists of multiple senders and multiple receivers and may also contain intermediate processing devices, creating a complex causal structure. We consider the case where two senders $A_1$ and $A_2$ are given the inputs $x_1\in\Xmc_1$ and $x_2\in\Xmc_2$ respectively. After the information flows through the network, two independent receivers output the values $y_1\in\Ymc_1$ and $y_2\in\Ymc_2$, respectively. \rev{}{Note that $M_i$ is used as a placeholder for nodes in the final network layer, which we refer to as the measurement layer.} Although, we focus on networks with two senders and two receivers, we consider many underlying causal structures (see Fig.~\ref{fig:multipoint_network_dags}).

\begin{figure*}
    \centering
    \resizebox{0.7\textwidth}{!}{%
    \begin{tabular}{c c c}
        (a) Interference $\IF(\sigarb{\Xmc_1\Xmc_2}{\dv}{\Ymc_1,\Ymc_2})$  & & (b) Compressed Interference $\CIF(\sigarb{\Xmc_1\Xmc_2}{\dv}{\Ymc_1,\Ymc_2})$  \\
        \hfill \\
        \begin{tikzpicture}
            \node[terminal] (x1) at (0,1.15) {$x_1$};
            \node[terminal] (x2) at (0,-1.15) {$x_2$};    
            \node[dev] (A1) at (1.4,1.15) {$P^{A_1}_{a_1|x_1}$};
            \node[dev] (A2) at (1.4,-1.15) {$P^{A_2}_{a_2|x_2}$};
            \node[dev] (B) at (2.8, 0) {$P^B_{b_1b_2|a_1a_2}$};
            \node[dev] (C1) at (4.2,1.15) {$P^{C_1}_{y_1|b_1}$};
            \node[dev] (C2) at (4.2,-1.15) {$P^{C_2}_{y_2|b_2}$};
            \node[terminal] (y1) at (5.6, 1.15) {$y_1$};
            \node[terminal] (y2) at (5.6, -1.15) {$y_2$};

            \path (x1) \cedge (A1);
            \path (x2) \cedge (A2);
            \path (A1) \cedge  node[el, above=3pt, xshift=4pt] {$a_1$} (B);
            \path (A2) \cedge node[el, below=3pt, xshift=4pt] {$a_2$} (B);
            \path (B) \cedge  node[el, above=3pt, xshift=-4pt] {$b_1$} (C1);
            \path (B) \cedge node[el, below=3pt, xshift=-4pt] {$b_2$} (C2);
            \path (C1) \cedge (y1);
            \path (C2) \cedge (y2);
        \end{tikzpicture} & & \begin{tikzpicture}
            \node[terminal] (x1) at (0,1.15) {$x_1$};
            \node[terminal] (x2) at (0,-1.15) {$x_2$};    
            \node[dev] (A1) at (1.4,1.15) {$P^{A_1}_{a_1|x_1}$};
            \node[dev] (A2) at (1.4,-1.15) {$P^{A_2}_{a_2|x_2}$};
            \node[dev] (B) at (2.8, 0) {$P^B_{b|a_1a_2}$};
            \node[dev] (C) at (5, 0) {$P^C_{c_1c_2|b}$};
            \node[dev] (D1) at (6.4,1.15) {$P^{D_1}_{y_1|c_1}$};
            \node[dev] (D2) at (6.4,-1.15) {$P^{D_2}_{y_2|c_2}$};
            \node[terminal] (y1) at (7.8, 1.15) {$y_1$};
            \node[terminal] (y2) at (7.8, -1.15) {$y_2$};

            \path (x1) \cedge (A1);
            \path (x2) \cedge (A2);
            \path (A1) \cedge  node[el, above=3pt, xshift=4pt] {$a_1$} (B);
            \path (A2) \cedge node[el, below=3pt, xshift=4pt] {$a_2$} (B);
            \path (B) \cedge node[el, above=3pt] {$b$} (C);
            \path (C) \cedge  node[el, above=3pt, xshift=-4pt] {$c_1$} (D1);
            \path (C) \cedge node[el, below=3pt, xshift=-4pt] {$c_2$} (D2);
            \path (D1) \cedge (y1);
            \path (D2) \cedge (y2);
        \end{tikzpicture} \\
        \hfill \\
        (c) Hourglass $\HG(\sigarb{\Xmc_1\Xmc_2}{\dv}{\Ymc_1,\Ymc_2})$  & & (d) Butterfly, $\BF(\sigarb{\Xmc_1\Xmc_2}{\dv}{\Ymc_1,\Ymc_2})$ \\
        \hfill \\
        \begin{tikzpicture}
            \node[terminal] (x1) at (0,1.15) {$x_1$};
            \node[terminal] (x2) at (0,-1.15) {$x_2$};    
            \node[dev] (A1) at (1.6,1.15) {$P^{A_1}_{a_1a_2|x_1}$};
            \node[dev] (A2) at (1.6,-1.15) {$P^{A_2}_{a_3a_4|x_2}$};
            \node[dev] (B1) at (4,1.15) {$P^{B_1}_{y_1|a_1a_3}$};
            \node[dev] (B2) at (4,-1.15) {$P^{B_2}_{y_2|a_2a_4}$};
            \node[terminal] (y1) at (5.6, 1.15) {$y_1$};
            \node[terminal] (y2) at (5.6, -1.15) {$y_2$};

            \path (x1) \cedge (A1);
            \path (x2) \cedge (A2);
            \path (A1) \cedge  node[el, above=3pt, xshift=-6pt] {$a_1$} (B1);
            \path (A1) \cedge  node[el, above=4pt, xshift=-24pt] {$a_2$} (B2);
            \path (A2) \cedge node[el, below=4pt, xshift=-24pt] {$a_3$} (B1);
            \path (A2) \cedge node[el, below=3pt, xshift=-6pt] {$a_4$} (B2);
            \path (B1) \cedge (y1);
            \path (B2) \cedge (y2);
        \end{tikzpicture} & & \begin{tikzpicture}
            \node[terminal] (x1) at (0,1.15) {$x_1$};
            \node[terminal] (x2) at (0,-1.15) {$x_2$};    
            \node[dev] (A1) at (1.6,1.15) {$P^{A_1}_{a_1a_2|x_1}$};
            \node[dev] (A2) at (1.6,-1.15) {$P^{A_2}_{a_3a_4|x_2}$};
            \node[dev] (B) at (2.8, 0) {$P^B_{b|a_1a_2}$};
            \node[dev] (C) at (4.8, 0) {$P^C_{c_1c_2|b}$};
            \node[dev] (D1) at (6.2,1.15) {$P^{D_1}_{y_1|a_1c_1}$};
            \node[dev] (D2) at (6.2,-1.15) {$P^{D_2}_{y_2|a_4c_2}$};
            \node[terminal] (y1) at (7.8, 1.15) {$y_1$};
            \node[terminal] (y2) at (7.8, -1.15) {$y_2$};
   
            \path (x1) \cedge (A1);
            \path (x2) \cedge (A2);
            \path (A1) \cedge  node[el, above=-6pt, xshift=-14pt] {$a_2$} (B);
            \path (A2) \cedge node[el, below=-6pt, xshift=-14pt] {$a_3$} (B);
            \path (B) \cedge node[el, above=3pt] {$b$} (C);
            \path (C) \cedge  node[el, below=2pt, xshift=4pt] {$c_1$} (D1);
            \path (C) \cedge node[el, above=2pt, xshift=4pt] {$c_2$} (D2);
            \path (D1) \cedge (y1);
            \path (D2) \cedge (y2);
            \path (A1) \cedge node[el, above=3pt, xshift=-36pt] {$a_1$} (D1);
            \path (A2) \cedge node[el, below=3pt, xshift=-36pt] {$a_4$} (D2);
        \end{tikzpicture} \\
    \end{tabular}
    }
    \caption{Classical multipoint network DAGs. (a) Interference (IF) network, (b) compressed interference (CIF) network, (c) hourglass (HG) network, and (d) butterfly (BF) network.}
    \label{fig:multipoint_network_dags}
\end{figure*}

\begin{figure*}
    \centering
    \resizebox{0.7\textwidth}{!}{%
    \begin{tabular}{c c c}
        (a) QC & (b) ERx QC & (c) ETx QC \\
        \hfill \\
        \begin{tikzpicture}
            \node[terminal] (x1) at (-0.15,1.25) {$x_1$};
            \node[terminal] (x2) at (-0.15,-1.25) {$x_2$};    
            \node[prep_dev] (A1) at (1.2,1.25) {$\rho^{A_1}_{x_1}$};
            \node[prep_dev] (A2) at (1.2,-1.25) {$\rho^{A_2}_{x_2}$};
            \node[proc_dev] (B) at (2.4, 0) {$\Emc^B$};
            \node[meas_dev] (C1) at (3.6,1.25) {$\Pi^{C_1}_{y_1}$};
            \node[meas_dev] (C2) at (3.6,-1.25) {$\Pi^{C_2}_{y_2}$};
            \node[terminal] (y1) at (4.8, 1.25) {$y_1$};
            \node[terminal] (y2) at (4.8, -1.25) {$y_2$};

            \path (x1) \cedge (A1);
            \path (x2) \cedge (A2);
            \path (A1) \qedge  node[el, above=3pt, xshift=4pt] {} (B);
            \path (A2) \qedge node[el, below=3pt, xshift=4pt] {} (B);
            \path (B) \qedge  node[el, above=3pt, xshift=-4pt] {} (C1);
            \path (B) \qedge node[el, below=3pt, xshift=-4pt] {} (C2);
            \path (C1) \cedge (y1);
            \path (C2) \cedge (y2);
        \end{tikzpicture} & \begin{tikzpicture}
            \node[terminal] (x1) at (-0.15,1.25) {$x_1$};
            \node[terminal] (x2) at (-0.15,-1.25) {$x_2$};    
            \node[prep_dev] (A1) at (1.2,1.25) {$\rho^{A_1}_{x_1}$};
            \node[prep_dev] (A2) at (1.2,-1.25) {$\rho^{A_2}_{x_2}$};
            \node[qsource] (Lambda) at (-0.15,0) {$\rho^{\Lambda}$};
            \node[proc_dev] (B) at (2.4, 0) {$\Emc^B$};
            \node[meas_dev] (C1) at (3.6,1.25) {$\Pi^{C_1}_{y_1}$};
            \node[meas_dev] (C2) at (3.6,-1.25) {$\Pi^{C_2}_{y_2}$};
            \node[terminal] (y1) at (4.8, 1.25) {$y_1$};
            \node[terminal] (y2) at (4.8, -1.25) {$y_2$};

            \path (x1) \cedge (A1);
            \path (x2) \cedge (A2);
            \path (Lambda) \qedge (C1);
            \path (Lambda) \qedge (C2);
            \path (A1) \qedge  node[el, above=3pt, xshift=4pt] {} (B);
            \path (A2) \qedge node[el, below=3pt, xshift=4pt] {} (B);
            \path (B) \qedge  node[el, above=3pt, xshift=-4pt] {} (C1);
            \path (B) \qedge node[el, below=3pt, xshift=-4pt] {} (C2);
            \path (C1) \cedge (y1);
            \path (C2) \cedge (y2);
        \end{tikzpicture} & \begin{tikzpicture}
            \node[terminal] (x1) at (-0.15,1.25) {$x_1$};
            \node[terminal] (x2) at (-0.15,-1.25) {$x_2$};    
            \node[proc_dev] (A1) at (1.2,1.25) {$\Emc^{A_1}_{x_1}$};
            \node[proc_dev] (A2) at (1.2,-1.25) {$\Emc^{A_2}_{x_2}$};
            \node[qsource] (Lambda) at (-0.15,0) {$\rho^{\Lambda}$};
            \node[proc_dev] (B) at (2.4, 0) {$\Emc^B$};
            \node[meas_dev] (C1) at (3.6,1.25) {$\Pi^{C_1}_{y_1}$};
            \node[meas_dev] (C2) at (3.6,-1.25) {$\Pi^{C_2}_{y_2}$};
            \node[terminal] (y1) at (4.8, 1.25) {$y_1$};
            \node[terminal] (y2) at (4.8, -1.25) {$y_2$};

            \path (x1) \cedge (A1);
            \path (x2) \cedge (A2);
            \path (Lambda) \qedge (A1);
            \path (Lambda) \qedge (A2);
            \path (A1) \qedge  node[el, above=3pt, xshift=4pt] {} (B);
            \path (A2) \qedge node[el, below=3pt, xshift=4pt] {} (B);
            \path (B) \qedge  node[el, above=3pt, xshift=-4pt] {} (C1);
            \path (B) \qedge node[el, below=3pt, xshift=-4pt] {} (C2);
            \path (C1) \cedge (y1);
            \path (C2) \cedge (y2);
        \end{tikzpicture} \\
    \end{tabular}%
    }
    \caption{Quantum resource configurations for multipoint network DAGs in Fig.~\ref{fig:multipoint_network_dags}. (a) Quantum communication. (b) Quantum communication with entanglement-assisted receivers. (c) Quantum communication with entanglement-assisted senders.}
    \label{fig:multipoint_network_quantum_dags}
\end{figure*}

\begin{table}[b!]
    \centering
    \resizebox{\columnwidth}{!}{%
    \begin{tabular}{|c|c|}
        \hline
         \vspace{-9pt}& \\
         \textbf{Set} & \textbf{Behavior Decomposition}  \\
         \vspace{-9pt}& \\
         \hhline{|=|=|}
         \vspace{-9pt}& \\
         a) $\Cbb^{\IF}_{}$ & $\Pbf^{\IF} = (\Pbf^{C_1}\otimes\Pbf^{C_2})\Pbf^B(\Pbf^{A_1}\otimes \Pbf^{A_2})$ \\
         \vspace{-9pt}& \\
         \hline 
         \vspace{-9pt}& \\
         b) $\Cbb^{\CIF}_{}$ & $\Pbf^{\IF} = (\Pbf^{D_1}\otimes\Pbf^{D_2})\Pbf^{C}\Pbf^B(\Pbf^{A_1}\otimes \Pbf^{A_2})$ \\
        \vspace{-9pt} & \\
        \hline
        \vspace{-9pt}& \\
        c) $\Cbb^{\HG}_{}$ & $\Pbf^{\HG} = \begin{matrix} (\Pbf^{B_1}\otimes\Pbf^{B_2})(\Ibb_2\otimes \Vbf^{\leftrightarrow}\otimes \Ibb_2)\times\\
        \times(\Pbf^{A_1}\otimes\Pbf^{A_2})\end{matrix}$ \\
        \vspace{-9pt}& \\
        \hline
        \vspace{-9pt}& \\
        d) $\Cbb^{\BF}_{}$ & 
        $\Pbf^{\BF} = \begin{matrix} (\Pbf^{D_1}\otimes\Pbf^{D_2})(\Ibb_2\otimes \Pbf^{C}\Pbf^B\otimes \Ibb_2)\times\\
        \times(\Pbf^{A_1}\otimes \Pbf^{A_2})\\
        \vspace{-12pt}
        \end{matrix}$ \\
        \hline
    \end{tabular}%
    }
    \caption{Classical multipoint network polytope definitions (see Fig.~\ref{fig:multipoint_network_dags} for DAGs). a) Interference (IF) network. b) Compressed interference (CIF) network. c) Hourglass (HG) network. d) Butterfly (BF) network. The identity matrices $\Ibb_2$ represent communication through an intermediate layer. In the hourglass network, an explicit swap $\Vbf^{\leftrightarrow}$ is needed to model the communication from $A_1 \to B_2$ and $A_2\to B_1$.  }
\end{table}

\rev{
In Section~\ref{section:multipoint_nonclassicality_witnesses}, we discuss how to utilize simulation games to efficiently derive related facet inequalities when standard facet enumeration techniques fail to scale. In Section~\ref{section:multipoint_violations}, we present quantum violations and noise robustness for numerous simulation games and related facet inequalities.
Overall, we find that quantum communication is sufficient to achieve nonclassicality in all considered multisender networks where the strongest violations are found in multiaccess networks. Furthermore, we find that entanglement-assisted senders tend to have stronger violations than entanglement-assisted receivers. 
}{}

\subsubsection{Multipoint Network Nonclassicality Witnesses}\label{section:multipoint_nonclassicality_witnesses}

For multipoint networks, it is difficult to enumerate the set of vertices $\Vbb^{\Net}$ of the classical network polytope  due to the large number of independent devices and the associated exponential growth in the dimension of the network behavior. Given the memory constraints of a typical laptop computer, we are able to enumerate $\Vbb^{\Net}$ in the simplest nontrivial case  where $|\Xmc_1|=|\Xmc_2|=|\Ymc_1|=|\Ymc_2|=3$ for each of the networks in Fig.~\ref{fig:multipoint_network_dags}.
Using this set of vertices, we obtain both simulation games and facet inequalities.

Recall that a simulation game $(\gamma, \Vbf)$ can be obtained from any deterministic behavior that is excluded from the classical network polytope such that $\Vbf \notin \Cbb^{\Net}$. These simulation games can be obtained from simple logical and arithmetic operations where a complete list is provided in Table~\ref{table:games_for_multipoint_communication_networks}.
To derive a selection of facet inequalities, we use the linear program in Eq.~\eqref{eq:facet_linear_program}. For each network DAG and deterministic test behavior in Table~\ref{table:games_for_multipoint_communication_networks}, we obtain a nonclassicality witness and verify it to be a facet inequality $(\gamma, \Fbf)$ of the $\Cbb^{\Net}$ polytope. Note that each computed facet inequality is unique to the network although the same deterministic test behaviors are the considered across all networks. To avoid confusion, we use $\Fbf$ to denote facet inequalities and $\Vbf$ to denote simulation games where the superscript labels from Table~\ref{table:games_for_multipoint_communication_networks} are used in both cases. In Appendix~\ref{section:multipoint_facet_inequalities}, we present the facet inequalities  obtained for each of the considered multipoint communication networks.

\begin{table}[t!]
    \centering
    \resizebox{\columnwidth}{!}{%
    \begin{tabular}{|c|c|c|}
        \hline
        \textbf{Name} & \textbf{Sym.} & \textbf{Definition}\\
        \hhline{|=|=|=|}
        \vspace{-9pt}&&\\
        $\begin{matrix}
            \text{Multiplication} \\ \text{(0,1,2)}
        \end{matrix}$ & $\Vbf^{\times_0}$ & $V^{\times_0}_{\yv|\xv} = \left\{\begin{matrix} 1 & if \; \yv = x_1 \times x_2\\ 0 & otherwise \\ \end{matrix} \right.$   \\
        \vspace{-9pt}&&\\
        \hline
        \vspace{-9pt}&&\\
        $\begin{matrix}
            \text{Multiplication} \\ \text{(1,2,3)}
        \end{matrix}$ & $\Vbf^{\times_1}$ & $V^{\times_1}_{\yv|\xv} = \left\{\begin{matrix} 1 & if \; \yv = x_1 \times x_2\\ 0 & otherwise \\ \end{matrix} \right.$   \\
        \vspace{-9pt}&&\\
        \hline
        \vspace{-9pt}&&\\
        Swap & $\Vbf^{\leftrightarrow}$ & $V^{\leftrightarrow}_{\yv|\xv} = \left\{ \begin{matrix} 1 & if\;y_1=x_2\; and\; y_2=x_1 \\ 0 & otherwise \\ \end{matrix}\right.$ \\
        \vspace{-9pt}&&\\
        \hline
        \vspace{-9pt}&&\\
        $\begin{matrix}\text{Addition}\\  \text{(0,1,2)}\end{matrix}$ & $\Vbf^{+}$ & $V^{+}_{\yv|\xv} = \left\{ \begin{matrix} 1 & if\; \yv = x_1 + x_2 \\ 0 & otherwise \\ \end{matrix}\right.$ \\
        \vspace{-9pt}&&\\
        \hline
        \vspace{-9pt}&&\\
        Equality & $\Vbf^=$ & $V^=_{y|x_1,x_2} = 1- \delta_{x_1,x_2}$ \\
        \vspace{-9pt}&&\\
        \hline
        \vspace{-9pt}&&\\
        Comparison & $\Vbf^{\gtrless}$ & $V^{\gtrless}_{\yv|\xv} = \left\{ \begin{matrix} \delta_{y_1,0}\delta_{y_2,0} & if\; x_1=x_2  \\ \delta_{y_1,1}\delta_{y_2,2} & if \; x_1<x_2 \\ \delta_{y_1,2}\delta_{y_2,1} & if x_1 > x_2 \\ \end{matrix}\right.$ \\
        \vspace{-9pt}&&\\
        \hline
        \vspace{-9pt}&&\\
        Permutation & $\Vbf^{\pi}$ & $V^{\pi}_{\yv|\xv} = \left\{ \begin{matrix} \delta_{y_1,0}\delta_{y_2,x_2} & if\; x_1=0 \\ \delta_{y_1,1}\delta_{y_2,(x_2+2) \% 3} & if \; x_1=1 \\ \delta_{y_1,2}\delta_{y_2,(x_2+1) \% 3} & if \; x_1=2 \\ \end{matrix}\right.$ \\
        \vspace{-9pt}&&\\
        \hline
        \vspace{-9pt}&&\\
        Difference & $\Vbf^-$ & $V^{-}_{\yv|\xv} = \left\{ \begin{matrix} 1 & if\; y_1=y_2 = |x_1 - x_2| \\ 0 & otherwise \\ \end{matrix}\right.$\\
        \vspace{-9pt}&&\\
        \hline
        \vspace{-9pt}&&\\
        $\begin{matrix}\text{Communication} \\ \text{Value}\\\vspace{-9pt}\end{matrix}$ & $\Vbf^{\text{CV}}$ & $V^{\text{CV}}_{\yv|\xv} = \delta_{\yv,\xv}$ \\
        \hline
    \end{tabular}%
    }
    \caption{Simulation games for $\Net(\sigarb{3,3}{\dv}{3,3})$ multipoint communication networks. Each deterministic behavior $\Vbf\in\Vbb_{\YNet|\XNet}$ cannot be simulated by the networks given in Fig.~\ref{fig:multipoint_network_dags}. Note that $a\% b \equiv a \mod b$.}
    \label{table:games_for_multipoint_communication_networks}
\end{table}

\subsubsection{Numerical Quantum Violations of Classical Bounds in Multipoint Networks}\label{section:multipoint_violations}

For each multipoint communication network shown in Fig.~\ref{fig:multipoint_network_dags} we consider three quantum resource configurations: unassisted quantum communication ($\Qbb^{\Net}$), entanglement-assisted senders ($\Qbb^{\Net}_{\ETx}$), and entanglement-assisted receivers ($\Qbb^{\Net}_{\ERx}$) (see Fig.~\ref{fig:multipoint_network_quantum_dags}).
For each network, our variational ans\"{a}tze encode a broad set of free operations, including CPTP maps and POVMs (see Section~\ref{section:simulating_quantum_networks}). Since we consider $|\Ymc_1|=|\Ymc_2|=3$, we apply a deterministic post-processing map that takes each local two-qubit measurement result to a trit. 
We find that the post-processing map $\Vbf^1$ is sufficient to achieve the max violation in most cases, however, certain nonclassicality witnesses can achieve larger violations when $\Vbf^2$ is used as a post-processing map (see Eq.~\eqref{eq:postprocesssingmaps}),
\begin{equation}\label{eq:postprocesssingmaps}
    \text{\footnotesize $\Vbf^{1} = \begin{bmatrix}
        1 & 0 & 0 & 1 \\ 0 & 1 & 0 & 0 \\ 0 & 0 & 1 & 0 \\
    \end{bmatrix} \quad \text{or} \quad \Vbf^{2} = \begin{bmatrix}
        1 & 0 & 0 & 0 \\ 0 & 1 & 1 & 0 \\ 0 & 0 & 0 & 1 \\
    \end{bmatrix}$}.
\end{equation}

\begin{figure*}
    \centering
    \includegraphics[width=\textwidth]{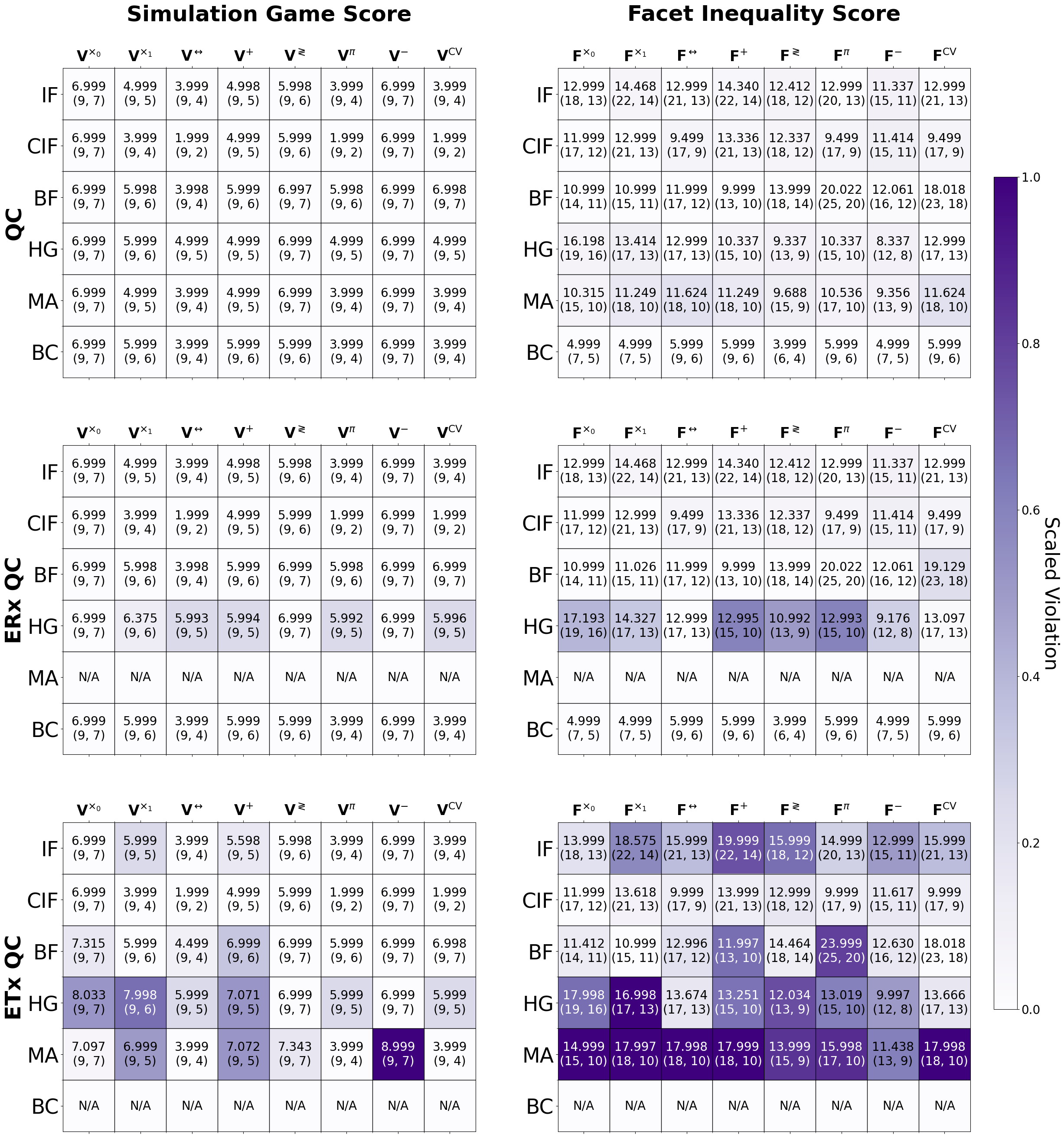}
    \caption{Quantum violation of classicality in $\sigarb{3,3}{\dv}{3,3}$  multipoint networks. For the interference (IF), compressed interference (CIF), butterfly (BF), hourglass (HG), multiaccess (MA), and broadcast (BC) networks we consider quantum communication (QC), entanglement-assisted receiver quantum communication (ERx QC), and entanglement-assisted transmitter quantum communication (ETx QC) resource configurations as shown in Fig.~\ref{fig:multipoint_network_quantum_dags}. The column of each plot corresponds to a different nonclassicality witness while each row corresponds to a different network DAG. The top number in each cell shows the largest numerical violation obtained via variational optimization. The lower tuple, $(\hat{\gamma}, \gamma)$, shows the largest possible score $\hat{\gamma}$ and the classical bound $\gamma$ for each linear black box game. In the left-hand column of plots, we show the quantum violations are listed for the simulation games listed in Table~\ref{table:games_for_multipoint_communication_networks} while the right-hand column of plots shows the quantum violations of the classical network polytope facet inequalities listed in Appendix~\ref{section:multipoint_facet_inequalities}. In both cases, the cells are shaded according to the scaled violation.  }
    \label{fig:quantum_violations_in_33-33_multipoint_communication_networks}
\end{figure*}

\begin{figure*}
    \centering
    \includegraphics[width=\textwidth]{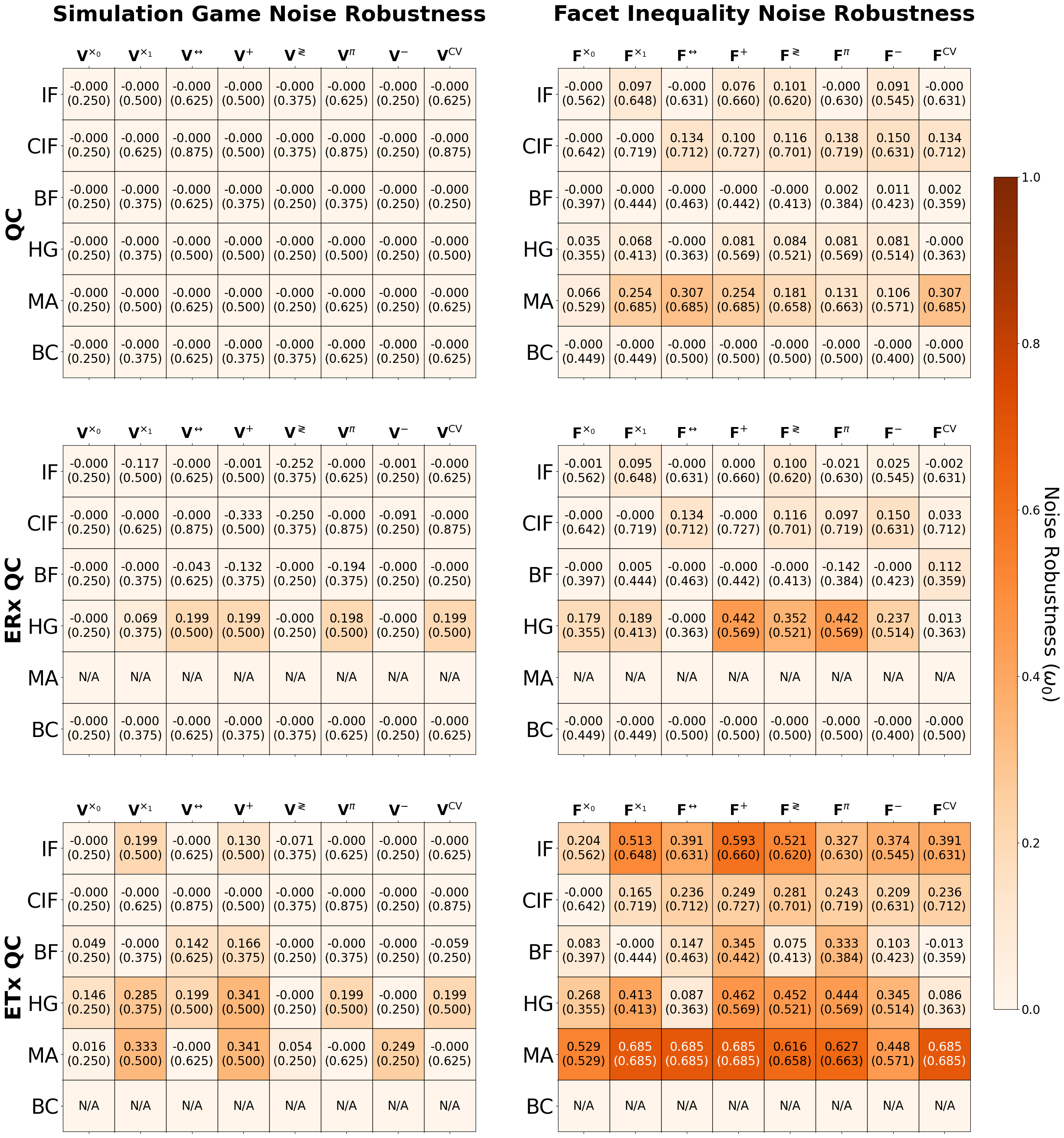}
    \caption{Noise robustness of $\sigarb{3,3}{\dv}{3,3}$  multipoint networks. For the interference (IF), compressed interference (CIF), butterfly (BF), hourglass (HG), multiaccess (MA), and broadcast (BC) networks we consider quantum communication (QC), entanglement-assisted receiver quantum communication (ERx QC), and entanglement-assisted transmitter quantum communication (ETx QC) resource configurations as shown in Fig.~\ref{fig:multipoint_network_quantum_dags}. The column of each plot corresponds to a different nonclassicality witness while each row corresponds to a different network DAG. The top number in each cell shows the noise robustness of the largest optimized violation. The lower number in each cell shows the noise robustness of the largest violation. In the left-hand column of  plots, the noise robustness is listed for the simulation games listed in Table~\ref{table:games_for_multipoint_communication_networks} while the right-hand column of plots lists the noise robustness of the facet inequalities listed in Appendix~\ref{section:multipoint_facet_inequalities}. In both cases, the cells are shaded according to their noise robustness. }
    \label{fig:noise_robustness_in_33-33_multipoint_communication_networks}
\end{figure*}

In Fig.~\ref{fig:quantum_violations_in_33-33_multipoint_communication_networks}, we plot the violation obtained using our VQO methods. We find that all considered quantum resource configurations can produce nonclassical behaviors in multipoint communication networks. 
When unassisted quantum signaling is considered (first row Fig.~\ref{fig:quantum_violations_in_33-33_multipoint_communication_networks}), we find violations of facet inequalities $(\gamma, \Fbf)$ in all networks except the broadcast network, but we find no violations in any of the considered simulation games $(\gamma, \Vbf)$. Thus, we demonstrate that unassisted quantum communication is sufficient for nonclassicality in multisender networks.

Entanglement-assisted receivers (see Fig.~\ref{fig:multipoint_network_quantum_dags}.b) are considered in the second row of Fig.~\ref{fig:quantum_violations_in_33-33_multipoint_communication_networks}. We find that the $\ERx$ resource configuration only provides advantage over unassisted quantum signaling in the hourglass (HG) and butterfly (BF) networks. Stronger violations are found in the hourglass network, which shows an advantage in the communication value game $(5, \Vbf^{\CV}_{3,3\to 3,3})$. This indicates that the entanglement-assisted receivers can increase the network's classical communication capacity. In the butterfly network, the violation of the facet inequality $\Fbf^{\CV}$ is interesting because it shows a numerical example where a larger score is achieved for entanglement-assisted receivers than entanglement-assisted senders.
This suggests that the two sets, $\Qbb^{\Net}_{\ETx}$ and $\Qbb^{\Net}_{\ERx}$, could have mutually excluded regions, which we leave as an interesting open problem.

Entanglement-assisted senders (see Fig.~\ref{fig:multipoint_network_quantum_dags}.c) are considered in the third row of Fig.~\ref{fig:quantum_violations_in_33-33_multipoint_communication_networks}.  We find that entanglement-assisted senders broadly achieve larger violations than unassisted quantum signaling. The  strongest violations with the greatest noise robustness are achieved by the multiaccess network, $\MA(\sigarb{3,3}{2,2}{9})$. In most examples, entanglement-assisted senders (ETx) achieve stronger violations than entanglement assisted receivers (ERx), while our findings for the butterfly network violation of $\Fbf^{\CV}$ suggest the possibility that neither resource configuration can fully simulate the other.

In Fig.~\ref{fig:noise_robustness_in_33-33_multipoint_communication_networks}, we plot the noise robustness for each of the violations. Remarkably, the largest noise robustness observed in this work has a critical noise parameter of about $\omega_0 \approx 0.685$, and is attributed to a facet inequality of the multiaccess network $\MA(\sigarb{3,3}{2,2}{9})$ with entanglement-assisted senders. Furthermore, these facet inequalities show greater noise robust than their corresponding simulation games, highlighting a benefit of using facet inequalities to witness nonclassicality. Therefore, we find that entanglement-assisted senders are more robust to noise than entanglement-assisted receivers, while the noise robustness of quantum communication is quite small in comparison. 

Overall, we provide numerical examples of communication advantages in quantum multipoint networks. We show that entangled senders and receivers are able to demonstrate advantages in simulation games, implying that the deterministic tasks in Table~\ref{table:games_for_multipoint_communication_networks} are performed with greater success using quantum resources. We also identify a large number of example violations of the facet inequalities of the classical network polytope. These violations pave the way for self-testing methods for network causal structures and resource configurations.

\section{Discussion}
\label{section:discussion}

\rev{
This work investigates the communication advantage offered by quantum resources in communication networks.
We develop a hardware agnostic and operational framework for realizing communication advantages in simulated and deployed quantum networks. 
In this framework, we quantify communication advantage in terms of nonclassicality \cite{Bowles2015_nonclassicality_communication_networks}, a phenomenon in which quantum resources produce correlations that cannot be reproduced using a similar amount of classical communication and GSR.
To maximize the communication advantage, we maximize the nonclassicality of a quantum network over its available communication resources using VQO methods \cite{doolittle_2023_vqo_nonlocality,doolittle_qNetVO}.

To overcome computational challenges in obtaining nonclassicality witnesses, we introduce \textit{simulation games} (see Section~\ref{section:simulation_games}), which can be efficiently constructed, have concrete operational interpretations, and have classical bounds that can be estimated using our VQO methods (see Section~\ref{section:operational_framework}).
Our methods can be scaled on quantum computers or applied on quantum networking hardware to automatically establish and maintain communication protocols in noisy networks.
In essence, we provide a practical approach for realizing quantum communication advantages that supports both experimental and theoretical efforts in quantum networking. 

Applying our operational framework, we numerically survey the quantum resource configurations that enable nonclassicality in communication networks.
We obtain many novel examples of network nonclassicality, and when applicable, our results are consistent with prior works. We find that classical violations in multiaccess networks are the strongest and most robust to noise. Furthermore, we find that entanglement-assisted senders enable stronger classical violations than entanglement-assisted receivers.
Our results inform three key insights regarding the conditions under which nonclassicality occurs in communication-constrained networks:
\begin{enumerate}
    \item Entanglement shared between devices is sufficient for nonclassicality in communication networks.
    \item Unassisted quantum communication is sufficient for nonclassicality whenever there are multiple independent senders.
    \item Entanglement shared between devices is necessary for nonclassicality in broadcast networks.
\end{enumerate}
Each of these these insights are rigorously demonstrated by our numerical analysis, whereas the third statement is proven by Theorem~\ref{thm:broadcast_classicality}, which shows that unassisted quantum broadcast networks are classically simulable. This foundational result in quantum information theory was numerically identified and proven using our framework, showing its merit. 

Fundamentally, our results demonstrate novel examples in which the classical concepts of locality, causality, and/or realism break down. Future experiments could apply these nonclassical behaviors to verify quantum theory in complex communication scenarios, or to self-test, certify, or verify quantum communication resources and network causal structures. In particular, the maximal violation of certain nonclassicality witnesses requires POVM measurements, CPTP maps, or entanglement-assistance to be encoded into our variational ans\"{a}tze, implying that the violation requires these resources and operations. Therefore, nonclassicality witnesses can test the presence of particular quantum resources or operations, or be used to infer a network's underlying causal structure and resource configuration by ruling out networks that are incompatible with the observed communication behavior. 

Our work can be extended in several directions. 
First, our framework can be scaled to larger networks by using high-performance computing and by developing more efficient classical algorithms for obtaining classical bound and violations.
Second, our examples of nonclassicality can be generalized, characterized, demonstrated, or applied in communication protocols.  
Third, NISQ processors can scale our framework to communication networks that cannot be efficiently simulated on classical processors. 
Fourth, our operational framework can be adapted to quantum network hardware and used to automate tasks in quantum resource certification and calibration.
}{}

\begin{table}[ht]
    \centering
    \resizebox{\columnwidth}{!}{%
    \def\arraystretch{1.6}
    \begin{tabular}{|c|c|}
        \hline 
        \textbf{Acronym} & \textbf{Description} \\
        \hhline{|=|=|}
        DAG & Directed Acyclic Graph \\
        \hline
        POVM & Positive Operator-Valued Measure \\
        \hline
        CPTP & Completely-Positive Trace-Preserving \\
        \hline
        QC & Quantum Communication \\
        \hline
        CC & Classical Communication \\
        \hline
        LOCC & Local Operations \& Classical Communication \\
        \hline
        LOQC & Local Operations \& Quantum Communication \\
        \hline
        GSR & Globally Shared Randomness \\
        \hline
        EA & Entanglement-Assisted \\
        \hline
        GEA & Globally Entanglement-Assisted \\
        \hline
        ETx & Entanglement-Assisted Senders \\
        \hline
        ERx & Entanglement-Assisted Receivers \\
        \hline
        VQO & Variational Quantum Optimization \\
        \hline
        CV & Communication Value \\
        \hline 
        BCV & Broadcast Communication Value\\
        \hline
        RAC & Random Access Coding \\
        \hline 
        PM & Prepare and Measure \\
        \hline 
        MA & Multiaccess \\
        \hline 
        BC & Broadcast \\
        \hline 
        IF & Interference \\
        \hline 
        CIF & Compressed Interference \\
        \hline 
        BF & Butterfly \\
        \hline 
        HG & Hourglass \\
        \hline

    \end{tabular}%
    }
    \caption{\linespread{1}\selectfont{\small A list of acronyms used in this work.}}
\end{table}

\subsection{Supplemental Code}\label{section:Supplemental_Code}

To make our work accessible, transparent, and reproducible we provide our supporting software and data in a public GitHub repository  \cite{supplemental_software}. 
To compute classical network polytope vertices and facet inequalities we use the BellScenario.jl Julia package \cite{BellScenario.jl}, which integrates with the Polytope Representation Transformation Algorithm (PoRTA) \cite{PORTA} via the the XPORTA.jl Julia wrapper \cite{doolittle_xporta} exposed through Polyhedra.jl interface \cite{legat2021polyhedra}. The HiGHS \cite{huangfu2018highs_parallelizing} mathematical programming solver exposed via the Julia Mathematical Programming toolbox (JuMP.jl) \cite{JuMP.jl-2017} is used to solve the linear program in Eq.~\eqref{eq:facet_linear_program}.
To maximize quantum network nonclassicality we apply the Quantum Network Variational Optimizer (QNetVO) \cite{doolittle_qNetVO}, which utilizes the PennyLane ecosystem \cite{bergholm2018pennylane}.

\subsection*{Acknowledgements}

This research was supported by a grant through the IBM-Illinois Discovery Accelerator Institute and by Aliro Technologies, Inc.

\bibliography{references}

\begin{thebibliography}{79}
\providecommand{\natexlab}[1]{#1}
\providecommand{\url}[1]{\texttt{#1}}
\expandafter\ifx\csname urlstyle\endcsname\relax
  \providecommand{\doi}[1]{doi: #1}\else
  \providecommand{\doi}{doi: \begingroup \urlstyle{rm}\Url}\fi

\bibitem[Kimble(2008)]{Kimble2008quantum_internet}
H.~J. Kimble.
\newblock The quantum internet.
\newblock \emph{Nature}, 453\penalty0 (7198):\penalty0 1023--1030, June 2008.
\newblock \doi{10.1038/nature07127}.

\bibitem[Meter(2012)]{vanmeter2012_quantum_networking}
Rodney~Van Meter.
\newblock Quantum networking and internetworking.
\newblock \emph{IEEE Network}, 26\penalty0 (4):\penalty0 59--64, 2012.
\newblock \doi{10.1109/MNET.2012.6246754}.

\bibitem[Wehner et~al.(2018)Wehner, Elkouss, and
  Hanson]{wehner2018quantum_internet}
Stephanie Wehner, David Elkouss, and Ronald Hanson.
\newblock Quantum internet: A vision for the road ahead.
\newblock \emph{Science}, 362\penalty0 (6412), 2018.
\newblock \doi{10.1126/science.aam9288}.

\bibitem[Singh et~al.(2021)Singh, Dev, Siljak, Joshi, and
  Magarini]{singh2021_quantum_internet}
Amoldeep Singh, Kapal Dev, Harun Siljak, Hem~Dutt Joshi, and Maurizio Magarini.
\newblock Quantum internet—applications, functionalities, enabling
  technologies, challenges, and research directions.
\newblock \emph{IEEE Communications Surveys \& Tutorials}, 23\penalty0
  (4):\penalty0 2218--2247, 2021.
\newblock \doi{10.1109/COMST.2021.3109944}.

\bibitem[Brassard(2003)]{Brassard2003_communication_complexity}
Gilles Brassard.
\newblock Quantum communication complexity.
\newblock \emph{Foundations of Physics}, 33\penalty0 (11):\penalty0
  1593–1616, 2003.
\newblock ISSN 0015-9018.
\newblock \doi{10.1023/a:1026009100467}.

\bibitem[Buhrman et~al.(2010)Buhrman, Cleve, Massar, and
  de~Wolf]{buhrman2010_communication_complexity_nonloclaity}
Harry Buhrman, Richard Cleve, Serge Massar, and Ronald de~Wolf.
\newblock Nonlocality and communication complexity.
\newblock \emph{Rev. Mod. Phys.}, 82:\penalty0 665--698, Mar 2010.
\newblock \doi{10.1103/RevModPhys.82.665}.

\bibitem[Bennett et~al.(2002)Bennett, Shor, Smolin, and
  Thapliyal]{Bennett2002_rev_shannon}
C.H. Bennett, P.W. Shor, J.A. Smolin, and A.V. Thapliyal.
\newblock Entanglement-assisted capacity of a quantum channel and the reverse
  shannon theorem.
\newblock \emph{{IEEE} Transactions on Information Theory}, 48\penalty0
  (10):\penalty0 2637--2655, October 2002.
\newblock \doi{10.1109/TIT.2002.802612}.

\bibitem[Winter(2002)]{Winter-2002a}
Andreas Winter.
\newblock Compression of sources of probability distributions and density
  operators.
\newblock \emph{arXiv preprint}, 2002.
\newblock \doi{10.48550/arXiv.quant-ph/0208131}.

\bibitem[Bennett et~al.(2014)Bennett, Devetak, Harrow, Shor, and
  Winter]{Bennett-2014a}
Charles~H. Bennett, Igor Devetak, Aram~W. Harrow, Peter~W. Shor, and Andreas
  Winter.
\newblock The quantum reverse shannon theorem and resource tradeoffs for
  simulating quantum channels.
\newblock \emph{IEEE Transactions on Information Theory}, 60\penalty0
  (5):\penalty0 2926--2959, 2014.
\newblock \doi{10.1109/TIT.2014.2309968}.

\bibitem[Cubitt et~al.(2011)Cubitt, Leung, Matthews, and
  Winter]{cubit2011_nonlocal_assisted_cc}
Toby~S. Cubitt, Debbie Leung, William Matthews, and Andreas Winter.
\newblock Zero-error channel capacity and simulation assisted by non-local
  correlations.
\newblock \emph{IEEE Transactions on Information Theory}, 57\penalty0
  (8):\penalty0 5509--5523, 2011.
\newblock \doi{10.1109/TIT.2011.2159047}.

\bibitem[Holevo(1973)]{holevo1973bounds}
Alexander~Semenovich Holevo.
\newblock Bounds for the quantity of information transmitted by a quantum
  communication channel.
\newblock \emph{Problemy Peredachi Informatsii}, 9\penalty0 (3):\penalty0
  3--11, 1973.
\newblock URL \url{https://www.mathnet.ru/eng/ppi903}.

\bibitem[Frenkel and
  Weiner(2015)]{Frenkel2015_classical_information_n-level_quantum_system}
P{\'{e}}ter~E. Frenkel and Mih{\'{a}}ly Weiner.
\newblock Classical information storage in an n-level quantum system.
\newblock \emph{Communications in Mathematical Physics}, 340\penalty0
  (2):\penalty0 563--574, September 2015.
\newblock \doi{10.1007/s00220-015-2463-0}.

\bibitem[Bowles et~al.(2015)Bowles, Brunner, and
  Paw\l{}owski]{Bowles2015_nonclassicality_communication_networks}
Joseph Bowles, Nicolas Brunner, and Marcin Paw\l{}owski.
\newblock Testing dimension and nonclassicality in communication networks.
\newblock \emph{Phys. Rev. A}, 92:\penalty0 022351, Aug 2015.
\newblock \doi{10.1103/PhysRevA.92.022351}.

\bibitem[Bell(1964)]{bell1964epr}
John~S Bell.
\newblock On the {E}instein {P}odolsky {R}osen paradox.
\newblock \emph{Physics Physique Fizika}, 1\penalty0 (3):\penalty0 195, 1964.
\newblock \doi{10.1103/PhysicsPhysiqueFizika.1.195}.

\bibitem[Brunner et~al.(2014)Brunner, Cavalcanti, Pironio, Scarani, and
  Wehner]{brunner2014nonlocality}
Nicolas Brunner, Daniel Cavalcanti, Stefano Pironio, Valerio Scarani, and
  Stephanie Wehner.
\newblock Bell nonlocality.
\newblock \emph{Rev. Mod. Phys.}, 86:\penalty0 419--478, Apr 2014.
\newblock \doi{10.1103/RevModPhys.86.419}.

\bibitem[Massar et~al.(2001)Massar, Bacon, Cerf, and Cleve]{Massar-2001a}
Serge Massar, Dave Bacon, Nicolas~J. Cerf, and Richard Cleve.
\newblock Classical simulation of quantum entanglement without local hidden
  variables.
\newblock \emph{Phys. Rev. A}, 63:\penalty0 052305, Apr 2001.
\newblock \doi{10.1103/PhysRevA.63.052305}.

\bibitem[Bacon and Toner(2003)]{bacon2003_bell_inequalities_aux_communication}
D.~Bacon and B.~F. Toner.
\newblock Bell inequalities with auxiliary communication.
\newblock \emph{Phys. Rev. Lett.}, 90:\penalty0 157904, Apr 2003.
\newblock \doi{10.1103/PhysRevLett.90.157904}.

\bibitem[Toner and Bacon(2003)]{Toner2003_simulating_bell_correlations}
B.~F. Toner and D.~Bacon.
\newblock Communication cost of simulating bell correlations.
\newblock \emph{Phys. Rev. Lett.}, 91:\penalty0 187904, Oct 2003.
\newblock \doi{10.1103/PhysRevLett.91.187904}.

\bibitem[Regev and Toner(2010)]{Regev-2010a}
Oded Regev and Ben Toner.
\newblock Simulating quantum correlations with finite communication.
\newblock \emph{SIAM Journal on Computing}, 39\penalty0 (4):\penalty0
  1562--1580, 2010.
\newblock \doi{10.1137/080723909}.

\bibitem[Maxwell and Chitambar(2014)]{Maxwell2014_bell_inequality_aux_comm}
Katherine Maxwell and Eric Chitambar.
\newblock Bell inequalities with communication assistance.
\newblock \emph{Phys. Rev. A}, 89:\penalty0 042108, Apr 2014.
\newblock \doi{10.1103/PhysRevA.89.042108}.

\bibitem[Brask and Chaves(2017)]{Brask2017_bell_scenario_comm}
J~B Brask and R~Chaves.
\newblock Bell scenarios with communication.
\newblock \emph{Journal of Physics A: Mathematical and Theoretical},
  50\penalty0 (9):\penalty0 094001, January 2017.
\newblock ISSN 1751-8121.
\newblock \doi{10.1088/1751-8121/aa5840}.

\bibitem[Zambrini~Cruzeiro and Gisin(2019)]{ZambriniCruzeiro2019}
Emmanuel Zambrini~Cruzeiro and Nicolas Gisin.
\newblock Bell inequalities with one bit of communication.
\newblock \emph{Entropy}, 21\penalty0 (2):\penalty0 171, February 2019.
\newblock ISSN 1099-4300.
\newblock \doi{10.3390/e21020171}.

\bibitem[Alimuddin et~al.(2023)Alimuddin, Chakraborty, Sidhardh, Patra, Sen,
  Chowdhury, Naik, and Banik]{Alimuddin-2023a}
Mir Alimuddin, Ananya Chakraborty, Govind~Lal Sidhardh, Ram~Krishna Patra,
  Samrat Sen, Snehasish~Roy Chowdhury, Sahil~Gopalkrishna Naik, and Manik
  Banik.
\newblock Advantage of hardy's nonlocal correlation in reverse zero-error
  channel coding.
\newblock \emph{Phys. Rev. A}, 108:\penalty0 052430, Nov 2023.
\newblock \doi{10.1103/PhysRevA.108.052430}.

\bibitem[Frenkel and Weiner(2022)]{Frenkel2022ea_signaling_dim_violation}
P{\'{e}}ter~E. Frenkel and Mih{\'{a}}ly Weiner.
\newblock On entanglement assistance to a noiseless classical channel.
\newblock \emph{{Quantum}}, 6:\penalty0 662, March 2022.
\newblock ISSN 2521-327X.
\newblock \doi{10.22331/q-2022-03-01-662}.

\bibitem[Dall'Arno et~al.(2017)Dall'Arno, Brandsen, Tosini, Buscemi, and
  Vedral]{DallArno_no_hypersignaling_2017}
Michele Dall'Arno, Sarah Brandsen, Alessandro Tosini, Francesco Buscemi, and
  Vlatko Vedral.
\newblock No-hypersignaling principle.
\newblock \emph{Phys. Rev. Lett.}, 119:\penalty0 020401, Jul 2017.
\newblock \doi{10.1103/PhysRevLett.119.020401}.

\bibitem[Heinosaari and Kerppo(2019)]{Heinosaari-2019a}
Teiko Heinosaari and Oskari Kerppo.
\newblock Communication of partial ignorance with qubits.
\newblock \emph{Journal of Physics A: Mathematical and Theoretical},
  52\penalty0 (39):\penalty0 395301, September 2019.
\newblock ISSN 1751-8121.
\newblock \doi{10.1088/1751-8121/ab3ae4}.

\bibitem[Heinosaari et~al.(2020)Heinosaari, Kerppo, and
  Lepp\"{a}j\"{a}rvi]{heinosaari2020communication}
Teiko Heinosaari, Oskari Kerppo, and Leevi Lepp\"{a}j\"{a}rvi.
\newblock Communication tasks in operational theories.
\newblock \emph{Journal of Physics A: Mathematical and Theoretical},
  53\penalty0 (43):\penalty0 435302, October 2020.
\newblock ISSN 1751-8121.
\newblock \doi{10.1088/1751-8121/abb5dc}.

\bibitem[Poderini et~al.(2020)Poderini, Brito, Nery, Sciarrino, and
  Chaves]{Poderini2020_prep_meas_nonclassicality}
Davide Poderini, Samura\'{\i} Brito, Ranieri Nery, Fabio Sciarrino, and Rafael
  Chaves.
\newblock Criteria for nonclassicality in the prepare-and-measure scenario.
\newblock \emph{Phys. Rev. Res.}, 2:\penalty0 043106, Oct 2020.
\newblock \doi{10.1103/PhysRevResearch.2.043106}.

\bibitem[Doolittle and Chitambar(2021)]{doolittle_2021_certify_classical_cost}
Brian Doolittle and Eric Chitambar.
\newblock Certifying the classical simulation cost of a quantum channel.
\newblock \emph{Phys. Rev. Res.}, 3:\penalty0 043073, Oct 2021.
\newblock \doi{10.1103/PhysRevResearch.3.043073}.

\bibitem[Tavakoli et~al.(2021)Tavakoli, Pauwels, Woodhead, and
  Pironio]{Tavakoli-2021a}
Armin Tavakoli, Jef Pauwels, Erik Woodhead, and Stefano Pironio.
\newblock Correlations in entanglement-assisted prepare-and-measure scenarios.
\newblock \emph{PRX Quantum}, 2:\penalty0 040357, Dec 2021.
\newblock \doi{10.1103/PRXQuantum.2.040357}.

\bibitem[Renner et~al.(2023)Renner, Tavakoli, and Quintino]{Renner-2023a}
Martin~J. Renner, Armin Tavakoli, and Marco~T\'ulio Quintino.
\newblock Classical cost of transmitting a qubit.
\newblock \emph{Phys. Rev. Lett.}, 130:\penalty0 120801, Mar 2023.
\newblock \doi{10.1103/PhysRevLett.130.120801}.

\bibitem[Grudka et~al.(2015)Grudka, Horodecki, Horodecki, and
  W\'ojcik]{Grudka-2015a}
Andrzej Grudka, Micha\l{} Horodecki, Ryszard Horodecki, and Antoni W\'ojcik.
\newblock Nonsignaling quantum random access-code boxes.
\newblock \emph{Phys. Rev. A}, 92:\penalty0 052312, Nov 2015.
\newblock \doi{10.1103/PhysRevA.92.052312}.

\bibitem[Tavakoli et~al.(2015)Tavakoli, Hameedi, Marques, and
  Bourennane]{Tavakoli-2015a}
Armin Tavakoli, Alley Hameedi, Breno Marques, and Mohamed Bourennane.
\newblock Quantum random access codes using single $d$-level systems.
\newblock \emph{Phys. Rev. Lett.}, 114:\penalty0 170502, Apr 2015.
\newblock \doi{10.1103/PhysRevLett.114.170502}.

\bibitem[Heinosaari and Lepp\"{a}j\"{a}rvi(2022)]{Heinosaari2022}
Teiko Heinosaari and Leevi Lepp\"{a}j\"{a}rvi.
\newblock Random access test as an identifier of nonclassicality.
\newblock \emph{Journal of Physics A: Mathematical and Theoretical},
  55\penalty0 (17):\penalty0 174003, April 2022.
\newblock ISSN 1751-8121.
\newblock \doi{10.1088/1751-8121/ac5b91}.

\bibitem[Piveteau et~al.(2022)Piveteau, Pauwels, Håkansson, Muhammad,
  Bourennane, and Tavakoli]{Piveteau-2022a}
Amélie Piveteau, Jef Pauwels, Emil Håkansson, Sadiq Muhammad, Mohamed
  Bourennane, and Armin Tavakoli.
\newblock Entanglement-assisted quantum communication with simple measurements.
\newblock \emph{Nature Communications}, 13\penalty0 (1), December 2022.
\newblock ISSN 2041-1723.
\newblock \doi{10.1038/s41467-022-33922-5}.

\bibitem[Sakharwade et~al.(2023)Sakharwade, Studziński, Eckstein, and
  Horodecki]{Sakharwade-2023a}
Nitica Sakharwade, Michał Studziński, Michał Eckstein, and Paweł Horodecki.
\newblock Two instances of random access code in the quantum regime.
\newblock \emph{New Journal of Physics}, 25\penalty0 (5):\penalty0 053038, May
  2023.
\newblock ISSN 1367-2630.
\newblock \doi{10.1088/1367-2630/acd716}.

\bibitem[Lauand et~al.(2023)Lauand, Poderini, Nery, Moreno, Pollyceno, Rabelo,
  and Chaves]{Lauand2023_causal_structure_3_variables}
Pedro Lauand, Davide Poderini, Ranieri Nery, George Moreno, Lucas Pollyceno,
  Rafael Rabelo, and Rafael Chaves.
\newblock Witnessing nonclassicality in a causal structure with three
  observable variables.
\newblock \emph{PRX Quantum}, 4:\penalty0 020311, Apr 2023.
\newblock \doi{10.1103/PRXQuantum.4.020311}.

\bibitem[Cerezo et~al.(2021)Cerezo, Arrasmith, Babbush, Benjamin, Endo, Fujii,
  McClean, Mitarai, Yuan, Cincio, and Coles]{Cerezo2021_vqa}
M.~Cerezo, Andrew Arrasmith, Ryan Babbush, Simon~C. Benjamin, Suguru Endo,
  Keisuke Fujii, Jarrod~R. McClean, Kosuke Mitarai, Xiao Yuan, Lukasz Cincio,
  and Patrick~J. Coles.
\newblock Variational quantum algorithms.
\newblock \emph{Nature Reviews Physics}, 3\penalty0 (9):\penalty0 625–644,
  August 2021.
\newblock ISSN 2522-5820.
\newblock \doi{10.1038/s42254-021-00348-9}.

\bibitem[Doolittle et~al.(2023)Doolittle, Bromley, Killoran, and
  Chitambar]{doolittle_2023_vqo_nonlocality}
Brian Doolittle, Thomas~R. Bromley, Nathan Killoran, and Eric Chitambar.
\newblock Variational quantum optimization of nonlocality in noisy quantum
  networks.
\newblock \emph{IEEE Transactions on Quantum Engineering}, pages 1--28, 2023.
\newblock \doi{10.1109/TQE.2023.3243849}.

\bibitem[Doolittle and Bromley(2022)]{doolittle_qNetVO}
Brian Doolittle and Tom Bromley.
\newblock qnetvo: the quantum network variational optimizer.
\newblock March 2022.
\newblock \doi{10.5281/zenodo.6345834}.
\newblock URL \url{https://github.com/ChitambarLab/qNetVO}.

\bibitem[Doolittle(2023)]{doolittle_phd_thesis_2023}
Brian Doolittle.
\newblock \emph{Nonclassicality in Noisy Quantum Networks}.
\newblock Phd thesis, University of Illinois Urbana-Champaign, September 2023.
\newblock URL \url{https://hdl.handle.net/2142/121945}.

\bibitem[Chen et~al.(2023)Chen, Doolittle, Larson, Saleem, and
  Chitambar]{chen_2022_inferring_network_topology}
Daniel~T. Chen, Brian Doolittle, Jeffrey Larson, Zain~H. Saleem, and Eric
  Chitambar.
\newblock Inferring quantum network topology using local measurements.
\newblock \emph{PRX Quantum}, 4:\penalty0 040347, Dec 2023.
\newblock \doi{10.1103/PRXQuantum.4.040347}.

\bibitem[Ma et~al.(2021)Ma, Gokhale, Zheng, Zhou, Yu, Jiang, Maurer, and
  Chong]{Ma2021_noisy_parameter-shift}
Ziqi Ma, Pranav Gokhale, Tian-Xing Zheng, Sisi Zhou, Xiaofei Yu, Liang Jiang,
  Peter Maurer, and Frederic~T. Chong.
\newblock Adaptive circuit learning for quantum metrology.
\newblock In \emph{2021 IEEE International Conference on Quantum Computing and
  Engineering (QCE)}, pages 419--430, 2021.
\newblock \doi{10.1109/QCE52317.2021.00063}.

\bibitem[Heinosaari et~al.(2024)Heinosaari, Kerppo, Lepp\"aj\"arvi, and
  Pl\'avala]{heinosaari_unbounded_advantage2024}
Teiko Heinosaari, Oskari Kerppo, Leevi Lepp\"aj\"arvi, and Martin Pl\'avala.
\newblock Simple information-processing tasks with unbounded quantum advantage.
\newblock \emph{Phys. Rev. A}, 109:\penalty0 032627, Mar 2024.
\newblock \doi{10.1103/PhysRevA.109.032627}.

\bibitem[Tavakoli et~al.(2022)Tavakoli, Pozas-Kerstjens, Luo, and
  Renou]{Tavakoli-2022a}
Armin Tavakoli, Alejandro Pozas-Kerstjens, Ming-Xing Luo, and Marc-Olivier
  Renou.
\newblock Bell nonlocality in networks.
\newblock \emph{Reports on Progress in Physics}, 85\penalty0 (5):\penalty0
  056001, Mar 2022.
\newblock ISSN 1361-6633.
\newblock \doi{10.1088/1361-6633/ac41bb}.

\bibitem[Brito et~al.(2018)Brito, Amaral, and
  Chaves]{brito2018_nonlocality_trace_distance}
S.~G.~A. Brito, B.~Amaral, and R.~Chaves.
\newblock Quantifying bell nonlocality with the trace distance.
\newblock \emph{Phys. Rev. A}, 97:\penalty0 022111, Feb 2018.
\newblock \doi{10.1103/PhysRevA.97.022111}.

\bibitem[Ziegler(2012)]{Ziegler-2012a_lectures_on_polytopes}
G.M. Ziegler.
\newblock \emph{Lectures on Polytopes}.
\newblock Graduate Texts in Mathematics. Springer New York, 2012.
\newblock \doi{10.1007/978-1-4613-8431-1}.

\bibitem[Rosset et~al.(2014)Rosset, Bancal, and
  Gisin]{Rosset2014_classifying_bell_inequalities}
Denis Rosset, Jean-Daniel Bancal, and Nicolas Gisin.
\newblock Classifying 50 years of bell inequalities.
\newblock \emph{Journal of Physics A: Mathematical and Theoretical},
  47\penalty0 (42):\penalty0 424022, October 2014.
\newblock \doi{10.1088/1751-8113/47/42/424022}.

\bibitem[Christof and L{\"o}bel(1997)]{PORTA}
Thomas Christof and Andreas L{\"o}bel.
\newblock Porta, 1997.
\newblock URL \url{http://porta.zib.de/}.

\bibitem[Doolittle(2024)]{supplemental_software}
Brian Doolittle.
\newblock
  \url{https://github.com/ChitambarLab/nonclassicality-in-quantum-communication-networks-supplemental-code}
  (v0.2.0).
\newblock Feb 2024.
\newblock \doi{10.5281/zenodo.10780789}.

\bibitem[Pal and Ghosh(2015)]{Pal_2015_nonlocality_breaking}
Rajarshi Pal and Sibasish Ghosh.
\newblock Non-locality breaking qubit channels: the case for chsh inequality.
\newblock \emph{Journal of Physics A: Mathematical and Theoretical},
  48\penalty0 (15):\penalty0 155302, mar 2015.
\newblock \doi{10.1088/1751-8113/48/15/155302}.

\bibitem[Zhang et~al.(2020)Zhang, Bravo, Lorenz, and
  Chitambar]{Zhang_2020_nonlocality_breaking}
Yujie Zhang, Rodrigo~Araiza Bravo, Virginia~O Lorenz, and Eric Chitambar.
\newblock Channel activation of chsh nonlocality.
\newblock \emph{New Journal of Physics}, 22\penalty0 (4):\penalty0 043003, apr
  2020.
\newblock \doi{10.1088/1367-2630/ab7bef}.

\bibitem[Bergholm et~al.(2018)Bergholm, Izaac, Schuld, Gogolin, Ahmed, Ajith,
  Alam, Alonso-Linaje, AkashNarayanan, Asadi, et~al.]{bergholm2018pennylane}
Ville Bergholm, Josh Izaac, Maria Schuld, Christian Gogolin, Shahnawaz Ahmed,
  Vishnu Ajith, M~Sohaib Alam, Guillermo Alonso-Linaje, B~AkashNarayanan, Ali
  Asadi, et~al.
\newblock Pennylane: Automatic differentiation of hybrid quantum-classical
  computations.
\newblock \emph{arXiv preprint}, 2018.
\newblock \doi{10.48550/arXiv.1811.04968}.

\bibitem[Nielsen and Chuang(2009)]{nielsen_chuang2009}
Michael~A. Nielsen and Isaac~L. Chuang.
\newblock \emph{Quantum Computation and Quantum Information}.
\newblock Cambridge University Press, 2009.
\newblock \doi{10.1017/cbo9780511976667}.

\bibitem[Rumelhart et~al.(1986)Rumelhart, Hinton, and
  Williams]{rumelhart1986backprop}
David~E Rumelhart, Geoffrey~E Hinton, and Ronald~J Williams.
\newblock Learning representations by back-propagating errors.
\newblock \emph{Nature}, 323\penalty0 (6088):\penalty0 533--536, 1986.
\newblock \doi{https://doi.org/10.1038/323533a0}.

\bibitem[Schuld et~al.(2019)Schuld, Bergholm, Gogolin, Izaac, and
  Killoran]{schuld2019_parameter-shift}
Maria Schuld, Ville Bergholm, Christian Gogolin, Josh Izaac, and Nathan
  Killoran.
\newblock Evaluating analytic gradients on quantum hardware.
\newblock \emph{Phys. Rev. A}, 99:\penalty0 032331, Mar 2019.
\newblock \doi{10.1103/PhysRevA.99.032331}.

\bibitem[Mari et~al.(2021)Mari, Bromley, and
  Killoran]{Mari2021_higher-order_quantum_gradients}
Andrea Mari, Thomas~R. Bromley, and Nathan Killoran.
\newblock Estimating the gradient and higher-order derivatives on quantum
  hardware.
\newblock \emph{Phys. Rev. A}, 103:\penalty0 012405, Jan 2021.
\newblock \doi{10.1103/PhysRevA.103.012405}.

\bibitem[Wierichs et~al.(2022)Wierichs, Izaac, Wang, and
  Lin]{Wierichs2022_generall_parameter-shift_rule}
David Wierichs, Josh Izaac, Cody Wang, and Cedric Yen-Yu Lin.
\newblock General parameter-shift rules for quantum gradients.
\newblock \emph{Quantum}, 6:\penalty0 677, March 2022.
\newblock \doi{10.22331/q-2022-03-30-677}.

\bibitem[Kyriienko and Elfving(2021)]{Kyriienko2021_generalized_circuit_diff}
Oleksandr Kyriienko and Vincent~E. Elfving.
\newblock Generalized quantum circuit differentiation rules.
\newblock \emph{Phys. Rev. A}, 104:\penalty0 052417, Nov 2021.
\newblock \doi{10.1103/PhysRevA.104.052417}.

\bibitem[Ruder(2016)]{ruder2016overview}
Sebastian Ruder.
\newblock An overview of gradient descent optimization algorithms.
\newblock \emph{arXiv preprint}, 2016.
\newblock \doi{10.48550/arXiv.1609.04747}.

\bibitem[Kingma and Ba(2014)]{kingma2014adam}
Diederik~P Kingma and Jimmy Ba.
\newblock Adam: A method for stochastic optimization.
\newblock \emph{arXiv preprint}, 2014.
\newblock \doi{10.48550/arXiv.1412.6980}.

\bibitem[Bittel and Kliesch(2021)]{bittel2021_training_vqa_is_np-hard}
Lennart Bittel and Martin Kliesch.
\newblock Training variational quantum algorithms is np-hard.
\newblock \emph{Phys. Rev. Lett.}, 127:\penalty0 120502, Sep 2021.
\newblock \doi{10.1103/PhysRevLett.127.120502}.

\bibitem[Bennett and Wiesner(1992)]{bennet1992_dense_coding}
Charles~H. Bennett and Stephen~J. Wiesner.
\newblock Communication via one- and two-particle operators on
  einstein-podolsky-rosen states.
\newblock \emph{Phys. Rev. Lett.}, 69:\penalty0 2881--2884, Nov 1992.
\newblock \doi{10.1103/PhysRevLett.69.2881}.

\bibitem[Chitambar et~al.(2023)Chitambar, George, Doolittle, and
  Junge]{chitambar_2023_cv_channel}
Eric Chitambar, Ian George, Brian Doolittle, and Marius Junge.
\newblock The communication value of a quantum channel.
\newblock \emph{IEEE Transactions on Information Theory}, 69\penalty0
  (3):\penalty0 1660--1679, 2023.
\newblock \doi{10.1109/TIT.2022.3218540}.

\bibitem[Ambainis et~al.(2008)Ambainis, Leung, Mancinska, and
  Ozols]{ambainis2008quantum}
Andris Ambainis, Debbie Leung, Laura Mancinska, and Maris Ozols.
\newblock Quantum random access codes with shared randomness.
\newblock \emph{arXiv preprint}, 2008.
\newblock \doi{10.48550/arXiv.0810.2937}.

\bibitem[Gallego et~al.(2010)Gallego, Brunner, Hadley, and
  Ac{\'i}n]{gallego2010_dim_witnessing}
Rodrigo Gallego, Nicolas Brunner, Christopher Hadley, and Antonio Ac{\'i}n.
\newblock Device-independent tests of classical and quantum dimensions.
\newblock \emph{Phys. Rev. Lett.}, 105:\penalty0 230501, Nov 2010.
\newblock \doi{10.1103/PhysRevLett.105.230501}.

\bibitem[Paw\l{}owski and Winter(2012)]{pawlowski2012_eacc_qc_simulation}
Marcin Paw\l{}owski and Andreas Winter.
\newblock ``hyperbits'': The information quasiparticles.
\newblock \emph{Phys. Rev. A}, 85:\penalty0 022331, Feb 2012.
\newblock \doi{10.1103/PhysRevA.85.022331}.

\bibitem[Pauwels et~al.(2022)Pauwels, Pironio, Cruzeiro, and
  Tavakoli]{Tavakoli2022_ea_communications}
Jef Pauwels, Stefano Pironio, Emmanuel~Zambrini Cruzeiro, and Armin Tavakoli.
\newblock Adaptive advantage in entanglement-assisted communications.
\newblock \emph{Phys. Rev. Lett.}, 129:\penalty0 120504, Sep 2022.
\newblock \doi{10.1103/PhysRevLett.129.120504}.

\bibitem[Paw\l{}owski and Brunner(2011)]{pawlowski2011_sdi_qkd}
Marcin Paw\l{}owski and Nicolas Brunner.
\newblock Semi-device-independent security of one-way quantum key distribution.
\newblock \emph{Phys. Rev. A}, 84:\penalty0 010302, Jul 2011.
\newblock \doi{10.1103/PhysRevA.84.010302}.

\bibitem[Li et~al.(2011)Li, Yin, Wu, Zou, Wang, Chen, Guo, and
  Han]{li2011_sdi_qrng}
Hong-Wei Li, Zhen-Qiang Yin, Yu-Chun Wu, Xu-Bo Zou, Shuang Wang, Wei Chen,
  Guang-Can Guo, and Zheng-Fu Han.
\newblock Semi-device-independent random-number expansion without entanglement.
\newblock \emph{Phys. Rev. A}, 84:\penalty0 034301, Sep 2011.
\newblock \doi{10.1103/PhysRevA.84.034301}.

\bibitem[Moreno et~al.(2021)Moreno, Nery, de~Gois, Rabelo, and
  Chaves]{moreno2021_sdi_certification_entanglement_sd_coding}
George Moreno, Ranieri Nery, Carlos de~Gois, Rafael Rabelo, and Rafael Chaves.
\newblock Semi-device-independent certification of entanglement in superdense
  coding.
\newblock \emph{Phys. Rev. A}, 103:\penalty0 022426, Feb 2021.
\newblock \doi{10.1103/PhysRevA.103.022426}.

\bibitem[Tavakoli et~al.(2018)Tavakoli, Kaniewski, V\'ertesi, Rosset, and
  Brunner]{tavaokoli2018_pm_self-test}
Armin Tavakoli, J\ifmmode \mbox{\k{e}}\else~\k{e}\fi{}drzej Kaniewski, Tam\'as
  V\'ertesi, Denis Rosset, and Nicolas Brunner.
\newblock Self-testing quantum states and measurements in the
  prepare-and-measure scenario.
\newblock \emph{Phys. Rev. A}, 98:\penalty0 062307, Dec 2018.
\newblock \doi{10.1103/PhysRevA.98.062307}.

\bibitem[Popescu and Rohrlich(1994)]{Popescu1994}
Sandu Popescu and Daniel Rohrlich.
\newblock Quantum nonlocality as an axiom.
\newblock \emph{Foundations of Physics}, 24\penalty0 (3):\penalty0 379–385,
  March 1994.
\newblock ISSN 1572-9516.
\newblock \doi{10.1007/bf02058098}.

\bibitem[Clauser et~al.(1969)Clauser, Horne, Shimony, and Holt]{chsh1969}
John~F. Clauser, Michael~A. Horne, Abner Shimony, and Richard~A. Holt.
\newblock Proposed experiment to test local hidden-variable theories.
\newblock \emph{Phys. Rev. Lett.}, 23:\penalty0 880--884, Oct 1969.
\newblock \doi{10.1103/PhysRevLett.23.880}.

\bibitem[Doolittle(2020)]{BellScenario.jl}
Brian Doolittle.
\newblock \url{https://github.com/ChitambarLab/BellScenario.jl} (v0.1.3).
\newblock 2020.
\newblock \doi{10.5281/zenodo.10277572}.

\bibitem[Doolittle and Legat(2020)]{doolittle_xporta}
Brian Doolittle and Beno{\^i}t Legat.
\newblock \url{https://github.com/JuliaPolyhedra/XPORTA.jl} (v0.1.3), 2020.

\bibitem[Legat et~al.(2021)Legat, Deits, Goretkin, Koolen, Huchette, Oyama, and
  Forets]{legat2021polyhedra}
Beno{\^i}t Legat, Robin Deits, Gustavo Goretkin, Twan Koolen, Joey Huchette,
  Daisuke Oyama, and Marcelo Forets.
\newblock \url{https://github.com/JuliaPolyhedra/Polyhedra.jl} (v0.6.16).
\newblock 2021.
\newblock \doi{10.5281/zenodo.1214290}.

\bibitem[Huangfu and Hall(2018)]{huangfu2018highs_parallelizing}
Qi~Huangfu and JA~Julian Hall.
\newblock Parallelizing the dual revised simplex method.
\newblock \emph{Mathematical Programming Computation}, 10\penalty0
  (1):\penalty0 119--142, 2018.
\newblock \doi{10.1007/s12532-017-0130-5}.

\bibitem[Dunning et~al.(2017)Dunning, Huchette, and Lubin]{JuMP.jl-2017}
Iain Dunning, Joey Huchette, and Miles Lubin.
\newblock Jump: A modeling language for mathematical optimization.
\newblock \emph{SIAM Review}, 59\penalty0 (2):\penalty0 295--320, 2017.
\newblock \doi{10.1137/15M1020575}.

\end{thebibliography}

\appendix
\onecolumn

\newpage

\section{Linear Nonclassicality Witnesses for Bipartite Scenarios}

\begin{table*}[h!]
    \centering
    \resizebox{\textwidth}{!}{
    \begin{tabular}{c c c c}
         $2 \geq  \Fbf^a_{3\to 4} = \begin{bmatrix}
            1 & 0 & 0 \\
            1 & 0 & 0 \\
            0 & 1 & 0 \\
            0 & 0 & 1 \\
        \end{bmatrix}$ & $3 \geq \Fbf^c_{4 \to 4}= \begin{bmatrix}
            1 & 1 & 0 & 0 \\
            1 & 0 & 1 & 0 \\
            0 & 1 & 1 & 0 \\
            0 & 0 & 0 & 1 \\
        \end{bmatrix}$ &  $4 \geq \Fbf^e_{4\to 4}=\begin{bmatrix}
            2 & 0 & 0 & 0 \\
            0 & 2 & 0 & 0 \\
            0 & 0 & 1 & 1 \\
            1 & 1 & 1 & 0 \\
        \end{bmatrix}$  & $4 \geq \Fbf^g_{6 \to 4} = \begin{bmatrix}
            1 & 0 & 0 & 1 & 0 & 0 \\
            0 & 1 & 0 & 0 & 1 & 0 \\
            0 & 0 & 1 & 0 & 0 & 1 \\
            1 & 1 & 1 & 0 & 0 & 0 \\
        \end{bmatrix}$ \\
        \hfill \\
        $2 \geq \Fbf^b_{4\to 4} =\begin{bmatrix}
            1 & 0 & 0 & 0 \\
            0 & 1 & 0 & 0 \\
            0 & 0 & 1 & 0 \\
            0 & 0 & 0 & 1 \\
        \end{bmatrix}$
        & $4 \geq \Fbf^d_{3 \to 4} = \begin{bmatrix}
            2 & 0 & 0 \\
            0 & 2 & 0 \\
            0 & 0 & 2 \\
            1 & 1 & 1 \\
        \end{bmatrix}$ & $4 \geq \Fbf^f_{5\to 4}= \begin{bmatrix}
            2 & 0 & 0 & 0 & 0 \\
            0 & 1 & 0 & 1 & 0 \\
            0 & 0 & 1 & 0 & 1 \\
            1 & 1 & 1 & 0 & 0 \\
        \end{bmatrix}$ & $5 \geq \Fbf^h_{6 \to 4}= \begin{bmatrix}
            1 & 1 & 1 & 0 & 0 & 0 \\
            1 & 0 & 0 & 1 & 1 & 0 \\
            0 & 1 & 0 & 1 & 0 & 1 \\
            0 & 0 & 1 & 0 & 1 & 1 \\
        \end{bmatrix}$
    \end{tabular}
    }
    \caption{  Facet inequalities bounding the $d=2$ point-to-point network polytope \cite{doolittle_2021_certify_classical_cost}.
    }
    \label{Table:d2_signaling_dimension_witness}
\end{table*}

\begin{table*}[h!]
    \centering
    \resizebox{0.9\textwidth}{!}{
    \begin{tabular}{c}
    \begin{tabular}{l r}
    $4 \geq \Fbf^a_{\PM}= \begin{bmatrix}
        0 & 0 & 0 & 1 & 1 & 0 \\
        1 & 1 & 1 & 0 & 0 & 0 \\
    \end{bmatrix}$ & $5 \geq \Fbf^c_{\PM}= \begin{bmatrix}
        0 & 0 & 1 & 0 & 1 & 0 & 1 & 0 & 0 \\
        1 & 0 & 0 & 0 & 0 & 1 & 0 & 1 & 0 \\
    \end{bmatrix}$  \\
    \hfill \\
    $5 \geq \Fbf^b_{\PM} = \begin{bmatrix}
        0 & 0 & 1 & 0 & 1 & 0 & 0 & 0 & 1 \\
        1 & 0 & 0 & 1 & 0 & 0 & 0 & 1 & 0 \\
    \end{bmatrix}$ & $7 \geq \Fbf_{\PM}^d = \begin{bmatrix}
        0 & 0 & 1 & 0 & 1 & 0 & 0 & 1 & 1 \\
        1 & 1 & 0 & 1 & 0 & 1 & 1 & 0 & 0 \\
    \end{bmatrix}$\\
    \hfill\\
    \end{tabular}\\
    \begin{tabular}{c}
        $8 \geq \Fbf^e_{\PM} = \begin{bmatrix}
            0 & 0 & 1 & 1 & 0 & 0 & 0 & 0 & 0 & 0 & 0 & 0 & 0 & 0 & 0 & 0 & 1 & 0 & 0 & 0 & 0 & 0 & 0 & 0 \\
            0 & 0 & 0 & 0 & 0 & 1 & 0 & 0 & 0 & 0 & 1 & 1 & 0 & 0 & 0 & 1 & 0 & 1 & 1 & 1 & 0 & 0 & 0 & 0 \\
        \end{bmatrix}$
    \end{tabular} \\
    \hfill \\
    \hline
    \begin{tabular}{c c} \\
        $7 \geq \Vbf_{\PM}^=\begin{bmatrix}
            1 & 0 & 0 & 0 & 1 & 0 & 0 & 0 & 1 \\
            0 & 1 & 1 & 1 & 0 & 1 & 1 & 1 & 0 \\
        \end{bmatrix}$ & $6 \geq \Vbf^{\text{RAC}_2}_{\PM} = \begin{bmatrix}
            1 & 1 & 1 & 0 & 0 & 1 & 0 & 0 \\
            0 & 0 & 0 & 1 & 1 & 0 & 1 & 1 \\
        \end{bmatrix}$
    \end{tabular} \\
    \hfill \\
    \begin{tabular}{c}
        $18 \geq \Vbf_{\PM}^{\text{RAC}_3} = \begin{bmatrix}
            1 & 1 & 1 & 1 & 1 & 0 & 1 & 0 & 1 & 1 & 0 & 0 & 0 & 1 & 1 & 0 & 1 & 0 & 0 & 0 & 1 & 0 & 0 & 0 \\
            0 & 0 & 0 & 0 & 0 & 1 & 0 & 1 & 0 & 0 & 1 & 1 & 1 & 0 & 0 & 1 & 0 & 1 & 1 & 1 & 0 & 1 & 1 & 1 \\
        \end{bmatrix}$ \\
    \end{tabular}
    \end{tabular}
    }
    \caption{(Above Line) Nonclassicality witnesses for prepare-and-measure networks. $\Fbf^a_{\PM}$, qubit dimensionality witness for $\PM(\sigarb{3,2}{2}{2})$ network \cite{gallego2010_dim_witnessing}. $\Fbf^b_{\PM}$, $\Fbf^c_{\PM}$, and $\Fbf^d_{\PM}$, nonclassicality witnesses for $\PM(\sigarb{3,3}{2}{2})$ network. $\Fbf^e_{\PM}$, qubit dimensionality witness for $\PM(\sigarb{8,3}{2}{2})$. (Below Line) $\Vbf^=_{\PM}$, simulation game for evaluating whether two input trits are equal (see Table~\ref{table:multiaccess_simulation_games} for definition). $\Vbf^{\text{RAC}_2}$ and $\Vbf^{\text{RAC}_3}$, simulation games that respectively correspond to the two bit and three bit random access coding task $\text{RAC}_n$.}
    \label{Table:prepare_and_measure-nonclassicality-witnesses}
\end{table*}

\begin{table*}[h!]
    \centering
    \resizebox{0.7\textwidth}{!}{
    \begin{tabular}{c c}
       $2 \geq \Fbf^a_{\BT 2}= \begin{bmatrix}
           1 & 0 & 0 & 0 \\
           0 & 0 & 1 & 0 \\
           0 & 1 & 0 & 0 \\
           0 & 0 & 0 & 1 \\
       \end{bmatrix}$  & $7 \geq \Fbf^{c}_{\BT 3}= \begin{bmatrix}
            0 & 0 & 1 & 0 & 1 & 1 & 1 & 1 & 1 \\
            0 & 1 & 0 & 1 & 0 & 0 & 0 & 0 & 0 \\
            0 & 1 & 0 & 1 & 0 & 0 & 0 & 0 & 0 \\
            0 & 0 & 1 & 0 & 1 & 1 & 1 & 1 & 1 \\
       \end{bmatrix}$ \\
       \hfill \\
        $2 \geq \Fbf^{b}_{\BT 2} = \begin{bmatrix}
             1 & 0 & 0 & 0 \\
             0 & 0 & 1 & 0 \\
             1 & 0 & 0 & 1 \\
             0 & 0 & 0 & 1 \\
        \end{bmatrix}$ & $13 \geq \Fbf^{d}_{\BT 3} = \begin{bmatrix}
            1 & 2 & 0 & 2 & 1 & 2 & 0 & 2 & 1 \\
            0 & 0 & 2 & 0 & 0 & 0 & 2 & 0 & 0 \\
            0 & 0 & 2 & 0 & 0 & 0 & 2 & 0 & 0 \\
            1 & 2 & 0 & 2 & 1 & 2 & 0 & 2 & 1 \\
        \end{bmatrix}$
    \end{tabular}
    }
    \caption{Facet inequalities for the Bell scenario with signaling, also known as the Bacon-Toner ($\BT$) scenario \cite{bacon2003_bell_inequalities_aux_communication}. $\Fbf^a_{\BT 2}$ and $\Fbf^b_{\BT 2}$, facet inequalities for $\BT(\sigarb{2,2}{2}{2,2}) = \BT 2$ classical network. $\Fbf^c_{\BT 3}$ and $\Fbf^d_{\BT 3}$, facet inequalities for the $\BT(\sigarb{3,3}{2}{2,2}) = \BT 3$ classical network.}
    \label{Table:Bell_inequalities_aux_communication}
\end{table*}

\newpage

\section{Facet Inequalities for Multiaccess Networks}

\begin{table*}[ht]
    \centering
    \begin{tabular}{c}
    \begin{tabular}{l  r}
        $4 \geq \Fbf^1_{33\to 2} =\begin{bmatrix}
            1 & 1 & 0 & 1 & 0 & 0 & 0 & 0 & 0 \\
            0 & 0 & 0 & 0 & 1 & 0 & 1 & 0 & 0 \\
        \end{bmatrix}$  & $8 \geq \Fbf^8_{33\to 2} =  \begin{bmatrix}
            0 & 0 & 1 & 0 & 1 & 0 & 2 & 0 & 0 \\
            2 & 1 & 0 & 0 & 0 & 1 & 0 & 2 & 0 \\
        \end{bmatrix}$   \\
        \hfill \\
        $5 \geq \Fbf^2_{33\to 2} = \begin{bmatrix}
            0 & 0 & 0 & 0 & 0 & 1 & 1 & 0 & 0 \\
            1 & 1 & 0 & 1 & 0 & 0 & 0 & 1 & 0 \\
        \end{bmatrix}$ & $10 \geq \Fbf^9_{33\to 2} =  \begin{bmatrix}
            0 & 0 & 1 & 0 & 2 & 0 & 1 & 0 & 1 \\
            3 & 2 & 0 & 2 & 0 & 1 & 0 & 0 & 0 \\
        \end{bmatrix}$  \\
        \hfill \\
        $7 \geq \Fbf^3_{33\to 2} =  \begin{bmatrix}
            0 & 0 & 1 & 0 & 1 & 0 & 1 & 0 & 0 \\
            2 & 1 & 0 & 1 & 0 & 1 & 0 & 1 & 0 \\
        \end{bmatrix}$ & $11 \geq \Fbf^{10}_{33\to 2} =  \begin{bmatrix}
            0 & 0 & 2 & 0 & 1 & 0 & 2 & 0 & 0 \\
            3 & 1 & 0 & 1 & 0 & 2 & 0 & 2 & 0 \\
        \end{bmatrix}$  \\
        \hfill \\
        $7 \geq \Fbf^4_{33\to 2} = \begin{bmatrix}
            0 & 0 & 1 & 0 & 1 & 0 & 1 & 0 & 0 \\
            2 & 2 & 0 & 1 & 0 & 0 & 0 & 1 & 0 \\
        \end{bmatrix}$ & $14 \geq \Fbf^{11}_{33\to 2} =  \begin{bmatrix}
            0 & 0 & 2 & 0 & 2 & 0 & 2 & 0 & 0 \\
            3 & 1 & 0 & 1 & 0 & 3 & 0 & 3 & 1 \\
        \end{bmatrix}$   \\
        \hfill \\
        $6 \geq \Fbf^5_{33\to 2} =  \begin{bmatrix}
            0 & 0 & 0 & 0 & 1 & 1 & 1 & 0 & 1 \\
            1 & 1 & 0 & 1 & 0 & 0 & 0 & 1 & 0 \\
        \end{bmatrix}$ & $16 \geq \Fbf^{12}_{33\to 2} =  \begin{bmatrix}
            0 & 0 & 2 & 0 & 3 & 0 & 2 & 0 & 1 \\
            5 & 3 & 0 & 3 & 0 & 1 & 0 & 1 & 0 \\
        \end{bmatrix}$    \\   
        \hfill \\
        $8 \geq \Fbf^6_{33\to 2} = \begin{bmatrix}
            0 & 0 & 0 & 0 & 1 & 1 & 2 & 0 & 0 \\
            2 & 2 & 0 & 1 & 0 & 0 & 0 & 1 & 0 \\
        \end{bmatrix}$ & $17 \geq \Fbf^{13}_{33\to 2} = \begin{bmatrix}
            0 & 0 & 2 & 1 & 2 & 0 & 5 & 0 & 1 \\
            4 & 2 & 0 & 0 & 0 & 1 & 0 & 4 & 0 \\
        \end{bmatrix}$   \\
        \hfill \\
        $8 \geq \Fbf^7_{33\to 2} = \begin{bmatrix}
            0 & 0 & 0 & 0 & 2 & 0 & 1 & 0 & 1 \\
            2 & 2 & 0 & 1 & 0 & 1 & 0 & 0 & 0 \\
        \end{bmatrix}$ &  \\
    \end{tabular} \\
    \hfill\\
    \hline
    \hfill \\
    \begin{tabular}{l r}
    $6 \geq \Fbf^{14}_{33\to 2} =  \begin{bmatrix}
        0 &  0 &  0 &  0 &  1 &  0 &  1 &  0 & 0 \\
        1 &  1 &  0 &  1 &  0 &  1 &  0 &  0 & 1 \\
    \end{bmatrix}$ & $5 \geq \Fbf^{17}_{33\to 2} =  \begin{bmatrix}
        0 & 0 & 1 & 0 & 1 & 0 & 1 & 0 & 0 \\
        1 & 0 & 0 & 0 & 0 & 1 & 0 & 1 & 0 \\
    \end{bmatrix}$  \\
    \hfill \\
    $7 \geq \Fbf^{15}_{33\to 2} =  \begin{bmatrix}
        1 & 0 & 0 & 0 & 1 & 0 & 0 & 0 & 1 \\
        0 & 1 & 1 & 1 & 0 & 1 & 1 & 1 & 0 \\
    \end{bmatrix}$ & $9 \geq \Fbf^{18}_{33\to 2} =  \begin{bmatrix}
        0 & 0 & 2 & 0 & 1 & 0 & 2 & 0 & 0 \\
        2 & 1 & 0 & 1 & 0 & 1 & 0 & 1 & 0 \\
    \end{bmatrix}$  \\
    \hfill \\
     $5 \geq \Fbf^{16}_{33\to 2} = \begin{bmatrix}
        0 & 0 & 1 & 0 & 1 & 0 & 1 & 0 & 0 \\
        1 & 1 & 0 & 1 & 0 & 0 & 0 & 0 & 0 \\
    \end{bmatrix}$ & $11 \geq \Fbf^{19}_{33\to 2} =  \begin{bmatrix}
        0 & 0 & 1 & 1 & 0 & 0 & 2 & 0 & 2 \\
        2 & 1 & 0 & 0 & 1 & 2 & 0 & 1 & 0 \\
    \end{bmatrix}$  \\
    \end{tabular}
    \end{tabular}
    \caption{(Top) Complete set of thirteen facet inequalities for the multiaccess network polytope $\Cbb^{\MA(\sigarb{3,3}{2,2}{2})}_{}$ \cite{Bowles2015_nonclassicality_communication_networks}. (Bottom) Complete set of six facet inequalities for the multiaccess network polytope $ \Cbb^{\MA(\sigarb{3,3}{\{2,3\}}{2})}_{}$ as defined in Eq.\eqref{eq:ma_23/32_polytope}.}
    \label{Table:ma-33-dd-2-nonclassicality-witnesses}
\end{table*}

\newpage
\section{Facet Inequalities for Broadcast Networks}

\begin{table*}[h!]
    \centering
    \resizebox{\textwidth}{!}{%
    \begin{tabular}{c c}
        \begin{tabular}{l r}
            $2 \geq \Fbf_{\BC 3}^{a} = \begin{bmatrix}
            0 & 0 & 1\\0 & 0 & 0\\1 & 0 & 0\\0 & 0 & 0\\0 & 1 & 0\\1 & 0 & 0\\1 & 0 & 0\\1 & 0 & 0\\1 & 0 & 0 \\ 
            \end{bmatrix}$ &  $2 \geq \Fbf_{\BC 3}^{c}  = \begin{bmatrix}
                0 & 0 & 1\\0 & 0 & 0\\0 & 0 & 0\\0 & 0 & 0\\1 & 0 & 0\\0 & 1 & 0\\0 & 0 & 0\\0 & 1 & 0\\1 & 0 & 0
            \end{bmatrix}$ \\
        \hfill \\
            $2 \geq \Fbf_{\BC 3}^{b}= \begin{bmatrix}
                0 & 0 & 1\\0 & 0 & 1\\0 & 0 & 1\\0 & 1 & 0\\0 & 1 & 0\\0 & 1 & 0\\1 & 0 & 0\\1 & 0 & 0\\1 & 0 & 0
            \end{bmatrix}$ & $4 \geq \Fbf_{\BC 3}^{d}  = \begin{bmatrix}
                0 & 0 & 2\\0 & 0 & 1\\0 & 1 & 1\\0 & 2 & 0\\0 & 1 & 0\\0 & 1 & 1\\1 & 0 & 0\\2 & 0 & 0\\1 & 1 & 1\\
            \end{bmatrix}$ \\
        \end{tabular} & \begin{tabular}{l r}
            $2 \geq \Fbf_{\BC 4}^a = \begin{bmatrix}
                1 & 0 & 0 & 0 \\
                0 & 1 & 0 & 0 \\
                0 & 0 & 0 & 0 \\
                0 & 0 & 0 & 0 \\
                0 & 1 & 0 & 0 \\
                1 & 0 & 0 & 0 \\
                0 & 0 & 0 & 0 \\
                0 & 0 & 0 & 0 \\
                0 & 0 & 0 & 0 \\
                0 & 0 & 0 & 0 \\
                0 & 0 & 1 & 0 \\
                0 & 0 & 0 & 1 \\
                0 & 0 & 0 & 0 \\
                0 & 0 & 0 & 0 \\
                0 & 0 & 0 & 1 \\
                0 & 0 & 1 & 0 \\
            \end{bmatrix}$ & $8 \geq \Fbf_{\BC 4}^b = \begin{bmatrix}
                3 & 0 & 0 & 0\\0 & 3 & 0 & 0\\0 & 0 & 0 & 3\\1 & 1 & 0 & 2\\1 & 1 & 0 & 2\\1 & 1 & 0 & 2\\2 & 1 & 0 & 2\\1 & 3 & 0 & 2\\1 & 0 & 2 & 0\\0 & 1 & 2 & 0\\0 & 0 & 0 & 3\\1 & 1 & 0 & 2\\2 & 1 & 0 & 2\\1 & 2 & 0 & 2\\1 & 1 & 1 & 2\\2 & 2 & 1 & 2\\
            \end{bmatrix}$ \\
        \end{tabular}
    \end{tabular}%
    }
    \caption{Canonical facet inequalities for the classical broadcast network polytope $\Cbb^{\BC(\sigarb{3}{22}{33})}_{}$ labeled as $\BC 3$ and two facet inequalities for the classical broadcast network polytope $\Cbb^{\BC(\sigarb{4}{22}{44})}_{}$ labeled as $\BC 4$.}
    \label{Table:broadcast_X_22_YY_polytope}
\end{table*}

\newpage

\section{Facet Inequalities for Multipoint Networks}\label{section:multipoint_facet_inequalities}

\subsection{Interference Network}

\begin{table*}[h!]
    \centering
    \begin{tabular}{c c}
         $13 \geq \Fbf_{\IF}^{\times_0} = \begin{bmatrix}
             1 & 2 &  1 &  3 &  0 &  1 &  0 &  0 &  1 \\
             1 & 0 &  0 &  1 &  3 &  0 &  0 &  1 &  2 \\
             1 & 0 &  0 &  1 &  1 &  3 &  0 &  2 &  0 \\
             0 & 1 &  1 &  0 &  1 &  2 &  1 &  1 &  2 \\
             0 & 1 &  1 &  0 &  1 &  2 &  1 &  1 &  2 \\
             0 & 1 &  1 &  1 &  1 &  2 &  1 &  1 &  2 \\
             1 & 0 &  0 &  1 &  3 &  1 &  0 &  0 &  2 \\
             1 & 1 &  0 &  1 &  1 &  2 &  0 &  1 &  2 \\
             1 & 1 &  1 &  2 &  2 &  2 &  1 &  1 &  1 \\            
         \end{bmatrix}$ & $13 \geq  \Fbf_{\IF}^{\times_1} = \begin{bmatrix}
            3 &  0 &  1 &  0 &  2 &  1 &  1 &  0 &  1 \\
            0 &  3 &  0 &  2 &  0 &  0 &  1 &  0 &  1 \\
            0 &  0 &  4 &  0 &  2 &  1 &  2 &  0 &  0 \\
            2 &  1 &  2 &  0 &  2 &  1 &  0 &  1 &  1 \\
            2 &  1 &  2 &  1 &  1 &  2 &  0 &  1 &  1 \\
            2 &  1 &  2 &  0 &  0 &  3 &  0 &  2 &  0 \\
            2 &  1 &  2 &  0 &  2 &  1 &  0 &  0 &  1 \\
            2 &  1 &  2 &  0 &  2 &  1 &  0 &  0 &  1 \\
            2 &  2 &  3 &  1 &  1 &  2 &  1 &  1 &  0 \\  
         \end{bmatrix}$  \\
         & \\
         $14 \geq \Fbf_{\IF}^{+} = \begin{bmatrix}
             3 &  0 &  1 &  0 &  2 &  1 &  1 &  0 &  1 \\
             0 &  3 &  0 &  2 &  0 &  0 &  1 &  0 &  1 \\
             0 &  0 &  4 &  0 &  2 &  1 &  2 &  0 &  0 \\
             2 &  1 &  2 &  0 &  0 &  3 &  0 &  2 &  0 \\
             1 &  0 &  3 &  1 &  1 &  2 &  1 &  1 &  1 \\
             1 &  2 &  2 &  1 &  1 &  2 &  0 &  1 &  1 \\
             2 &  1 &  2 &  0 &  2 &  1 &  0 &  0 &  1 \\
             1 &  1 &  3 &  0 &  2 &  1 &  1 &  0 &  1 \\
             2 &  2 &  3 &  1 &  1 &  2 &  1 &  1 &  0 \\
         \end{bmatrix}$ & $11 \geq \Fbf_{\IF}^{-} = \begin{bmatrix}
             2 &  0 &  0 &  0 &  2 &  0 &  1 &  0 &  1 \\
             0 &  0 &  2 &  1 &  1 &  2 &  1 &  0 &  1 \\
             1 &  0 &  2 &  0 &  0 &  2 &  1 &  0 &  1 \\
             0 &  0 &  2 &  1 &  1 &  2 &  1 &  0 &  1 \\
             0 &  0 &  2 &  2 &  0 &  2 &  0 &  1 &  0 \\
             0 &  0 &  2 &  2 &  1 &  2 &  0 &  0 &  1 \\
             1 &  0 &  2 &  0 &  0 &  2 &  1 &  0 &  1 \\
             0 &  0 &  2 &  2 &  1 &  2 &  0 &  0 &  1 \\
             1 &  1 &  3 &  1 &  1 &  2 &  1 &  0 &  0 \\           
         \end{bmatrix}$ \\
         & \\
         $12 \geq \Fbf_{\IF}^{\gtrless} = \begin{bmatrix}
            2 &  0 &  0 &  0 &  3 &  0 &  0 &  0 &  1 \\
            0 &  1 &  1 &  1 &  0 &  3 &  1 &  1 &  0 \\
            1 &  1 &  1 &  0 &  0 &  3 &  0 &  1 &  0 \\
            0 &  2 &  1 &  0 &  2 &  2 &  1 &  0 &  1 \\
            1 &  1 &  1 &  0 &  0 &  3 &  1 &  1 &  0 \\
            0 &  3 &  1 &  0 &  1 &  3 &  1 &  0 &  0 \\
            1 &  2 &  1 &  1 &  1 &  2 &  0 &  0 &  1 \\
            0 &  1 &  2 &  2 &  0 &  1 &  1 &  1 &  0 \\
            1 &  2 &  2 &  1 &  2 &  2 &  1 &  0 &  0 \\    
         \end{bmatrix}$ & $13 \geq \Fbf_{\IF}^{\pi} = \begin{bmatrix}
             3 &  0 &  0 &  0 &  1 &  1 &  0 &  1 &  1 \\
             0 &  3 &  2 &  1 &  0 &  1 &  0 &  0 &  1 \\
             0 &  0 &  4 &  1 &  1 &  1 &  0 &  1 &  0 \\
             1 &  1 &  2 &  1 &  2 &  0 &  0 &  1 &  1 \\
             1 &  2 &  2 &  1 &  0 &  3 &  0 &  1 &  0 \\
             1 &  2 &  2 &  2 &  0 &  0 &  0 &  0 &  1 \\
             1 &  2 &  2 &  1 &  1 &  1 &  0 &  1 &  1 \\
             0 &  1 &  3 &  0 &  1 &  2 &  1 &  0 &  0 \\
             2 &  2 &  3 &  1 &  1 &  2 &  0 &  1 &  0 \\
         \end{bmatrix}$ \\
         & \\
         $13 \geq \Fbf_{\IF}^{\leftrightarrow} = \begin{bmatrix}
            3 &  0 &  0 &  0 &  2 &  0 &  1 &  0 &  1 \\
            1 &  1 &  2 &  2 &  0 &  0 &  1 &  0 &  1 \\
            0 &  1 &  3 &  0 &  1 &  1 &  2 &  0 &  0 \\
            0 &  3 &  0 &  2 &  0 &  0 &  0 &  1 &  1 \\
            1 &  1 &  2 &  0 &  2 &  0 &  0 &  1 &  1 \\
            1 &  0 &  3 &  1 &  0 &  1 &  0 &  2 &  0 \\
            0 &  0 &  4 &  1 &  1 &  0 &  1 &  1 &  0 \\
            1 &  2 &  2 &  0 &  0 &  2 &  1 &  1 &  0 \\
            2 &  2 &  3 &  1 &  1 &  1 &  1 &  1 &  0 \\          
         \end{bmatrix}$ & $13 \geq \Fbf_{\IF}^{\CV} = \begin{bmatrix}
             3 &  0 &  0 &  0 &  2 &  0 &  1 &  0 &  1 \\
             0 &  3 &  0 &  2 &  0 &  0 &  0 &  1 &  1 \\
             0 &  0 &  4 &  1 &  1 &  0 &  1 &  1 &  0 \\
             1 &  1 &  2 &  2 &  0 &  0 &  1 &  0 &  1 \\
             1 &  1 &  2 &  0 &  2 &  0 &  0 &  1 &  1 \\
             1 &  2 &  2 &  0 &  0 &  2 &  1 &  1 &  0 \\
             0 &  1 &  3 &  0 &  1 &  1 &  2 &  0 &  0 \\
             1 &  0 &  3 &  1 &  0 &  1 &  0 &  2 &  0 \\
             2 &  2 &  3 &  1 &  1 &  1 &  1 &  1 &  0 \\
         \end{bmatrix}$ \\
    \end{tabular}
    \caption{Derived nonclassicality witnesses for the interference network. Each inequality $(\gamma, \Gbf)$ is expressed as $\gamma \geq \Gbf$.}
    \label{tab:interference_network_games}
\end{table*}

\newpage
\subsection{Compressed Interference Network}

\begin{table*}[h!]
    \centering
    \begin{tabular}{c c}
         $12 \geq \Fbf_{\CIF}^{\times_0} = \begin{bmatrix}
             1 &  2 &  2 &  1 &  0 &  0 &  0 &  0 &  1 \\
             0 &  0 &  1 &  0 &  3 &  0 &  1 &  2 &  1 \\
             0 &  0 &  0 &  0 &  0 &  3 &  1 &  3 &  0 \\
             0 &  1 &  2 &  0 &  1 &  2 &  1 &  2 &  1 \\
             0 &  1 &  2 &  0 &  1 &  2 &  1 &  2 &  1 \\
             0 &  1 &  2 &  0 &  1 &  2 &  1 &  2 &  1 \\
             0 &  1 &  2 &  0 &  1 &  2 &  1 &  2 &  1 \\
             0 &  1 &  2 &  0 &  1 &  2 &  1 &  2 &  1 \\
             1 &  1 &  2 &  0 &  2 &  2 &  1 &  2 &  0 \\          
         \end{bmatrix}$ & $13 \geq  \Fbf_{\CIF}^{\times_1} = \begin{bmatrix}
             4 &  0 &  1 &  1 &  0 &  2 &  0 &  0 &  1 \\
             1 &  3 &  0 &  3 &  0 &  0 &  0 &  0 &  1 \\
             0 &  0 &  3 &  0 &  2 &  1 &  1 &  0 &  0 \\
             2 &  1 &  1 &  2 &  2 &  1 &  0 &  0 &  1 \\
             2 &  1 &  2 &  2 &  1 &  2 &  0 &  0 &  1 \\
             2 &  1 &  1 &  2 &  0 &  3 &  0 &  1 &  0 \\
             2 &  1 &  2 &  2 &  1 &  2 &  0 &  0 &  1 \\
             2 &  1 &  2 &  2 &  1 &  2 &  0 &  0 &  1 \\
             3 &  2 &  2 &  2 &  1 &  2 &  0 &  0 &  0 \\
         \end{bmatrix}$  \\
         & \\
         $13 \geq \Fbf_{\CIF}^{+} = \begin{bmatrix}
            4 &  0 &  1 &  1 &  0 &  2 &  0 &  0 &  1 \\
            1 &  3 &  0 &  3 &  0 &  0 &  0 &  0 &  1 \\
            0 &  0 &  3 &  0 &  2 &  1 &  1 &  0 &  0 \\
            2 &  1 &  1 &  2 &  0 &  3 &  0 &  1 &  0 \\
            2 &  1 &  2 &  2 &  1 &  2 &  0 &  0 &  1 \\
            2 &  1 &  2 &  2 &  1 &  2 &  0 &  0 &  1 \\
            2 &  1 &  2 &  2 &  1 &  2 &  0 &  0 &  1 \\
            2 &  1 &  2 &  2 &  1 &  2 &  0 &  0 &  1 \\
            3 &  2 &  2 &  2 &  1 &  2 &  0 &  0 &  0 \\
         \end{bmatrix}$ & $11 \geq \Fbf_{\CIF}^{-} = \begin{bmatrix}
             2 &  0 &  0 &  0 &  2 &  0 &  0 &  0 &  1 \\
             0 &  1 &  3 &  1 &  1 &  2 &  0 &  0 &  1 \\
             0 &  1 &  3 &  1 &  1 &  2 &  0 &  0 &  1 \\
             0 &  1 &  3 &  1 &  1 &  2 &  0 &  0 &  1 \\
             0 &  1 &  3 &  2 &  0 &  2 &  0 &  1 &  0 \\
             0 &  1 &  3 &  1 &  1 &  2 &  0 &  0 &  1 \\
             0 &  1 &  3 &  1 &  1 &  2 &  0 &  0 &  1 \\
             0 &  1 &  3 &  1 &  1 &  2 &  0 &  0 &  1 \\
             1 &  2 &  3 &  1 &  1 &  2 &  0 &  0 &  0 \\      
         \end{bmatrix}$ \\
         & \\
         $12 \geq \Fbf_{\CIF}^{\gtrless} = \begin{bmatrix}
             3 &  0 &  0 &  0 &  2 &  0 &  0 &  0 &  1 \\
             1 &  0 &  3 &  2 &  1 &  2 &  0 &  0 &  1 \\
             1 &  0 &  3 &  2 &  1 &  2 &  0 &  0 &  1 \\
             1 &  0 &  3 &  2 &  1 &  2 &  0 &  0 &  1 \\
             1 &  0 &  3 &  2 &  1 &  2 &  0 &  0 &  1 \\
             0 &  2 &  3 &  1 &  0 &  3 &  0 &  0 &  0 \\
             1 &  0 &  3 &  2 &  1 &  2 &  0 &  0 &  1 \\
             0 &  0 &  3 &  3 &  0 &  2 &  0 &  1 &  0 \\
             2 &  1 &  3 &  2 &  1 &  2 &  0 &  0 &  0 \\  
         \end{bmatrix}$ & $9 \geq \Fbf_{\CIF}^{\pi} = \begin{bmatrix}
             3 &  0 &  0 &  0 &  0 &  1 &  0 &  0 &  1 \\
             0 &  3 &  0 &  0 &  0 &  1 &  0 &  0 &  1 \\
             0 &  1 &  2 &  0 &  1 &  0 &  0 &  1 &  0 \\
             1 &  1 &  0 &  1 &  2 &  0 &  0 &  0 &  1 \\
             1 &  1 &  0 &  1 &  0 &  2 &  0 &  1 &  0 \\
             0 &  2 &  0 &  2 &  1 &  0 &  0 &  0 &  1 \\
             1 &  1 &  1 &  1 &  1 &  1 &  0 &  0 &  1 \\
             0 &  1 &  0 &  0 &  1 &  2 &  1 &  0 &  0 \\
             2 &  2 &  1 &  1 &  1 &  1 &  0 &  0 &  0 \\
         \end{bmatrix}$ \\
         & \\
         $9 \geq \Fbf_{\CIF}^{\leftrightarrow} = \begin{bmatrix}
             3 &  0 &  0 &  0 &  0 &  1 &  0 &  0 &  1 \\
             0 &  2 &  0 &  2 &  1 &  0 &  0 &  0 &  1 \\
             0 &  1 &  0 &  0 &  1 &  2 &  1 &  0 &  0 \\
             0 &  3 &  0 &  0 &  0 &  1 &  0 &  0 &  1 \\
             1 &  1 &  0 &  1 &  2 &  0 &  0 &  0 &  1 \\
             1 &  1 &  0 &  1 &  0 &  2 &  0 &  1 &  0 \\
             0 &  1 &  2 &  0 &  1 &  0 &  0 &  1 &  0 \\
             1 &  1 &  0 &  1 &  0 &  2 &  0 &  1 &  0 \\
             2 &  2 &  1 &  1 &  1 &  1 &  0 &  0 &  0 \\         
         \end{bmatrix}$ & $9 \geq \Fbf_{\CIF}^{\CV} = \begin{bmatrix}
             3 &  0 &  0 &  0 &  0 &  1 &  0 &  0 &  1 \\
             0 &  3 &  0 &  0 &  0 &  1 &  0 &  0 &  1 \\
             0 &  1 &  2 &  0 &  1 &  0 &  0 &  1 &  0 \\
             0 &  2 &  0 &  2 &  1 &  0 &  0 &  0 &  1 \\
             1 &  1 &  0 &  1 &  2 &  0 &  0 &  0 &  1 \\
             1 &  1 &  0 &  1 &  0 &  2 &  0 &  1 &  0 \\
             0 &  1 &  0 &  0 &  1 &  2 &  1 &  0 &  0 \\
             1 &  1 &  0 &  1 &  0 &  2 &  0 &  1 &  0 \\
             2 &  2 &  1 &  1 &  1 &  1 &  0 &  0 &  0 \\
         \end{bmatrix}$ \\
    \end{tabular}
    \caption{Derived nonclassicality witnesses for the compressed interference network. Each inequality $(\gamma, \Gbf)$ is expressed as $\gamma \geq \Gbf$.}
    \label{tab:compressed_interference_network_games}
\end{table*}

\newpage
\subsection{Butterfly Network}

\begin{table*}[h!]
    \centering
    \begin{tabular}{c c}
         $11 \geq \Fbf_{\BF}^{\times_0} = \begin{bmatrix}
             1 &  3 &  1 &  1 &  0 &  0 &  1 &  0 &  0 \\
             0 &  0 &  0 &  1 &  2 &  0 &  1 &  0 &  1 \\
             0 &  1 &  0 &  0 &  0 &  1 &  0 &  2 &  0 \\
             0 &  2 &  1 &  0 &  1 &  1 &  0 &  1 &  1 \\
             0 &  1 &  0 &  1 &  2 &  0 &  0 &  0 &  2 \\
             0 &  2 &  1 &  1 &  1 &  1 &  0 &  1 &  1 \\
             0 &  2 &  1 &  0 &  1 &  1 &  0 &  1 &  1 \\
             0 &  0 &  0 &  1 &  1 &  0 &  1 &  0 &  1 \\
             1 &  2 &  1 &  1 &  1 &  1 &  1 &  1 &  1 \\         
         \end{bmatrix}$ & $11 \geq  \Fbf_{\BF}^{\times_1} = \begin{bmatrix}
             2 &  0 &  1 &  1 &  0 &  0 &  0 &  0 &  0 \\
             0 &  1 &  1 &  3 &  0 &  0 &  0 &  0 &  0 \\
             0 &  0 &  1 &  1 &  0 &  0 &  2 &  0 &  0 \\
             1 &  0 &  1 &  0 &  1 &  1 &  0 &  1 &  0 \\
             0 &  1 &  1 &  1 &  0 &  2 &  0 &  1 &  0 \\
             0 &  0 &  1 &  1 &  0 &  2 &  0 &  2 &  0 \\
             1 &  1 &  1 &  1 &  1 &  1 &  0 &  0 &  1 \\
             0 &  1 &  0 &  2 &  1 &  1 &  1 &  0 &  1 \\
             1 &  1 &  1 &  2 &  1 &  1 &  1 &  1 &  1 \\
         \end{bmatrix}$  \\
         & \\
         $10 \geq \Fbf_{\BF}^{+} = \begin{bmatrix}
             1 &  0 &  1 &  0 &  1 &  1 &  1 &  1 &  0 \\
             0 &  2 &  0 &  1 &  0 &  0 &  0 &  0 &  2 \\
             1 &  1 &  2 &  0 &  1 &  1 &  1 &  1 &  1 \\
             0 &  1 &  1 &  0 &  0 &  2 &  1 &  1 &  0 \\
             0 &  1 &  0 &  0 &  0 &  0 &  0 &  0 &  2 \\
             0 &  1 &  2 &  0 &  0 &  1 &  1 &  1 &  1 \\
             1 &  0 &  1 &  0 &  1 &  1 &  1 &  1 &  1 \\
             0 &  1 &  0 &  1 &  0 &  0 &  0 &  0 &  2 \\
             1 &  1 &  2 &  0 &  1 &  1 &  1 &  1 &  1 \\
         \end{bmatrix}$ & $12 \geq \Fbf_{\BF}^{-} = \begin{bmatrix}
             3 &  0 &  0 &  0 &  2 &  0 &  0 &  1 &  1 \\
             1 &  0 &  0 &  1 &  1 &  0 &  0 &  1 &  1 \\
             2 &  0 &  0 &  0 &  1 &  0 &  0 &  0 &  1 \\
             2 &  0 &  1 &  0 &  0 &  2 &  0 &  1 &  0 \\
             1 &  0 &  1 &  2 &  0 &  1 &  0 &  2 &  0 \\
             2 &  1 &  1 &  1 &  1 &  2 &  0 &  1 &  0 \\
             1 &  0 &  0 &  0 &  0 &  1 &  0 &  1 &  1 \\
             0 &  0 &  1 &  1 &  0 &  2 &  0 &  1 &  1 \\
             2 &  1 &  2 &  1 &  1 &  2 &  1 &  1 &  0 \\    
         \end{bmatrix}$ \\
         & \\
         $14 \geq \Fbf_{\BF}^{\gtrless} = \begin{bmatrix}
             3 &  0 &  0 &  0 &  2 &  0 &  1 &  0 & 1 \\
             2 &  1 &  0 &  0 &  1 &  0 &  1 &  1 & 1 \\
             2 &  1 &  1 &  0 &  0 &  1 &  0 &  1 & 0 \\
             1 &  1 &  2 &  0 &  0 &  2 &  0 &  1 & 1 \\
             0 &  2 &  2 &  0 &  0 &  2 &  0 &  1 & 1 \\
             0 &  3 &  2 &  0 &  0 &  3 &  0 &  0 & 0 \\
             1 &  1 &  2 &  0 &  1 &  1 &  1 &  1 & 0 \\
             1 &  2 &  2 &  1 &  1 &  1 &  1 &  1 & 0 \\
             2 &  2 &  3 &  1 &  1 &  2 &  1 &  1 & 0 \\
         \end{bmatrix}$ & $20 \geq \Fbf_{\BF}^{\pi} = \begin{bmatrix}
             2 &  1 &  0 &  1 &  3 &  0 &  1 &  1 &  2 \\
             1 &  5 &  2 &  0 &  0 &  2 &  1 &  1 &  0 \\
             0 &  3 &  3 &  1 &  3 &  3 &  0 &  2 &  0 \\
             1 &  2 &  0 &  1 &  4 &  0 &  1 &  0 &  2 \\
             0 &  3 &  1 &  1 &  1 &  3 &  1 &  1 &  0 \\
             0 &  3 &  3 &  2 &  1 &  2 &  0 &  1 &  0 \\
             1 &  0 &  0 &  1 &  3 &  0 &  1 &  1 &  1 \\
             0 &  3 &  1 &  0 &  0 &  2 &  2 &  1 &  1 \\
             2 &  3 &  3 &  1 &  3 &  3 &  1 &  2 &  0 \\
         \end{bmatrix}$ \\
         & \\
         $12 \geq \Fbf_{\BF}^{\leftrightarrow} = \begin{bmatrix}
             2 &  0 &  1 &  0 &  0 &  1 &  0 &  1 &  0 \\
             0 &  0 &  1 &  2 &  0 &  1 &  0 &  1 &  0 \\
             1 &  0 &  1 &  1 &  1 &  1 &  1 &  1 &  1 \\
             0 &  2 &  0 &  0 &  0 &  1 &  1 &  0 &  0 \\
             0 &  0 &  1 &  0 &  2 &  0 &  1 &  0 &  0 \\
             0 &  1 &  1 &  1 &  1 &  1 &  0 &  1 &  1 \\
             0 &  0 &  3 &  0 &  0 &  2 &  0 &  1 &  0 \\
             0 &  0 &  2 &  0 &  0 &  3 &  0 &  1 &  0 \\
             1 &  1 &  2 &  1 &  1 &  2 &  1 &  1 &  1 \\     
         \end{bmatrix}$ & $18 \geq \Fbf_{\BF}^{\CV} = \begin{bmatrix}
             3 &  0 &  1 &  0 &  3 &  1 &  0 &  0 &  0 \\
             1 &  3 &  0 &  0 &  3 &  1 &  0 &  0 &  0 \\
             0 &  0 &  3 &  0 &  4 &  1 &  0 &  0 &  1 \\
             1 &  0 &  1 &  3 &  2 &  1 &  0 &  0 &  0 \\
             3 &  1 &  1 &  1 &  3 &  0 &  0 &  0 &  0 \\
             3 &  1 &  2 &  0 &  0 &  2 &  0 &  0 &  1 \\
             0 &  0 &  1 &  0 &  3 &  1 &  2 &  1 &  1 \\
             2 &  0 &  1 &  0 &  0 &  1 &  2 &  1 &  1 \\
             4 &  1 &  2 &  1 &  4 &  1 &  1 &  0 &  0 \\
         \end{bmatrix}$ \\
    \end{tabular}
    \caption{Derived nonclassicality witnesses for the butterfly network. Each inequality $(\gamma, \Gbf)$ is expressed as $\gamma \geq \Gbf$.}
    \label{tab:butterfly_network_games}
\end{table*}

\newpage
\subsection{Hourglasss Network}

\begin{table*}[h!]
    \centering
    \begin{tabular}{c c}

     $16 \geq \Fbf_{\HG}^{\times_0} = \begin{bmatrix}
        2 & 2 & 2 & 2 & 1 & 0 & 2 & 0 & 2 \\
        1 & 1 & 1 & 1 & 2 & 0 & 1 & 1 & 2 \\
        0 & 0 & 0 & 0 & 0 & 2 & 0 & 2 & 0 \\
        1 & 1 & 1 & 1 & 1 & 1 & 2 & 1 & 2 \\
        1 & 2 & 1 & 1 & 2 & 0 & 1 & 0 & 3 \\
        1 & 2 & 1 & 1 & 1 & 1 & 1 & 1 & 2 \\
        0 & 0 & 1 & 1 & 2 & 0 & 1 & 0 & 2 \\
        1 & 1 & 1 & 2 & 2 & 1 & 2 & 1 & 2 \\
        1 & 2 & 2 & 2 & 2 & 1 & 2 & 1 & 2 \\
    \end{bmatrix}$ &
    $13 \geq \Fbf^{\times_1}_{\HG} = \begin{bmatrix}
        3 & 0 & 0 & 1 & 1 & 0 & 0 & 0 & 1 \\
        0 & 3 & 0 & 2 & 0 & 0 & 0 & 0 & 1 \\
        1 & 1 & 2 & 1 & 0 & 0 & 1 & 1 & 1 \\
        2 & 1 & 0 & 0 & 1 & 1 & 0 & 1 & 1 \\
        0 & 2 & 0 & 1 & 0 & 1 & 0 & 1 & 1 \\
        1 & 1 & 1 & 0 & 0 & 2 & 1 & 2 & 0 \\
        2 & 0 & 0 & 1 & 1 & 1 & 0 & 1 & 1 \\
        0 & 2 & 0 & 1 & 1 & 0 & 0 & 1 & 1 \\
        2 & 2 & 2 & 1 & 1 & 1 & 1 & 1 & 1 \\
    \end{bmatrix}$ \\
    \hfill \\
    $13 \geq \Fbf^{\leftrightarrow}_{\HG} = \begin{bmatrix}
        2 & 0 & 0 & 0 & 1 & 0 &  0 &  0 &  1 \\
        0 & 0 & 0 & 2 & 0 & 0 &  0 &  0 &  1 \\
        0 & 1 & 0 & 0 & 1 & 0 &  2 &  2 &  1 \\
        0 & 2 & 0 & 1 & 0 & 0 &  0 &  0 &  1 \\
        0 & 0 & 0 & 0 & 2 & 0 &  0 &  0 &  1 \\
        1 & 0 & 0 & 1 & 0 & 0 &  2 &  2 &  1 \\
        0 & 0 & 2 & 1 & 1 & 2 &  0 &  0 &  1 \\
        1 & 1 & 2 & 0 & 0 & 2 &  0 &  0 &  1 \\
        1 & 1 & 2 & 1 & 1 & 2 &  2 &  2 &  0 \\
    \end{bmatrix}$ & 
    $10 \geq \Fbf^{+}_{\HG} = \begin{bmatrix}
        2 & 0 & 0 & 0 & 0 & 2 & 0 & 1 & 0 \\
        0 & 2 & 0 & 2 & 1 & 0 & 0 & 0 & 1 \\
        0 & 1 & 2 & 1 & 2 & 1 & 1 & 0 & 0 \\
        2 & 0 & 0 & 0 & 0 & 2 & 0 & 1 & 0 \\
        0 & 2 & 0 & 1 & 1 & 1 & 0 & 0 & 1 \\
        0 & 1 & 2 & 1 & 1 & 1 & 1 & 0 & 0 \\
        2 & 0 & 0 & 0 & 0 & 2 & 0 & 1 & 0 \\
        0 & 1 & 0 & 2 & 1 & 0 & 0 & 0 & 1 \\
        1 & 1 & 2 & 1 & 2 & 1 & 1 & 0 & 0 \\
    \end{bmatrix}$ \\
    \hfill \\
    $9 \geq \Fbf^{\gtrless}_{\HG} = \begin{bmatrix}
        2 &  0 &  0 & 0 & 2 & 0 & 0 & 0 & 1 \\
        0 &  0 &  1 & 1 & 1 & 1 & 0 & 1 & 0 \\
        1 &  0 &  0 & 0 & 1 & 2 & 1 & 0 & 0 \\
        1 &  0 &  0 & 0 & 0 & 2 & 0 & 0 & 1 \\
        0 &  0 &  1 & 1 & 0 & 2 & 0 & 1 & 0 \\
        1 &  1 &  1 & 0 & 1 & 2 & 1 & 0 & 0 \\
        1 &  0 &  0 & 0 & 1 & 1 & 0 & 0 & 1 \\
        0 &  0 &  1 & 2 & 0 & 1 & 0 & 1 & 0 \\
        1 &  1 &  1 & 1 & 1 & 2 & 1 & 0 & 0 \\
    \end{bmatrix}$ &
    $10 \geq \Fbf^{\pi}_{\HG} = \begin{bmatrix}
        2 & 0 & 0 & 1 & 1 & 1 & 0 & 0 & 1 \\
        0 & 2 & 0 & 0 & 0 & 2 & 1 & 0 & 0 \\
        1 & 0 & 2 & 1 & 1 & 1 & 0 & 1 & 0 \\
        1 & 0 & 0 & 1 & 2 & 0 & 0 & 0 & 1 \\
        0 & 2 & 0 & 0 & 0 & 2 & 1 & 0 & 0 \\
        1 & 1 & 2 & 2 & 1 & 1 & 0 & 1 & 0 \\
        1 & 0 & 0 & 1 & 2 & 0 & 0 & 0 & 1 \\
        0 & 2 & 0 & 0 & 0 & 2 & 1 & 0 & 0 \\
        1 & 1 & 2 & 2 & 1 & 1 & 0 & 1 & 0 \\
    \end{bmatrix}$ \\
    \hfill \\
    $8 \geq \Fbf^-_{\HG} = \begin{bmatrix}
        2 & 0 &  0 &  0  &  1 &  0 &  0 &  0 &  1 \\
        1 & 0 &  0 &  1  &  0 &  1 &  0 &  0 &  1 \\
        1 & 0 &  0 &  0  &  0 &  1 &  0 &  0 &  1 \\
        1 & 0 &  0 &  1  &  0 &  1 &  0 &  0 &  1 \\
        0 & 0 &  1 &  2  &  0 &  1 &  0 &  1 &  0 \\
        0 & 0 &  1 &  1  &  0 &  2 &  0 &  0 &  0 \\
        1 & 0 &  0 &  0  &  0 &  1 &  0 &  0 &  1 \\
        0 & 0 &  1 &  1  &  0 &  2 &  0 &  0 &  0 \\
        1 & 1 &  1 &  1  &  0 &  2 &  1 &  0 &  0 \\
    \end{bmatrix}$ &
    $13 \geq \Fbf^{\CV}_{\HG} = \begin{bmatrix}
        2 & 0 & 0 & 0 & 1 &  0 &  0 &  0 &  1 \\
        0 & 2 & 0 & 1 & 0 &  0 &  0 &  0 &  1 \\
        0 & 0 & 2 & 1 & 1 &  2 &  0 &  0 &  1 \\
        0 & 0 & 0 & 2 & 0 &  0 &  0 &  0 &  1 \\
        0 & 0 & 0 & 0 & 2 &  0 &  0 &  0 &  1 \\
        1 & 1 & 2 & 0 & 0 &  2 &  0 &  0 &  1 \\
        0 & 1 & 0 & 0 & 1 &  0 &  2 &  2 &  1 \\
        1 & 0 & 0 & 1 & 0 &  0 &  2 &  2 &  1 \\
        1 & 1 & 2 & 1 & 1 &  2 &  2 &  2 &  0 \\
    \end{bmatrix}$ \\
    \hfill \\
    \end{tabular}
    \caption{ Facet inequalities for the hourglass (HG) network. Each inequality $(\gamma, \Fbf)$ is presented as $\gamma \geq \Fbf$. }
\end{table*}

\newpage
\subsection{Multiaccess Network}

\begin{table*}[h!]
    \centering
    \begin{tabular}{c c}
     $10 \geq \Fbf_{\MA}^{\times_0} = \begin{bmatrix}
            0 & 2 & 1 & 2 & 0 & 0 & 0 & 0 & 1 \\
            0 & 0 & 0 & 0 & 3 & 0 & 0 & 0 & 2 \\
            0 & 0 & 0 & 0 & 1 & 2 & 0 & 2 & 0 \\
            0 & 0 & 0 & 0 & 3 & 0 & 0 & 0 & 2 \\
            0 & 0 & 0 & 0 & 3 & 0 & 0 & 0 & 2 \\
            0 & 0 & 0 & 0 & 3 & 0 & 0 & 0 & 2 \\
            0 & 0 & 0 & 0 & 3 & 0 & 0 & 0 & 2 \\
            0 & 0 & 0 & 0 & 3 & 0 & 0 & 0 & 2 \\
            0 & 1 & 1 & 1 & 2 & 1 & 1 & 1 & 1 \\
    \end{bmatrix}$ &
    $10 \geq \Fbf^{\times_1}_{\MA} = \begin{bmatrix}
            2 & 0 & 0 & 0 & 2 & 0 & 0 & 0 & 2 \\
            0 & 2 & 0 & 2 & 0 & 0 & 0 & 0 & 2 \\
            0 & 0 & 2 & 0 & 2 & 0 & 2 & 0 & 0 \\
            2 & 0 & 0 & 0 & 2 & 0 & 0 & 0 & 2 \\
            2 & 0 & 0 & 0 & 2 & 0 & 0 & 0 & 2 \\
            2 & 0 & 0 & 0 & 0 & 2 & 0 & 2 & 0 \\
            2 & 0 & 0 & 0 & 2 & 0 & 0 & 0 & 2 \\
            2 & 0 & 0 & 0 & 2 & 0 & 0 & 0 & 2 \\
            1 & 1 & 1 & 1 & 1 & 1 & 1 & 1 & 1 \\
    \end{bmatrix}$ \\
    \hfill \\
    $10 \geq \Fbf^{\leftrightarrow}_{\MA} = \begin{bmatrix}
            2 & 0 & 0 & 0 & 2 & 0 & 0 & 0 & 2 \\
            0 & 2 & 0 & 2 & 0 & 0 & 0 & 0 & 2 \\
            0 & 2 & 0 & 0 & 0 & 2 & 2 & 0 & 0 \\
            0 & 2 & 0 & 2 & 0 & 0 & 0 & 0 & 2 \\
            2 & 0 & 0 & 0 & 2 & 0 & 0 & 0 & 2 \\
            2 & 0 & 0 & 0 & 0 & 2 & 0 & 2 & 0 \\
            0 & 0 & 2 & 2 & 0 & 0 & 0 & 2 & 0 \\
            2 & 0 & 0 & 0 & 0 & 2 & 0 & 2 & 0 \\
            1 & 1 & 1 & 1 & 1 & 1 & 1 & 1 & 1 \\
    \end{bmatrix}$ & 
    $10 \geq \Fbf^{+}_{\MA} = \begin{bmatrix}
        2 & 0 & 0 & 0 & 2 & 0 & 0 & 0 & 2 \\
        0 & 2 & 0 & 2 & 0 & 0 & 0 & 0 & 2 \\
        0 & 0 & 2 & 0 & 2 & 0 & 2 & 0 & 0 \\
        2 & 0 & 0 & 0 & 0 & 2 & 0 & 2 & 0 \\
        2 & 0 & 0 & 0 & 2 & 0 & 0 & 0 & 2 \\
        2 & 0 & 0 & 0 & 2 & 0 & 0 & 0 & 2 \\
        2 & 0 & 0 & 0 & 2 & 0 & 0 & 0 & 2 \\
        2 & 0 & 0 & 0 & 2 & 0 & 0 & 0 & 2 \\
        1 & 1 & 1 & 1 & 1 & 1 & 1 & 1 & 1 \\
    \end{bmatrix}$ \\
    \hfill \\
    $9 \geq \Fbf^{\gtrless}_{\MA} = \begin{bmatrix}
        2 & 0 & 0 & 0 & 2 & 0 & 0 & 0 & 1 \\
        1 & 1 & 0 & 1 & 0 & 1 & 0 & 1 & 1 \\
        1 & 1 & 0 & 1 & 0 & 1 & 0 & 1 & 1 \\
        1 & 1 & 0 & 1 & 0 & 1 & 0 & 1 & 1 \\
        1 & 1 & 0 & 1 & 0 & 1 & 0 & 1 & 1 \\
        0 & 2 & 0 & 0 & 0 & 2 & 1 & 0 & 0 \\
        1 & 1 & 0 & 1 & 0 & 1 & 0 & 1 & 1 \\
        0 & 0 & 1 & 2 & 0 & 0 & 0 & 2 & 0 \\
        1 & 1 & 1 & 1 & 1 & 1 & 1 & 1 & 0 \\
    \end{bmatrix}$ &
    $10 \geq \Fbf^{\pi}_{\MA} = \begin{bmatrix}
        2 & 0 & 0 & 0 & 2 & 0 & 0 & 0 & 2 \\
        0 & 2 & 0 & 1 & 0 & 0 & 1 & 0 & 2 \\
        0 & 0 & 2 & 1 & 1 & 0 & 1 & 1 & 0 \\
        1 & 0 & 0 & 0 & 2 & 0 & 1 & 0 & 2 \\
        1 & 1 & 0 & 0 & 0 & 2 & 1 & 1 & 0 \\
        0 & 2 & 0 & 2 & 0 & 0 & 0 & 0 & 2 \\
        1 & 0 & 0 & 0 & 2 & 0 & 1 & 0 & 2 \\
        0 & 1 & 1 & 0 & 1 & 1 & 2 & 0 & 1 \\
        1 & 1 & 1 & 1 & 1 & 1 & 1 & 1 & 1 \\
    \end{bmatrix}$ \\
    \hfill \\
    $9 \geq \Fbf^-_{\MA} = \begin{bmatrix}
        2 & 0 & 0 & 0 & 2 & 0 & 0 & 0 & 1 \\
        1 & 0 & 1 & 1 & 0 & 2 & 0 & 0 & 1 \\
        1 & 0 & 1 & 1 & 0 & 2 & 0 & 0 & 1 \\
        1 & 0 & 1 & 1 & 0 & 2 & 0 & 0 & 1 \\
        0 & 0 & 1 & 2 & 0 & 1 & 0 & 1 & 0 \\
        1 & 0 & 1 & 1 & 0 & 2 & 0 & 0 & 1 \\
        1 & 0 & 1 & 1 & 0 & 2 & 0 & 0 & 1 \\
        1 & 0 & 1 & 1 & 0 & 2 & 0 & 0 & 1 \\
        1 & 1 & 1 & 1 & 1 & 2 & 1 & 0 & 0 \\
    \end{bmatrix}$ &
    $10 \geq \Fbf^{\CV}_{\MA} = \begin{bmatrix}
        2 & 0 & 0 & 0 & 2 & 0 & 0 & 0 & 2 \\
        0 & 2 & 0 & 2 & 0 & 0 & 0 & 0 & 2 \\
        0 & 0 & 2 & 2 & 0 & 0 & 0 & 2 & 0 \\
        0 & 2 & 0 & 2 & 0 & 0 & 0 & 0 & 2 \\
        2 & 0 & 0 & 0 & 2 & 0 & 0 & 0 & 2 \\
        2 & 0 & 0 & 0 & 0 & 2 & 0 & 2 & 0 \\
        0 & 2 & 0 & 0 & 0 & 2 & 2 & 0 & 0 \\
        2 & 0 & 0 & 0 & 0 & 2 & 0 & 2 & 0 \\
        1 & 1 & 1 & 1 & 1 & 1 & 1 & 1 & 1 \\
    \end{bmatrix}$ \\
    \hfill \\
    \end{tabular}
    \caption{ Facet inequalities for multiaccess network $\text{MA}(\sigarb{3,3}{2,2}{9})$. Each inequality $(\gamma, \Fbf)$ is presented as $\gamma \geq \Fbf$. }
\end{table*}

\newpage
\subsection{Broadcast Network}

\begin{table*}[h!]
    \centering
    \begin{tabular}{c c}
     $5 \geq \Fbf_{\BC}^{\times_0} = \begin{bmatrix}
        1 & 0 & 0 & 1 & 0 & 0 & 0 & 0 & 0 \\
        0 & 0 & 0 & 0 & 2 & 0 & 0 & 0 & 0 \\
        0 & 0 & 0 & 0 & 0 & 2 & 0 & 0 & 0 \\
        1 & 0 & 0 & 0 & 0 & 0 & 0 & 0 & 1 \\
        0 & 0 & 0 & 0 & 1 & 0 & 0 & 0 & 1 \\
        0 & 0 & 0 & 0 & 0 & 2 & 0 & 0 & 1 \\
        1 & 0 & 0 & 0 & 1 & 1 & 0 & 0 & 0 \\
        1 & 0 & 0 & 0 & 1 & 1 & 0 & 0 & 0 \\
        1 & 0 & 0 & 0 & 1 & 2 & 0 & 0 & 0 \\
    \end{bmatrix}$ &
    $5 \geq \Fbf^{\times_1}_{\BC} = \begin{bmatrix}
        2 & 0 & 0 & 0 & 0 & 0 & 0 & 0 & 0 \\
        0 & 2 & 0 & 0 & 0 & 0 & 0 & 0 & 0 \\
        0 & 0 & 2 & 0 & 0 & 0 & 0 & 0 & 0 \\
        1 & 0 & 0 & 0 & 0 & 1 & 0 & 0 & 0 \\
        0 & 1 & 0 & 0 & 0 & 1 & 0 & 0 & 0 \\
        0 & 0 & 2 & 0 & 0 & 1 & 0 & 0 & 0 \\
        1 & 1 & 1 & 0 & 0 & 0 & 0 & 0 & 0 \\
        1 & 1 & 1 & 0 & 0 & 0 & 0 & 0 & 0 \\
        1 & 1 & 2 & 0 & 0 & 0 & 0 & 0 & 0 \\
    \end{bmatrix}$ \\
    \hfill \\
    $6 \geq \Fbf^{\leftrightarrow}_{\BC} = \begin{bmatrix}
        2 & 0 & 0 & 0 & 0 & 0 & 0 & 0 & 0 \\
        1 & 0 & 0 & 1 & 0 & 0 & 0 & 0 & 0 \\
        1 & 1 & 2 & 0 & 0 & 0 & 0 & 0 & 0 \\
        0 & 2 & 0 & 0 & 0 & 0 & 0 & 0 & 0 \\
        0 & 1 & 0 & 0 & 1 & 0 & 0 & 0 & 0 \\
        1 & 1 & 2 & 0 & 0 & 0 & 0 & 0 & 0 \\
        0 & 0 & 3 & 0 & 0 & 0 & 0 & 0 & 0 \\
        0 & 0 & 3 & 0 & 0 & 0 & 0 & 0 & 0 \\
        1 & 1 & 3 & 0 & 0 & 0 & 0 & 0 & 0 \\
    \end{bmatrix}$ & 
    $6 \geq \Fbf^{+}_{\BC} = \begin{bmatrix}
        2 & 0 & 0 & 0 & 0 & 0 & 0 & 0 & 0 \\
        0 & 1 & 0 & 1 & 0 & 0 & 0 & 0 & 0 \\
        0 & 0 & 3 & 0 & 0 & 0 & 0 & 0 & 0 \\
        1 & 0 & 0 & 0 & 0 & 1 & 0 & 0 & 0 \\
        0 & 1 & 0 & 0 & 0 & 0 & 0 & 0 & 1 \\
        0 & 0 & 3 & 0 & 0 & 0 & 0 & 0 & 0 \\
        1 & 1 & 2 & 0 & 0 & 0 & 0 & 0 & 0 \\
        1 & 1 & 2 & 0 & 0 & 0 & 0 & 0 & 0 \\
        1 & 1 & 3 & 0 & 0 & 0 & 0 & 0 & 0 \\
    \end{bmatrix}$ \\
    \hfill \\
    $4 \geq \Fbf^{\gtrless}_{\BC} = \begin{bmatrix}
        1 & 0 & 0 & 0 & 0 & 0 & 0 & 0 & 0 \\
        0 & 0 & 0 & 2 & 0 & 0 & 0 & 0 & 0 \\
        0 & 2 & 0 & 0 & 0 & 0 & 0 & 0 & 0 \\
        0 & 0 & 0 & 0 & 0 & 0 & 0 & 0 & 1 \\
        0 & 0 & 0 & 2 & 0 & 0 & 0 & 0 & 0 \\
        0 & 2 & 0 & 0 & 0 & 0 & 0 & 0 & 0 \\
        0 & 1 & 0 & 1 & 0 & 0 & 0 & 0 & 0 \\
        0 & 1 & 0 & 2 & 0 & 0 & 0 & 0 & 0 \\
        0 & 2 & 0 & 1 & 0 & 0 & 0 & 0 & 0 \\
    \end{bmatrix}$ &
    $6 \geq \Fbf^{\pi}_{\BC} = \begin{bmatrix}
        2 & 0 & 0 & 0 & 0 & 0 & 0 & 0 & 0 \\
        0 & 2 & 0 & 0 & 0 & 0 & 0 & 0 & 0 \\
        0 & 0 & 3 & 0 & 0 & 0 & 0 & 0 & 0 \\
        1 & 0 & 0 & 0 & 1 & 0 & 0 & 0 & 0 \\
        0 & 1 & 0 & 0 & 0 & 1 & 0 & 0 & 0 \\
        0 & 0 & 3 & 0 & 0 & 0 & 0 & 0 & 0 \\
        1 & 1 & 2 & 0 & 0 & 0 & 0 & 0 & 0 \\
        1 & 1 & 2 & 0 & 0 & 0 & 0 & 0 & 0 \\
        1 & 1 & 3 & 0 & 0 & 0 & 0 & 0 & 0 \\
    \end{bmatrix}$ \\
    \hfill \\
    $5 \geq \Fbf^-_{\BC} = \begin{bmatrix}
        1 & 0 & 0 & 0 & 1 & 0 & 0 & 0 & 0 \\
        0 & 0 & 0 & 0 & 0 & 0 & 0 & 0 & 1 \\
        0 & 0 & 2 & 0 & 0 & 0 & 0 & 0 & 0 \\
        0 & 0 & 0 & 0 & 0 & 0 & 0 & 0 & 1 \\
        0 & 2 & 0 & 0 & 0 & 0 & 0 & 0 & 0 \\
        0 & 0 & 2 & 0 & 0 & 0 & 0 & 0 & 0 \\
        0 & 0 & 2 & 0 & 0 & 0 & 0 & 0 & 0 \\
        0 & 0 & 2 & 0 & 0 & 0 & 0 & 0 & 0 \\
        1 & 1 & 2 & 0 & 0 & 0 & 0 & 0 & 0 \\
    \end{bmatrix}$ &
    $6 \geq \Fbf^{\CV}_{\BC} = \begin{bmatrix}
        2 & 0 & 0 & 0 & 0 & 0 & 0 & 0 & 0 \\
        0 & 2 & 0 & 0 & 0 & 0 & 0 & 0 & 0 \\
        0 & 0 & 3 & 0 & 0 & 0 & 0 & 0 & 0 \\
        1 & 0 & 0 & 1 & 0 & 0 & 0 & 0 & 0 \\
        0 & 1 & 0 & 0 & 1 & 0 & 0 & 0 & 0 \\
        0 & 0 & 3 & 0 & 0 & 0 & 0 & 0 & 0 \\
        1 & 1 & 2 & 0 & 0 & 0 & 0 & 0 & 0 \\
        1 & 1 & 2 & 0 & 0 & 0 & 0 & 0 & 0 \\
        1 & 1 & 3 & 0 & 0 & 0 & 0 & 0 & 0 \\
    \end{bmatrix}$ \\
    \hfill \\
    \end{tabular}
    \caption{ Facet inequalities for the broadcast network $\BC(\sigarb{9}{2,2}{3,3})$. Each inequality $(\gamma, \Fbf)$ is presented as $\gamma \geq \Fbf$. }
\end{table*}

\end{document}